\documentclass[11pt,a4paper]{article}
\usepackage{alltt}
\usepackage{amscd}
\usepackage{amsfonts}
\usepackage{amsmath}
\usepackage{amssymb,latexsym,cite}
\usepackage{amsthm} 
\usepackage[english]{babel}
\usepackage{booktabs}
\usepackage{bbm}
\usepackage{bibentry}
\usepackage[makeroom]{cancel}
\usepackage{caption}
\usepackage{chngcntr}

\usepackage[pdfencoding=auto, psdextra]{hyperref}
\hypersetup{urlcolor=blue, colorlinks=false} 
\usepackage{cleveref}
    \crefformat{section}{\S#2#1#3}
    \crefformat{subsection}{\S#2#1#3}
    \crefformat{subsubsection}{\S#2#1#3}
    \crefrangeformat{section}{\S\S#3#1#4 to~#5#2#6}
    \crefmultiformat{section}{\S\S#2#1#3}{ and~#2#1#3}{, #2#1#3}{ and~#2#1#3}

\usepackage{changepage}
\usepackage{chngcntr}
\usepackage{color}
\usepackage{enumitem}
\usepackage{etoolbox}
\usepackage{faktor}
\usepackage{fancyhdr}
\usepackage[T1]{fontenc}
\usepackage{geometry}
\usepackage{graphicx}
\graphicspath{ {.} } 

\definecolor{dred}{rgb}{1,.5,.5}

\usepackage{imakeidx}
\usepackage[utf8]{inputenc}
\usepackage{lmodern}
\usepackage{mathdots}
\usepackage{mathrsfs} 
\usepackage{mathtools}
\usepackage{microtype}
\usepackage{multirow}

\usepackage{pgffor}  
\usepackage{pdflscape}
\usepackage{pgfplots}
    \pgfplotsset{compat=1.16}
\usepackage{simplewick}
\usepackage{siunitx}
\usepackage{shuffle}
\usepackage{slashed}
\usepackage{stmaryrd} 
\usepackage{tabu}
\usepackage{tabularx}
\usepackage[most]{tcolorbox}
\usepackage{tensor}
\usepackage[normalem]{ulem}
\usepackage{verbatim}  
\usepackage{wrapfig} 
\usepackage{xcolor}
\usepackage{xfrac}
\usepackage{xspace}
\usepackage[matrix,arrow,color]{xy}

\usepackage{tikz}
\usepackage{float} 

\pgfdeclarelayer{nodelayer}
\pgfdeclarelayer{edgelayer}
\pgfsetlayers{edgelayer,nodelayer,main}
\tikzstyle{dashed edge}=[<->, dashed]
\tikzstyle{blue pointer}=[->, draw=blue]
\tikzstyle{squiggle}=[->, decorate, decoration={snake,amplitude=.4mm,segment length=2mm, post length=0mm,pre length=0mm}]
\tikzstyle{dashed ->}=[->, dashed]
\tikzstyle{block}= [rounded corners, draw, thick, rectangle, minimum height = 1em, minimum width = 3em]

\tikzstyle{none}=[inner sep=0pt]
\usepackage{tikz}
\usepackage{tikz-cd}
\usetikzlibrary{decorations.markings}
\usepackage{tkz-euclide}
\usepackage[tikz]{bclogo}                   

\usetikzlibrary{arrows,shapes}
\usetikzlibrary{shapes.misc}
\usetikzlibrary{trees}
\usetikzlibrary{decorations}
\usetikzlibrary{matrix,arrows}              
\usetikzlibrary{positioning}                
\usetikzlibrary{calc,through}               
\usetikzlibrary{decorations.pathreplacing}  

\usetikzlibrary{decorations.pathmorphing}   
\usetikzlibrary{decorations.markings}
\usetikzlibrary{intersections}              
\usetikzlibrary{patterns,math}

\tikzset{
    A/.style={decorate, dotted, ultra thick,decoration={snake, segment length=5mm, amplitude=1.5mm}},
    scal/.style={postaction = {decorate}, densely dotted, thick, decoration={markings,mark=at position .8 with {\arrow{>}}}},
    scalreverse/.style={postaction = {decorate}, densely dotted, thick, decoration={markings,mark=at position .8 with {\arrow{<}}}},
    scalnoarrow/.style={postaction = {decorate}, densely dotted},
    sfield/.style={postaction = {decorate},line width = 2pt, dash pattern={on 5pt off 2pt on 1pt off 3pt}},
    Aplus/.style={decorate, line width= 0.8pt, decoration={snake, segment length=2mm, amplitude=1mm}, draw},
    vector/.style={decorate, decoration={snake}, draw}, 
    scalplus/.style={postaction={decorate}, line width= 0.8pt, decoration={markings,mark=at position .7 with {\arrow{<}}}},
    scalplusreverse/.style={postaction={decorate}, line width= 0.8pt, decoration={markings,mark=at position .7 with {\arrow{>}}}},
    scalplusnoarrow/.style={postaction={decorate}, line width= 0.8pt},
    antisfield/.style={line width = 2pt},
    cross/.style={cross out, draw=black, minimum size=2*(#1-\pgflinewidth), inner sep=0pt, outer sep=0pt},
    cross/.default={1pt}
}

\def\mathcolor#1#{\@mathcolor{#1}}
\def\@mathcolor#1#2#3{%
  \protect\leavevmode
  \begingroup
    \color#1{#2}#3%
  \endgroup
}

\newenvironment{myitemize}{\begin{itemize}[itemsep=-0.05cm, leftmargin=*, topsep=0.1cm]}{\end{itemize}}

\def\slasha#1{\setbox0=\hbox{$#1$}#1\hskip-\wd0\hbox to\wd0{\hss\sl/\/\hss}}

\def\periodb#1{\setbox0=\hbox{$#1$}#1\hskip-\wd0\hbox to\wd0{-}}

\newcommand{\ii}{\mathrm{i}}            
\newcommand{\e}{\mathrm{e}}             

\newcommand{\BVL}{{\mathsf\Delta}_{\textrm{\tiny BV}}}
\newcommand{\CA}{\mathcal{A}}               

\newcommand{\CCL}{\mathscr{L}}

\newcommand{\CF}{\mathcal{F}}

\newcommand{\CJ}{\mathcal{J}}

\newcommand{\CO}{\mathcal{O}}

\newcommand{\CS}{\mathcal{S}}

\newcommand{\CT}{\mathcal{T}}

\newcommand{\tte}{\texttt{e}}
\newcommand{\ttv}{\texttt{v}}

\def\swone{{\textrm{\tiny (1)}}}
\def\swzero{{\textrm{\tiny (0)}}}
\def\swtwo{{\textrm{\tiny (2)}}}
\def\swthree{{\textrm{\tiny (3)}}}
\def\swfour{{\textrm{\tiny (4)}}}
\def\swk{{\textrm{\tiny (k)}}}

\def\hodge{{\textrm{\tiny H}}}

\def\hodge{{\textrm{\tiny H}}}

\def\BV{{\textrm{\tiny BV}}}
\def\BRST{{\textrm{\tiny BRST}}}

\def\tv{{\textrm{\tiny $V$}}}
\def\tv1{{\textrm{\tiny $V[1]$}}}

\newcommand{\mbf}[1]{{\boldsymbol {#1} }}

\newcommand{\Sym}{\mathrm{Sym}}

\newcommand{\swap}[3]{\sfR_{#3}(#1)\star\sfR^{#3}(#2)}

\newcommand{\FR}{\mathbbm{R}}               
\newcommand{\FC}{\mathbbm{C}}               
\newcommand{\RZ}{\mathbbm{Z}}               

\newcommand{\dd}{\mathrm{d}}                

\newcommand{\sU}{\mathsf{U}}                

\newcommand{\sG}{\mathsf{G}}

\newcommand{\sH}{\mathsf{H}}

\newcommand{\sP}{\mathsf{P}}

\newcommand{\sD}{\mathsf{D}}

\renewcommand{\comment}[1]{}                  
\def\tyng(#1){\hbox{\tiny$\yng(#1)$}}           
\def\tyoung(#1){\hbox{\tiny$\young(#1)$}}           

\newcommand{\beq}{\begin{eqnarray}}
\newcommand{\eeq}{\end{eqnarray}}

\newcommand{\sfh}{{\sf h}}

\newcommand{\sfR}{\mathsf{R}}

\definecolor{outrageousorange}{rgb}{1.0, 0.43, 0.29}

\newcommand{\midwedge}{\text{\Large$\wedge$}}

\def\RR{{\mathcal R}}

\def\beq{\begin{equation}}
\def\bee{\begin{equation}}
\def\eeq{\end{equation}}
\def\bea{\begin{eqnarray}}
\def\eea{\end{eqnarray}}
\def\ba{\begin{align}}
\def\ea{\end{align}}

\renewcommand{\thefootnote}{\fnsymbol{footnote}}
\theoremstyle{plain}

\theoremstyle{definition}

\numberwithin{equation}{section}
\catcode`@=12

\begin{document}

    \begin{center}
        
        
        \baselineskip=24pt
        
        {\LARGE\bf Braided Scalar Quantum Electrodynamics}
        
        \baselineskip=14pt
        
        \vspace{5mm}
        
{\large\bf Marija Dimitrijevi\'c \'Ciri\'c}${}^{\,(a)\,,\,}$\footnote{Email: \ {\tt
dmarija@ipb.ac.rs}} \ \ \ \ \ {\large\bf Biljana Nikoli\'c}${}^{\,(a)\,,\,}$\footnote{Email: \ {\tt
biljana@ipb.ac.rs}}  \\[2mm] {\large\bf Voja Radovanovi\'c}${}^{\,(a)\,,\,}$\footnote{Email: \ {\tt
rvoja@ipb.ac.rs}} \ \ \ \ \ {\large\bf Richard J. Szabo}${}^{\,(b)\,,\,}$\footnote{Email: \ {\tt R.J.Szabo@hw.ac.uk}} \ \ \ \ \ {\large\bf Guillaume Trojani}${}^{\,(b)\,,\,}$\footnote{Email: \ {\tt gt43@hw.ac.uk}}
\\[6mm]

\noindent  ${}^{(a)}$ {\it Faculty of Physics, University of
Belgrade}\\ {\it Studentski trg 12, 11000 Beograd, Serbia}
\\[3mm]

\noindent  ${}^{(b)}$ {\it Department of Mathematics, Heriot-Watt University\\ Colin Maclaurin Building,
Riccarton, Edinburgh EH14 4AS, U.K.}\\ and {\it Maxwell Institute for
Mathematical Sciences, Edinburgh, U.K.}
\\[10mm]

\end{center}
    
    \begin{abstract}
        \noindent
We formulate scalar electrodynamics in the braided $L_\infty$-algebra formalism and study its perturbative expansion in the algebraic framework of Batalin-Vilkovisky quantization. We confirm that UV/IR mixing is absent at one-loop order in this noncommutative field theory, and that the non-anomalous Ward-Takahashi identities for the braided gauge symmetry are satisfied.
    \end{abstract}
    
    
{\baselineskip=10pt
    \tableofcontents}
    
\bigskip

\setcounter{page}{1}

\setcounter{footnote}{0}
\renewcommand{\thefootnote}{\arabic{footnote}}
\newcommand{\del}{\partial}

\section{Introduction and Summary}
\label{sec:Intro}

Braided field theories were introduced in~\cite{DimitrijevicCiric:2021jea,Giotopoulos:2021ieg} as a novel alternative framework for organising the symmetries and dynamics of noncommutative field theories, motivated by considerations from noncommutative gravity~\cite{Ciric:2020eab,Szabo:2022edp}. Their quantization was developed in~\cite{Nguyen:2021rsa,DimitrijevicCiric:2023hua,Bogdanovic:2024jnf} where it was shown that the non-locality due to noncommutativity is manifested in correlation functions in a very different manner than in the standard noncommutative field theories (reviewed in e.g.~\cite{Douglas:2001ba,Szabo:2001kg,Hersent:2022gry}). Similarly to the earlier systematic approaches of Oeckl~\cite{Oeckl:1999zu,Oeckl:2000eg,Sasai:2007me}, as well as of Grosse and Lechner~\cite{Grosse:2007vr,Grosse:2008dk}, braided quantum field theory already exhibits non-local effects at the level of free fields through a noncommutative deformation of Wick's Theorem. A main new impetus is the ability to incorporate theories with gauge symmetries, contrary to the original works which were limited to scalar field theories.

At the level of interacting quantum fields, in all calculations performed thus far, up to two-loop order and three-point multiplicity, the braided theories have been found to be free of the notorious problem of UV/IR mixing~\cite{Minwalla:1999px} which obstructs renormalisability of the standard models (this feature was anticipated earlier by~\cite{Balachandran:2005pn,Bu:2006ha,Fiore:2007vg}, however through somewhat adhoc heuristic reasonings). This has sparked the exciting prospect that these non-local models may be well-defined quantum field theories, without the need of modifying their ultraviolet and infrared behaviours~\cite{Langmann:2002cc} as in the perturbatively renormalisable Grosse-Wulkenhaar model~\cite{Grosse:2004yu,Rivasseau:2005bh,Gurau:2005gd}.

By their very definition, braided quantum field theories can only be handled using the framework of homotopy algebras: there is no `braided space' of fields, hence no notions of functional differentiation or path integration and the standard quantization methods do not apply. Even in the commutative case, this algebraic perspective expresses Feynman diagram techniques in a manner that is mathematically precise and conceptually clear, in contrast to the usual textbook approaches (e.g.~path integral techniques) whose mathematical foundations are not well-defined. This is not to say that it completely avoids the analytic complications of infinite-dimensional spaces of fields and the standard loop divergences of quantum field theory, but it computes correlation functions and scattering amplitudes in a purely algebraic setting. In the braided noncommutative case, the power of the formalism is even more striking: as a comparison of the results of the naive approach of~\cite{CiricDimitrijevic:2022eei} with the systematic approach of~\cite{DimitrijevicCiric:2023hua} shows, braided quantum field theory does not follow the traditional Feynman rules of perturbation theory.

At the classical level, braided field theories are defined by deforming the $L_\infty$-algebra which organises the symmetries and dynamics of a given field theory to a novel homotopy algebraic structure called a `braided $L_\infty$-algebra' in~\cite{DimitrijevicCiric:2021jea,Giotopoulos:2021ieg}. As $L_\infty$-algebras are the natural algebraic structure underlying the Batalin-Vilkovisky (BV) formalism (see e.g.~\cite{Jurco:2018sby,Jurco:2019bvp} for reviews), the quantization of braided field theories is captured by a braided version of BV theory and the homological perturbation lemma, originally developed in~\cite{Nguyen:2021rsa}. The purpose of this paper is to explore a new example of braided quantum field theory with gauge symmetry that originates in an $L_\infty$-algebra which is more general than a differential graded Lie algebra. At the classical level, the examples of~\cite{DimitrijevicCiric:2021jea,Giotopoulos:2021ieg} demonstrate that these noncommutative field theories differ most profoundly from the standard versions. Our aim here is to explore the extent to which differences occur at the quantum level.

In the following we treat in detail the model of braided scalar quantum electrodynamics (QED), which is the simplest theory of this type, possessing braided $\sU(1)$ gauge symmetry. From a physical perspective, scalar QED models the type of interactions between weak hypercharge gauge bosons and Higgs bosons in the Standard Model of particle physics; it also provides a simple venue for spontaneous symmetry breaking and the Higgs mechanism in other contexts. From a technical standpoint, its $L_\infty$-structure is more complicated than the spinor version of QED considered in~\cite{DimitrijevicCiric:2023hua} which originates in a differential graded Lie algebra, as it contains a non-zero ternary bracket organising the quartic interaction between scalars and photons. While we shall indeed find a rich classical theory whose gauge symmetry and dynamics drastically differ from both ordinary scalar QED as well as its usual noncommutative deformation with star-gauge symmetry, braided BV quantization produces the same qualitative results as the calculations of~\cite{DimitrijevicCiric:2023hua,Bogdanovic:2024jnf}: we confirm that UV/IR mixing is absent in correlation functions up to one-loop order and three-point multiplicity.

As a novel additional chapter in the story of braided quantum field theory, in this paper we shall also derive algebraic Ward-Takahashi identities satisfied by the correlation functions of braided scalar QED, which are a consequence of the braided gauge invariance of the theory. Our approach is based on the underlying local Becchi-Rouet-Stora-Tyutin (BRST) cohomology within the braided BV formalism and the homological perturbation lemma. These identities are intimate relatives of the algebraic braided Schwinger-Dyson equations derived in~\cite{Bogdanovic:2024jnf}. Our derivation is new, and since it does not rely on the traditional path integral techniques, it is also of broader interest as further development within the homotopy algebraic formulation of general quantum field theories.

Analogously to the standard situation in textbook quantum field theory (see e.g.~\cite{Peskin:1995ev}), gauge invariance in braided scalar QED is manifested in three ways: 
\begin{myitemize}
\item At the classical level as a fundamental symmetry which is captured by the Maurer-Cartan theory for the underlying braided $L_\infty$-algebra. 
\item In the weak conservation law for the electric current that follows from the Noether identity for the braided $\sU(1)$ gauge symmetry and the Maurer-Cartan equations.
\item At the quantum level in the algebraic Ward-Takahashi identities which follow from homological perturbation theory and impose braided $\sU(1)$ gauge transformations on correlation functions.
\end{myitemize}
In this paper we develop all three complementary perspectives in detail, thus demystifying some of the unusual features that braided gauge symmetries are known to present~\cite{DimitrijevicCiric:2021jea,Giotopoulos:2021ieg} compared to ordinary gauge symmetries in field theory and star-gauge symmetries in noncommutative field theory.

\paragraph{Outline.}

The organisation and main results of the remainder of this paper are as follows. 

In \cref{sec:commutative_scalar_qed} we formulate classical scalar electrodynamics in the $L_\infty$-algebra formalism. By deforming the underlying homotopy algebra to a braided $L_\infty$-algebra, we define our classical model of braided scalar electrodynamics using Maurer-Cartan theory. We also derive the classical BV extension of the braided Maurer-Cartan functional.

In \cref{sec:braided_scalar_qed} develop the quantization of braided scalar electrodynamics using extended braided BV theory and the homological perturbation lemma. We derive the interaction vertices and introduce a diagrammatic notation for them as well as for the propagators of the theory. We then explain in detail how to compute correlation functions of polynomial observables in braided homological perturbation theory.

In \cref{sec:correlation_functions_in_braided_scalar_qed} we present explicit calculations of correlation functions using braided homological perturbation theory up to one-loop order. We demonstrate that all one-point functions vanish, as well as  that the photon and scalar propagators receive contributions from only planar diagrams. We also compute the scalar-scalar-photon vertex function, showing that it differs from its commutative counterpart only in a noncommutative phase factor depending on the external momenta. Altogether these results show that the correlators of braided scalar QED are free from UV/IR mixing at one-loop.

Finally, in \cref{sec:braided_gauge_symmetries} we derive our new algebraic Ward-Takahashi identities, which impose braided gauge transformations on correlation functions, and compare our approach with other treatments in the literature based on homotopy algebras. We start by exhibiting the braided BRST symmetry of our theory, and use it to derive the conservation law for correlators of the extended Noether current for the braided $\sU(1)$ symmetry in the BV formalism. We then use the braided homological perturbation lemma to derive anomalous Ward-Takahashi identities, giving an intrinsic interpretation of the anomaly in terms of the action of the BV Laplacian on the scalar field BRST functional, which vanishes in braided scalar QED. As explicit examples we derive the all loop order relation between the improper three-vertex function and the scalar propagator, as well as transversality of the exact photon vacuum polarisation tensor, checking them explicitly using our one-loop results. We also derive an explicit identity for general correlation functions involving arbitary numbers of external scalars and photons. 

\paragraph{Notation and Conventions.}

In this paper we work on four-dimensional Minkowski space $\FR^{1,3}$, on which ordinary scalar QED is perturbatively renormalisable.
We use standard coordinates $x=(x^\mu)=(x^0,x^1,x^2,x^{3})$ on $\FR^{1,3}$, where $x^0$ represents time, and the shorthand notation $\partial_\mu$ for the partial derivative $\frac\partial{\partial x^\mu}$. The standard Minkowski metric of signature $(+-- -)$ is denoted $\eta_{\mu\nu}$, and the inner product of two vectors $v=(v^\mu)$ and $w=(w^\mu)$ in $\FR^{1,3}$ is $v\cdot w  = \eta_{\mu\nu}\,v^\mu\,w^\nu= v_\mu\,w^\mu$; unless otherwise explicitly stated, implicit summation over repeated indices is always understood throughout this paper. We write $v^2:=v\cdot v$. The standard volume form on $\FR^{1,3}$ is
\begin{align}
\dd^4x = \dd x^0\wedge\dd x^1\wedge \mathrm{d}x^2\wedge\dd x^{3} \ ,
\end{align}
and $\square=\eta^{\mu\nu} \, \partial_\mu \, \partial_\nu=\partial_\mu\,\partial^\mu$ is the wave operator.
We denote by $\ast_{\hodge}$ the Hodge duality operator induced by the Minkowski metric $\eta_{\mu \nu}$.

We abbreviate momentum space integrals as 
\begin{align}
\int_{k,p,\dots} := \int_{(\FR^{1,3})^*}\,\frac{\dd^4k}{(2\pi)^4} \ \int_{(\FR^{1,3})^*}\,\frac{\dd^4p}{(2\pi)^4} \ \cdots \ .
\end{align}
Throughout we use the formal representation 
\begin{align}
\int_{\FR^{1,3}} \, \dd^4x \ \e^{\pm\,\ii\,k\cdot x} = (2\pi)^4 \, \delta(k)
\end{align}
of the Dirac distribution.

All vector spaces are defined over the ground field $\FR$ of real numbers. For any vector space $V$ and any integer $p\in\RZ$, elements of $V[p]$ are of degree~$-p$.

\paragraph{Acknowledgments.}

This article is based upon work from COST Actions CaLISTA CA21109 and THEORY-CHALLENGES CA22113 supported by COST (European Cooperation in Science and Technology). The work of {\sc M.D.C., B.N.} and  {\sc V.R.} is supported by Project 451-03-66/2024-03/200162 of the Serbian Ministry of Science,
Technological Development and Innovation. The work of {\sc G.T.} is supported by the Doctoral Training Partnership Award ST/T506114/1 from the UK Science and Technology Facilities Council. 

\section{Classical Braided Scalar Electrodynamics} 
\label{sec:commutative_scalar_qed}

In this section we construct our novel model of braided scalar electrodynamics, starting from the well-known correspondence between perturbative classical field theories and $L_\infty$-algebras; see e.g.~\cite{Hohm:2017pnh,Jurco:2018sby} for classic expository accounts.
The $L_\infty$-algebra formulation of ordinary scalar quantum electrodynamics with a Higgs potential appears in~\cite{Elliott:2020uwn,Arvanitakis:2021ecw}, where spontaneous symmetry breaking and the Higgs mechanism are described using homotopy transfer.
The $L_\infty$-algebra description of the non-abelian version, i.e. scalar quantum chromodynamics, is considered by~\cite{Gomez2021} in the context of tree-level scattering amplitudes and the perturbiner expansion in the corresponding minimal model.

\subsection{Scalar Electrodynamics in the $L_\infty$-Algebra Formalism} 
\label{sub:classical_scalar_qed}

We construct the cyclic $L_\infty$-algebra $\mathscr{L} = (L, \ell_1,\ell_2,\ell_3, \langle -,- \rangle)$ of classical scalar electrodynamics  whose Maurer-Cartan functional is the standard action functional given by
\begin{align}\label{eq:scalarqedaction}
    {S} := \int_{\FR^{1,3}} \, \mathrm{d}^4x  \ 
    D_\mu\bar\phi\, D^\mu\phi - m^2\,\bar\phi\,\phi - \frac{1}{4}\,F_{\mu\nu}\,F^{\mu\nu} \ .
\end{align}
The matter field is a complex scalar $\phi$ of mass $m$ which is minimally coupled to a $\mathsf{U}(1)$ gauge field $A_\mu$ through the covariant derivatives 
\begin{equation}
\begin{split}\label{eq:commutativecovariantderivative}
    D_\mu\phi & = \del_\mu\phi - \mathrm{i}\,g\, A_\mu\,\phi \quad,\quad D_\mu\bar\phi = \del_\mu\bar\phi + \mathrm{i}\,g\,\bar\phi\, A_\mu \ , 
\end{split}
\end{equation}
where $g$ is the electromagnetic coupling constant. 
The kinetic terms come with the field strength tensor 
\begin{align} \label{eq:fieldstrengthMaxwell}
F_{\mu\nu} = \del_\mu A_\nu - \del_\nu A_\mu \ .
\end{align}

The action functional is invariant under infinitesimal gauge transformations with respect to a real parameter $ c$, which are defined as
\begin{equation}
\begin{split}\label{eq:commutative gauge transformations}
\delta_ c\phi & = \mathrm{i}\, c\,\phi \quad , \quad \delta_ c\bar\phi = - \mathrm{i}\,\bar\phi\,  c \quad, \quad \delta_ c A_\mu = \tfrac{1}{g}\,\del_\mu c\ .
\end{split}
\end{equation}
The covariant derivatives satisfy $\delta_c (D_\mu \phi) = \mathrm{i}\, c\, D_\mu \phi$ and similarly for their complex conjugates.

The  graded vector space $L=L^0\oplus L^1\oplus L^2\oplus L^3$ underlying the $L_\infty$-algebra $\CCL$ has homogeneous components
\begin{align}
L^0 = L^3 = \Omega^0(\mathbbm{R}^{1,3}) \quad , \quad L^1 = L^2 = \Omega^0(\mathbbm{R}^{1,3}) \oplus \Omega^0(\mathbbm{R}^{1,3}) \oplus \Omega^1(\mathbbm{R}^{1,3}) \ ,
\end{align}
which respectively organise the gauge symmetry, fields, equations of motion and Noether identities 
\begin{equation}\label{VSpace} 
 c \ \in \ L^0 \ ,\quad {\cal A} = \left( \begin{array}{c}
\phi\\
\bar\phi\\
A
\end{array}\right) \ \in \ L^1 \ , \quad \mathcal{A}^+ = \left( \begin{array}{c}
\bar\phi^+\\
\phi^+\\
A^+
\end{array}\right) \ \in \ L^2 \ , \quad c^+ \ \in \ L^3 \ .
\end{equation}
Here we used the isomorphism $\FR^2\simeq\FC$ to identify $\Omega^0(\mathbbm{R}^{1,3}) \oplus \Omega^0(\mathbbm{R}^{1,3})\simeq \Omega^0(\FR^{1,3})\otimes_\FR \FC$ as the real vector spaces of complex scalar fields~$(\phi,\bar\phi\,)$ and $(\bar\phi^+,\phi^+)$, regarded as zero-forms on $\FR^{1,3}$. The gauge field $A$ is regarded as a one-form and its field strength $F$ as a two-form on $\FR^{1,3}$.

The complete list of non-zero $L_\infty$-brackets $\ell_n : \midwedge^n L \longrightarrow L[2-n]$ is given by
\begin{subequations}
\begin{align} \label{eq:ell1}
    \ell_1( c) & = \left(\begin{array}{c}
        0\\
        0\\
        \frac{1}{g}\,\del_\mu c
    \end{array}\right) \ ,\quad \ell_1(\CA) = \left(\begin{array}{c}
        -(\square+m^2)\,\bar\phi\\
        -(\square+m^2)\,\phi\\
        \square A^\mu - \del^\mu\,\del_\rho A^\rho
    \end{array}\right) \ ,\quad\ell_1({\cal A}^+) = \frac{1}{g}\,\del_\mu A^{+ \mu} \ , 
\end{align}
\begin{align}
    \ell_2 ( c, {\cal A} ) & = \left(\begin{array}{c}
        \mathrm{i}\, c\,\phi\\
        -\mathrm{i}\,\bar\phi\,  c\\
         0
    \end{array}\right) \ ,\quad\ell_2 ( c, \mathcal{A}^+ ) = \left(\begin{array}{c}
        \mathrm{i}\, c\, \bar\phi^+\\
        -\mathrm{i}\,\phi^+\,  c\\
         0
    \end{array}\right) \ ,\quad\ell_2 ( c, c^+ ) = 0  \ ,
\end{align}
\begin{align}
    \ell_2(\mathcal{A}_1, \mathcal{A}_2) & = \mathrm{i}\,g\,
    \left(\begin{array}{c}
    \bar\phi_1\,\del_\mu A_2^\mu + \bar\phi_2\, \del_\mu A_1^\mu + 2\,\del_\mu\bar\phi_1\, A_2^\mu + 2\,\del_\mu\bar\phi_2\, A_1^\mu\\[1ex]
    -\del_\mu A_1^\mu\, \phi_2 - \del_\mu A_2^\mu\, \phi_1 - 2\, A_1^\mu\,\del_\mu\phi_2 - 2\,A_2^\mu\,\del_\mu\phi_1\\[1ex]
    \phi_1\,\del^\mu\bar\phi_2 + \phi_2\,\del^\mu\bar\phi_1 - \del^\mu\phi_1\,\bar\phi_2 - \del^\mu\phi_2\,\bar\phi_1
    \end{array}\right)  \ ,
\end{align}
\begin{align}
    \ell_2(\mathcal{A}, \mathcal{A}^+) & = -\mathrm{i}\,\phi^+\, \bar\phi + \mathrm{i}\, \bar\phi^+\, \phi \ ,
\end{align}
\begin{align}
\begin{split}
   &  \ell_3(\mathcal{A}_1, \mathcal{A}_2, \mathcal{A}_3) \\[4pt] & \qquad = -g^2\,{\small \left(\begin{array}{c}
    \bar\phi_1\, A_{2\mu}\, A_3^\mu + \bar\phi_2\, A_{1\mu}\, A_3^\mu + \bar\phi_1\, A_{3\mu}\, A_2^\mu + \bar\phi_3\, A_{1\mu}\, A_2^\mu + \bar\phi_2\, A_{3\mu}\,A_1^\mu + \bar\phi_3\, A_{2\mu}\, A_1^\mu\\[1ex]
    A_{1\mu}\, A_2^\mu\, \phi_3 + A_{2\mu}\,A_1^\mu\,\phi_3 + A_{1\mu}\, A_3^\mu\,\phi_2 + A_{3\mu}\,A_1^\mu\, \phi_2 + A_{2\mu}\, A_3^\mu\, \phi_1 + A_{3\mu}\,A_2^\mu\,\phi_1\\[1ex]
    2\,\big(\phi_1\,\bar\phi_2\, A_3^\mu + \phi_2\, \bar\phi_1\, A_3^\mu
    + \phi_1\,\bar\phi_3\, A_2^\mu + \phi_2\, \bar\phi_3\, A_1^\mu
    +  \phi_3\,\bar\phi_1\, A_2^\mu + \phi_3\,\bar\phi_2\, A_1^\mu \big)
    \end{array}\right) } \normalsize \ .
    \end{split}
\end{align}
\end{subequations}

The cyclic inner product $\langle -,- \rangle : L \otimes L \longrightarrow \mathbbm{R}[-3]$ is defined by
\begin{align}\label{CyclPairingScalarQED} 
\begin{split}
\langle  c,c^+\rangle = \int_{\FR^{1,3}} \,  c \ast_\hodge c^+ \quad , \quad
\langle {\cal A}, \mathcal{A}^+ \rangle = 
\int_{\FR^{1,3}} \, A\wedge\ast_\hodge\, {A}^+ +  \bar\phi^+\ast_\hodge \phi 
+ \bar\phi\ast_\hodge \phi^{+}  \ .
\end{split}
\end{align}
Using these brackets and pairing one easily checks that the Maurer-Cartan functional
\begin{align} \label{eq:MCaction}
\CS(\CCL) := \tfrac1{2!}\,\langle\CA,\ell_1(\CA)\rangle - \tfrac1{3!}\,\langle\CA,\ell_2(\CA,\CA)\rangle - \tfrac1{4!}\,\langle\CA,\ell_3(\CA,\CA,\CA)\rangle
\end{align}
reproduces the classical action functional \eqref{eq:scalarqedaction} of scalar electrodynamics: $\CS(\CCL) = S$.

Maurer-Cartan theory also recovers the curvature $\CF_\CA\in L^2$ of the generalized gauge field $\CA\in L^1$ given by
\begin{align}\label{eq:fieldstrength}
\CF_\CA := \ell_1(\CA)-\tfrac1{2!}\,\ell_2(\CA,\CA)-\tfrac1{3!}\,\ell_3(\CA,\CA,\CA) \ .
\end{align}
It is covariant under gauge variations $\delta_c \mathcal{A} = \ell_1(c) + \ell_2(c,\mathcal{A}) $, i.e. $\delta_c\CF_\CA=\ell_2(c,\CF_\CA)$, and its components read explicitly
\begin{equation}
\begin{split}
    \CF_\phi &= -(\square + m^2)\,\bar\phi - \mathrm{i}\,g\, \partial_\mu(\bar\phi\, A^\mu) - \mathrm{i}\, g\, \partial_\mu \bar\phi\, A^\mu + g^2\, \bar\phi\, A_\mu\, A^\mu \ , \\[4pt]
    \CF_{\bar\phi} &= -(\square + m^2)\,\phi + \mathrm{i}\,g\, \partial_\mu(A^\mu\,\phi) + \mathrm{i}\, g\, A^\mu\, \partial_\mu \phi + g^2\, A_\mu\, A^\mu\, \phi \ , \\[4pt]
    \CF_{A}^\mu &= \square A^\mu - \partial^\mu \, \partial_\nu  A^\nu + \mathrm{i}\,g\, \partial^\mu \phi \, \bar\phi- \mathrm{i}\, g\,\phi\, \partial^\mu \bar\phi +2\, g^2\, \phi\, \bar\phi\, A^\mu \ .
\end{split}
\end{equation}
The Maurer-Cartan equations $\CF_\CA=0$ define the space of classical solutions $\CA$ to the field equations of scalar electrodynamics.

The Noether identity for the $\sU(1)$ gauge symmetry $\delta_c \mathcal{A} = \ell_1(c) + \ell_2(c,\mathcal{A}) $ is encoded in Maurer-Cartan theory by the off-shell Bianchi identity in $L^3$ for the curvature $\CF_\CA$ given by
\begin{equation}
\begin{split}\label{eq:commutativeNoethersqed}
    \mathsf{d}_\mathcal{A} \CF_\mathcal{A} := \ell_1(\CF_\CA)+\ell_2(\CA,\CF_\CA) = \tfrac{1}{g}\,\partial_\mu \CF_A^\mu  + \mathrm{i}\, \bar\phi\, \CF_{\bar\phi} - \mathrm{i}\, \CF_{\phi}\, \phi \ = \ 0 \ .
\end{split}
\end{equation}
After putting the photon on-shell, $\CF_A^\mu = 0$, this identity implies that 
\begin{equation}\label{eq:commutativeJ}
    J_\mu :=\mathrm{i} \, \big(
    \phi\, D_\mu \bar\phi 
    - D_\mu \phi \, \bar\phi\, \big) 
\end{equation}
is the weakly conserved Noether current for the global $\sU(1)$ symmetry of the complex scalar field from \eqref{eq:commutative gauge transformations}, i.e. $\partial_\mu J^\mu\approx0$, where $\approx$ indicates a weak equality that holds only up to equations of motion. It minimally couples to the gauge field $A_\mu$ in the action functional \eqref{eq:scalarqedaction}.

\subsection{Minimal Model of Scalar Electrodynamics} 
\label{sub:minimal_model_of_sqed}

The differential $\ell_1$ from \eqref{eq:ell1} defines the underlying cochain complex $\mathsf{Ch}(\mathscr{L}) = (L,\ell_1)$ of the $L_\infty$-algebra $\CCL$ of scalar electrodynamics. It reads explicitly
\begin{equation}\label{eq:scalarelectrocomplex}
\small
\hspace{-0.45cm}
\begin{tikzcd}[row sep=0ex,ampersand replacement=\&, column sep= 10ex]
\Omega^0(\mathbbm{R}^{1,3})\!\! \arrow[r,"\frac{1}{g}\,\mathrm{d}"] \& 
\!\!\Omega^1(\mathbbm{R}^{1,3})[-1]\!\! \arrow[rr,"\mathrm{d}^\dagger\, \mathrm{d}"] \& \& 
\!\!\Omega^1(\mathbbm{R}^{1,3})[-2]\!\! \arrow[r,"\frac{1}{g}\,\mathrm{d}^\dagger"] \& \!\!\Omega^0(\mathbbm{R}^{1,3})[-3] \\
\& \oplus \& \&  \oplus \& \\
  \& \begin{matrix} \Omega^0(\mathbbm{R}^{1,3})[-1] \\[1ex] \oplus \\[1ex] \Omega^0(\mathbbm{R}^{1,3})[-1] \end{matrix} 
  \ar[rr," \ { \ \ \Big(\begin{matrix} 0 \!\! & \!\! -\square- m^2 \\[-0.3ex] -\square -m^2 \!\! & \!\! 0  \end{matrix}\Big) \ } \ \ "] 
  \&  \& \begin{matrix} \Omega^0(\mathbbm{R}^{1,3})[-2] \\[1ex] \oplus \\[1ex] \Omega^0(\mathbbm{R}^{1,3})[-2] \end{matrix}  \& 
\end{tikzcd}
\normalsize
\end{equation}
where $\dd$ is the exterior differential and $\mathrm{d}^\dagger := \ast_{\hodge}\, \mathrm{d}\, \ast_{\hodge}$ is the codifferential. 

The cohomology of this complex defines a cyclic $L_\infty$-algebra $H^\bullet(\mathscr{L})$ quasi-isomorphic to $\CCL$ with underlying cochain complex $\mathsf{Ch}\big(H^\bullet(\CCL)\big) = \big(H^\bullet({L}),0\big)$, called the minimal model for the $L_\infty$-algebra $\CCL$~\cite{Jurco:2018sby,Gomez2021}.
It describes the free field theory, or equivalently the linearized Maurer-Cartan equations.
In particular, fields $(A^{\textrm{\tiny(0)}},\phi^{\textrm{\tiny(0)}},\bar\phi{}^{\textrm{\tiny(0)}})$ in the degree~$1$ cohomology
\begin{align}
H^1({L}) = \frac{\ker(\mathrm{d}^\dagger\,\mathrm{d})}{{\rm im}(\mathrm{d})} \ \oplus \ \ker\bigg(\begin{matrix} 0 & - \square- m^2\, \\ - \square - m^2  & 0  \end{matrix}\bigg)
\end{align}
correspond to the usual plane wave photon states $A^{\textrm{\tiny(0)}}$ of Maxwell theory on $\mathbbm{R}^{1,3}$ and the conventional description of a free massive complex scalar as states $(\phi^{\textrm{\tiny(0)}},\bar\phi{}^{\textrm{\tiny(0)}})$ satisfying the Klein-Gordon equation. 

At the level of cochain complexes, scalar electrodynamics and its minimal model fit into the homotopy equivalence data of a deformation retract
\begin{equation}
    \begin{tikzcd}
        (L, \ell_1) \arrow[two heads, "p"  ,start anchor={[yshift=1ex]}, end anchor={[yshift=1ex]}]{r}         
        \arrow[loop left, distance=2em, start anchor={[yshift=-1ex]west}, end anchor={[yshift=1ex]west}]{}{\mathsf{h} }
        & (H^\bullet({L}), 0)  \arrow[hook, "\iota", below ,start anchor={[yshift=-1ex]}, end anchor={[yshift=-1ex]}]{l}
    \end{tikzcd} \ ,
\end{equation}
where $p \circ \iota = \mathrm{id}_{H^\bullet({L})}$ and the map $\iota \circ p$ decomposes into
\begin{equation}\label{eq:hodge-kodaira}
    \iota \circ p = \mathrm{id}_{{L}} + \ell_1 \circ\mathsf{h} + \mathsf{h} \circ\ell_1 \ .
\end{equation}
The homogeneous components of the contracting homotopy $\mathsf{h} : L \longrightarrow {L}[-1]$ are denoted by $\mathsf{h}^{\swk} : L^k \longrightarrow L^{k-1}$. The degree~$2$ component $\mathsf{h}^{\swtwo} = - \mathsf{G}_{\scriptscriptstyle \textrm{F}}:L^2\longrightarrow L^1$ is given by the {Feynman propagators} $\sG_{\scriptscriptstyle \rm F}$ of scalar electrodynamics, which satisfy $\mathsf{G}_{\scriptscriptstyle\rm F} \circ \ell_1 = \mathrm{id}_{L^1}$ and $\ell_1 \circ \mathsf{G}_{\scriptscriptstyle\rm F} = \mathrm{id}_{L^2}$ on  triplets of fields.

The contracting homotopy $\mathsf{h}$ is not unique. A convenient choice is obtained starting from the massive scalar Feynman propagator, which is the degree 0 map $\mathsf{G}: \Omega^0(\mathbbm{R}^{1,3}) \longrightarrow \Omega^{0}(\mathbbm{R}^{1,3})$ given by
\begin{equation}
    \mathsf{G} = - \frac{1}{\square + m^2} \quad  , \quad 
    \tilde{\mathsf{G}}(k)= \frac{1}{k^2 - m^2} \ .
\end{equation}
In the second expression we wrote the momentum space representation, i.e. the eigenvalues of $\mathsf{G}$ when acting on plane wave eigenfunctions of the form $\mathrm{e}^{\pm\, \mathrm{i}\, k \cdot x}$. These should be understood throughout as defined using the standard Feynman $\ii\,\epsilon$-prescription, but for brevity we do not write it explicitly.

Similarly, the massless scalar Feynman propagator $\mathsf{D} : \Omega^0(\mathbbm{R}^{1,3}) \longrightarrow \Omega^0(\mathbbm{R}^{1,3})$ is given by
\begin{equation}
    \mathsf{D} = \frac{1}{\square} \ \quad , \quad \mathsf{\tilde{D}}(k) = - \frac{1}{k^2} \ .
\end{equation}
We extend this to a map $\mathsf{D} : \Omega^1(\mathbbm{R}^{1,3}) \longrightarrow \Omega^1(\mathbbm{R}^{1,3})$ using the $\Omega^0(\mathbbm{R}^{1,3})$-module structure of the vector space of one-forms.
Finally, introduce the projector $\mathsf{\Pi} : \Omega^1(\mathbbm{R}^{1,3}) \longrightarrow \Omega^1(\mathbbm{R}^{1,3})$ onto the image of the Maxwell operator $\dd^\dag\, \mathrm{d}$, with matrix elements
\begin{equation}
\begin{split}\label{eq:projector}
\mathsf{\Pi}_{\mu\nu} = \eta_{\mu\nu}-\frac{\partial_\mu\,\partial_\nu}{\square} \quad , \quad    \tilde{\mathsf{\Pi}}_{\mu \nu}(k) = \eta_{\mu \nu} - \frac{k_\mu\, k_\nu}{k^2} \ .
\end{split}
\end{equation}

Then the components of the contracting homotopy satisfying \eqref{eq:hodge-kodaira} are given by
\begin{equation}
\begin{split}\label{eq:propagators}
    \mathsf{h}^{\swone}(\CA) =
    -g\, \mathsf{D} \, \dd^\dag A
    \quad , \quad
    \mathsf{h}^{\swtwo}(\mathcal{A}^+) = -
    \begin{pmatrix}
         \mathsf{G} \, \phi^+ \\
        \mathsf{G} \, \bar\phi^+\\
        \mathsf{D} \, {\mathsf{\Pi}}\, A^+ 
    \end{pmatrix} \quad , \quad
    \mathsf{h}^{\swthree}(c^+) = -g
    \begin{pmatrix} 0 \\ 0 \\ 
         \mathsf{D} \, \mathrm{d}\, c^+
    \end{pmatrix} \ .
\end{split}
\end{equation}
In terms of coordinate functions
\begin{equation}
\begin{split}\label{eq:contracting hom on coordfunc}
    \mathsf{h}^{\swone}(A_\mu) =  - g\,\frac{\partial^\mu}{\square}  A_\mu\ , \
    \mathsf{h}^{\swtwo}
    \begin{pmatrix}
        \bar\phi^+ \\
        \phi^+ \\
        A^{+ \mu}
    \end{pmatrix} 
    =
    \begin{pmatrix} \displaystyle
        \frac{1}{ \square + m^2}\,\phi^+ \\[2ex] \displaystyle
        \frac{1}{\square + m^2}\, \bar\phi^+ \\[2ex] \displaystyle 
        -\frac{1}{\square}\,\Big(\delta^{\mu}_\nu - \frac{\partial^\mu \, \partial_\nu }{\square}\Big)\, A^{+ \nu}
    \end{pmatrix} 
    \ , \
    \mathsf{h}^{\swthree}(c^+) &= - g\, \frac{\partial_\mu}{\square}\, c^+ \ .
\end{split}
\end{equation}

Gauge fixing is performed by enforcing $A \in \mathrm{ker}(\mathsf{h}^{\swone})$. The choice of contracting homotopy above with $\mathsf{h}^{\swone} : \Omega^1(\mathbbm{R}^{1,3}) \longrightarrow \Omega^0(\mathbbm{R}^{1,3})$ defines the Lorenz gauge.

\subsection{Scalar Electrodynamics in the Braided $L_\infty$-Algebra Formalism} 
\label{sub:braided_qed}

We will now deform the cyclic $L_\infty$-algebra of scalar electrodynamics from \cref{sub:classical_scalar_qed} into a cyclic braided  $L_\infty$-algebra $\mathscr{L}^\star = (L, \ell_1^\star,\ell_2^\star,\ell_3^\star, \langle -,- \rangle_\star)$, according to the formalism introduced in~\cite{DimitrijevicCiric:2021jea,Giotopoulos:2021ieg}. The underlying graded vector space $L$ is unchanged, while the structure maps of $\CCL^\star$ are deformed according to Drinfel'd twist deformation quantization. The Maurer-Cartan theory for the cyclic braided $L_\infty$-algebra $\CCL^\star$ defines a new noncommutative deformation of the theory from \cref{sub:classical_scalar_qed}, which we call \emph{braided scalar electrodynamics}. We describe its gauge symmetries and dynamics following~\cite{DimitrijevicCiric:2021jea,Giotopoulos:2021ieg}.

\paragraph{Drinfel'd Twist Deformation.}

As in~\cite{DimitrijevicCiric:2023hua}, in this paper we focus on the translation symmetries of scalar electrodynamics and work with the abelian Moyal-Weyl twist
\begin{equation}
\begin{split}\label{eq:F twist}
    \mathcal{F} = \exp\left(- \frac{\mathrm{i}}{2}\, \theta^{\mu\nu}\, {\pounds}_\mu \otimes {\pounds}_\nu\right) \ ,
\end{split}
\end{equation}
where $\theta$ is a constant bivector on $\FR^{1,3}$ and $\pounds_\mu:=\pounds_{\partial_\mu}$ are Lie derivatives along the holonomic frame of vector fields on $\FR^{1,3}$. The operator \eqref{eq:F twist} acts on tensor products (over $\FC$) of spaces of classical tensor fields (over $\Omega^0(\FR^{1,3})$). 

If $\mu:\CT\otimes\CT\longrightarrow\CT$ is a binary operation on a tensor space $\CT$, then the twist \eqref{eq:F twist} can be used to deform it into another binary operation
\begin{align}
\mu_\star := \mu\circ\CF^{-1} \ .
\end{align}
This preserves associativity properties of the operation $\mu$, but it generally deforms any commutativity properties to \emph{braided} commutativity. For example, if $\mu(T_1\otimes T_2)=\mu(T_2\otimes T_1)$ for $T_1,T_2\in\CT$, then
\begin{align}
\mu_\star(T_1\otimes T_2) = \mu_\star\circ\RR^{-1}(T_2\otimes T_1) = \mu_\star\big(\sfR_\alpha(T_2)\otimes\sfR^\alpha(T_1)\big) \ ,
\end{align}
where 
\begin{align}
\RR=\CF^{-2}
\end{align}
is the universal $R$-matrix and we use the standard notation $\RR^{-1} = \sfR_\alpha\otimes\sfR^\alpha$ (with summation understood). This deformation naturally extends to higher $n$-ary operations as well as to more general maps among combinations of different tensor spaces.

\paragraph{The Cyclic Braided $\mbf{L_\infty}$-Algebra ${\CCL^\star}$.}

Running the Drinfel'd twist machine on the brackets of \cref{sub:classical_scalar_qed}, the non-zero brackets $\ell_n^\star:\midwedge^nL\longrightarrow L[2-n]$ of the braided $L_\infty$-algebra $\CCL^\star$ are given by
\begin{subequations}\label{eq:braided brackets}
\begin{align} \label{eq:1-bracketBQED}
    \ell_1^\star( c) & = \left(\begin{array}{c}
        0\\
        0\\
        \frac{1}{g}\,\del_\mu c
    \end{array}\right) \quad,\quad\ell_1^\star({\cal A}) = \left(\begin{array}{c}
        -(\square+m^2)\,\bar\phi\\
        -(\square+m^2)\,\phi\\
        \square A^\mu - \del^\mu\,\del_\rho A^\rho
    \end{array}\right)\quad,\quad
    \ell_1^\star(\mathcal{A}^+) = \frac{1}{g}\,\del_\mu A^{+ \mu} \ ,
\end{align}
\begin{align}
\small
    \ell^\star_2 ( c, {\cal A} ) & ={\small \left(\begin{array}{c}
        \mathrm{i}\, c\star\phi\\
        -\mathrm{i}\,\sfR_\alpha (\bar\phi)\star\sfR^\alpha ( c)\\
         0
    \end{array}\right) } \normalsize\ ,\
    \ell^\star_2 ( c, \mathcal{A}^+ ) = {\small \left(\begin{array}{c}
        \mathrm{i}\,\sfR_\alpha (\bar\phi^+)\star\sfR^\alpha ( c)\\
        -\mathrm{i}\, c\star \phi^{+}\\
         0
    \end{array}\right) } \normalsize \ ,\ \ell^\star_2 ( c, c^+ ) = 0 \ ,
    \normalsize
\end{align}
\begin{align}\label{eq:2-bracket BQED}
    \ell_2^\star(\mathcal{A}_1, \mathcal{A}_2) & = \mathrm{i}\,g\,
    {\small \left(\begin{array}{c}
    \bar\phi_1\star\del_\mu A_2^\mu + \sfR_\alpha (\bar\phi_2)\star\sfR^\alpha (\del_\mu A_1^\mu) + 2\,\del_\mu\bar\phi_1\star A_2^\mu + 2\,\sfR_\alpha (\del_\mu\bar\phi_2)\star\sfR^\alpha (A_1^\mu)\\[1ex]
    -\del_\mu A_1^\mu\star\phi_2 - \sfR_\alpha (\del_\mu A_2^\mu)\star\sfR^\alpha (\phi_1) - 2\, A_1^\mu\star\del_\mu\phi_2 - 2\,\sfR_\alpha (A_2^\mu)\star\sfR^\alpha (\del_\mu\phi_1)\\[1ex]
    \phi_1\star\del^\mu\bar\phi_2 + \sfR_\alpha (\phi_2)\star\sfR^\alpha (\del^\mu\bar\phi_1) - \del^\mu\phi_1\star\bar\phi_2 - \sfR_\alpha (\del^\mu\phi_2)\star\sfR^\alpha (\bar\phi_1)
    \end{array}\right) } \normalsize \ ,
\end{align}
\begin{align} \label{eq:l2AA+}
    \ell_2^\star(\mathcal{A}, \mathcal{A}^+) 
    & = -\mathrm{i}\,\bar\phi \star \phi^{+} + \mathrm{i}\, \phi\star\bar\phi^+ \ ,
\end{align}
\begin{align}\label{eq:l_3}
    \ell_3^\star(\mathcal{A}_1, \mathcal{A}_2, \mathcal{A}_3) = -g^2\,{\small \left(\begin{array}{c}
    \bar\phi_1\star A_{2\mu}\star A_3^\mu + \swap{\bar\phi_2}{A_{1\mu}}{\alpha}\star A_3^\mu\\[1ex]
    +\, \bar\phi_1\star \swap{A_{3\mu}}{A_2^\mu}{\alpha} + \swap{\bar\phi_3}{A_{1\mu}\star A_2^\mu}{\alpha}\\[1ex]
    +\, \swap{\bar\phi_2\star A_{3\mu}}{A_1^\mu}{\alpha} + \swap{\bar\phi_3}{\swap{A_{2\mu}}{ A_1^\mu}{\beta}}{\alpha}\\
    \\
    A_{1\mu}\star A_2^\mu\star\phi_3 + \swap{A_{2\mu}}{A_1^\mu}{\alpha}\star\phi_3\\[1ex]
    +\, A_{1\mu}\star\swap{A_3^\mu}{\phi_2}{\alpha} + \swap{A_{3\mu}}{A_1^\mu\star\phi_2}{\alpha}\\[1ex]
    +\, \swap{A_{2\mu}\star A_3^\mu}{\phi_1}{\alpha} + \swap{A_{3\mu}}{\swap{A_2^\mu}{\phi_1}{\beta}}{\alpha}\\
    \\
    2\,\big[\phi_1\star\bar\phi_2\star A_3^\mu + \swap{\phi_2}{\bar\phi_1}{\alpha}\star A_3^\mu\\[1ex]
    +\, \phi_1\star\swap{\bar\phi_3}{A_2^\mu}{\alpha} + \swap{\phi_3}{\bar\phi_1\star A_2^\mu}{\alpha}\\[1ex]
    +\, \swap{\phi_2\star\bar\phi_3}{A_1^\mu}{\alpha} + \swap{\phi_3}{\swap{\bar\phi_2}{A_1^\mu}{\beta}}{\alpha}\big]
    \end{array}\right) } \normalsize \ .
\end{align}
\end{subequations}

The Moyal-Weyl twist deformation of the cyclic inner product \eqref{CyclPairingScalarQED} yields a pairing for $\CCL^\star$ which is strictly cyclic. It is given explicitly by
\begin{align}\label{CyclPairingBScalarQED} 
\begin{split}
\langle  c,c^+\rangle_\star = \int_{\FR^{3,1}} \,  c \wedge_\star\ast_\hodge\, c^+ \  , \
\langle {\cal A}, \mathcal{A}^+ \rangle_\star = \int_{\FR^{1,3}} \, 
A\wedge_\star\ast_\hodge\, A^+ 
+  \bar\phi^+ \wedge_\star \ast_\hodge\, \phi 
+ \bar\phi \wedge_\star \ast_\hodge\, \phi^{ +}  \ .
\end{split}
\end{align}

\paragraph{Braided Gauge Symmetry.}

The  (left) braided gauge transformations by $c \in L^0$ are given by~\cite{DimitrijevicCiric:2021jea,Giotopoulos:2021ieg} $\delta_c^\star\CA = \ell_1^\star(c)+\ell_2^\star(c,\CA)$, which read explicitly 
\begin{equation}
\begin{split}\label{eq:bsqedgaugetransfo}
    \delta^\star_c \phi = \mathrm{i}\, c\star\phi\quad , \quad 
    \delta^\star_c \bar\phi = -\mathrm{i}\,\sfR_\alpha (\bar\phi)\star\sfR^\alpha ( c)\quad , \quad
    \delta^\star_c A_\mu = \tfrac{1}{g}\,\partial_\mu c \ .
\end{split}
\end{equation}
The gauge variations close a trivial braided commutator algebra 
\begin{align}
[\delta^\star_{c_1}, \delta^\star_{c_2}]^\star := \delta_{c_1}^\star\circ\delta_{c_2}^\star - \delta_{\sfR_\alpha(c_2)}^\star\circ\delta_{\sfR^\alpha(c_1)}^\star = 0 \ , 
\end{align}
reflecting the abelian nature of the braided $\sU(1)$ gauge symmetry, or equivalently the vanishing brackets $\ell_2^\star(c_1,c_2)=0$.

\paragraph{Maurer-Cartan Equations.}

The braided curvature $\CF_\CA^\star\in L^2$ of a generalized gauge field $\CA\in L^1$ obtained from Maurer-Cartan theory is given by the formula \eqref{eq:fieldstrength} applied to the braided brackets in \eqref{eq:braided brackets}. It is covariant under braided gauge variations, i.e. $\delta_c^\star\CF_\CA^\star = \ell_2^\star(c,\CF_\CA^\star)$, and its components are given explicitly by
\begin{equation}
\begin{split}\label{eq:braidedfieldstrength}
    \CF_\phi^\star &= - (\square + m^2)\, \bar\phi - \tfrac{\ii\,g}{2} \,
    \big(  \bar\phi \star \partial_\mu A^\mu + \partial_\mu A^\mu \star \bar\phi + 2\, \partial_\mu \bar\phi \star A^\mu
    + 2\, A^\mu \star \partial_\mu \bar\phi
    \big) \\
    & \hspace{2cm} + \tfrac{g^2}{3} \,
    \big(\bar\phi \star A_\mu \star A^\mu + A_\mu \star \bar\phi \star A^\mu + 
    A_\mu \star A^\mu \star \bar\phi
    \big) \ ,
    \\[4pt]
    \CF_{\bar\phi}^\star &= - (\square + m^2)\, \phi + \tfrac{\ii\,g}{2}\, \big(
    \partial_\mu A^\mu \star \phi + \phi \star \partial_\mu A^\mu + 2\, A^\mu \star \partial_\mu \phi + 2\, \partial_\mu \phi \star A^\mu
    \big) \\
    & \hspace{2cm} + \tfrac{g^2}{3} \,
    \big(  A_\mu \star A^\mu \star \phi +  A_\mu \star \phi \star A^\mu + \phi \star A_\mu \star A^\mu\big) \ ,
    \\[4pt]
    \CF^{\star\mu}_{A} &= \square A^\mu - \partial^\mu\,\partial_\nu A^\nu 
    +\tfrac{\ii\,g}{2} \,\big(
    \partial^\mu \phi \star \bar\phi + \bar\phi \star \partial^\mu \phi - \phi \star \partial^\mu \bar\phi  - \partial^\mu \bar\phi \star \phi
    \big) \\
    & \hspace{2cm} +\tfrac{g^2}{3} \,
    \big(
    \phi \star \bar\phi \star A^\mu + \bar\phi \star \phi \star A^\mu + \phi \star A^\mu \star \bar\phi \\
    & \hspace{4cm} + \bar\phi \star A^\mu \star \phi
    + A^\mu \star \phi \star \bar\phi + A^\mu \star \bar\phi \star \phi
    \big)\ .
\end{split}
\end{equation}

The Maurer-Cartan equations $\CF_\CA^\star=0$ define the equations of motion for the scalar and gauge fields.
We can rewrite them in a more compact form by defining left and right covariant derivatives
\begin{equation} \label{eq:braidedcovderiv}
D_\mu^{\mbox{\tiny L}} \phi = \del_\mu\phi - \mathrm{i}\,g\, A_\mu\star\phi \quad ,\quad D_\mu^{\mbox{\tiny R}}\phi = \del_\mu\phi -\mathrm{i}\,g\, \phi\star A_\mu  
= \partial_\mu\phi - \ii\, g\, \mathsf{R}_\alpha ( A_\mu) \star \mathsf{R}^\alpha  (\phi) \ .
\end{equation}
They are braided gauge covariant, i.e. $\delta_c^\star(D^{\mbox{\tiny L,R}}_\mu \phi) = \mathrm{i}\, c \star D_\mu^{\mbox{\tiny L,R}} \phi$, and similarly for their complex conjugates. In the commutative limit $\theta=0$, they both reduce to the covariant derivatives introduced in \eqref{eq:commutativecovariantderivative}.
With them we present the equations of motion for braided scalar electrodynamics as
\begin{align}\label{eq:braidedcoveom}
\begin{split}
& \tfrac{1}{2}\,D_\mu^{\mbox{\tiny L}\,}D^{{\mbox{\tiny L}}\mu}\phi + \tfrac{1}{2}\,D_\mu^{\mbox{\tiny R}}\,D^{{\mbox{\tiny R}}\mu}\phi + m^2\,\phi
+ \tfrac{g^2}{6}\,\big( A_\mu\star A^\mu\star\phi + \phi\star A_\mu\star A^\mu -2\, A^\mu\star\phi\star A_\mu\big) \ = \ 0 \ , \\[4pt]
& \partial_\rho F^{\rho\mu} - \tfrac{\ii\,g}{4}\,\big[ \phi\star (D^{{\mbox{\tiny L}}\mu}\bar\phi + D^{{\mbox{\tiny R}}\mu} \bar\phi) + (D^{{\mbox{\tiny L}}\mu}\bar\phi + D^{{\mbox{\tiny R}}\mu}\bar\phi) \star \phi \\
& \hspace{2.5cm}
- \bar\phi\star (D^{{\mbox{\tiny L}}\mu}\phi + D^{{\mbox{\tiny R}}\mu}\phi )
- (D^{{\mbox{\tiny L}}\mu}\phi + D^{{\mbox{\tiny R}}\mu}\phi) \star \bar\phi \big] \\
& \hspace{4cm} + \tfrac{g^2}{12}\,\big( \phi\star \bar\phi\star A^\mu +  \bar\phi \star \phi \star A^\mu + A^\mu\star \phi\star \bar\phi 
+ A^\mu\star \bar\phi\star \phi \\
& \hspace{5.5cm} -2\, \phi\star A^\mu\star \bar\phi -2\, \bar\phi\star A^\mu\star \phi
\big) \ = \ 0 \ .
\end{split}
\end{align}

We stress that here the classical field strength \eqref{eq:fieldstrengthMaxwell} is invariant: $\delta_c^\star F_{\mu\nu}=0$. Despite the lack of manifest braided gauge covariance of the individual cubic terms proportional to $g^2$ in \eqref{eq:braidedcoveom}, their overall combinations are covariant. This is guaranteed by covariance of the braided curvature, and can also be checked by explicit calculation. This is a typical feature of gauge theories whose braided $L_\infty$-algebras involve higher brackets and thus do not define differential graded braided Lie algebras~\cite{DimitrijevicCiric:2021jea,Giotopoulos:2021ieg}. In the commutative limit $\theta=0$, the individual non-covariant terms cancel each other and we recover the equations of motion of scalar electrodynamics from \cref{sub:classical_scalar_qed}.

\paragraph{Braided Noether Identity.}

Braided field theories differ most profoundly from ordinary field theories and noncommutative field theories with star-gauge symmetry through their braided gauge symmetries. This is dually reflected in their realisations of the corresponding Noether identities~\cite{DimitrijevicCiric:2021jea,Giotopoulos:2021ieg}. The braided Noether identity is a Bianchi identity for the braided curvature $\CF_\CA^\star$ resulting from the homotopy Jacobi identities for the braided $L_\infty$-algebra $\CCL^\star$ in $L^3$. In the present case it reduces to
\begin{equation}
\begin{split}\label{eq:braided Noether ids}
    \mathsf{d}_\mathcal{A}^\star \CF^\star_{\mathcal{A}} &:= \ell_1^\star(\CF^\star_{\mathcal{A}}) 
    + \tfrac{1}{2!}\, \big(\ell_2^\star(\CF^\star_{\mathcal{A}}, \mathcal{A}) - \ell_2^\star(\mathcal{A}, \CF^\star_{\mathcal{A}}) \big) 
    + \tfrac{1}{3!}\, \ell_1^\star \big(\ell^\star_3(\mathcal{A},\CA,\CA)\big) \\
    & \hspace{2cm} + \tfrac{1}{2\cdot 2!}\, \big(
    \ell_2^\star (\ell_2^\star(\mathcal{A},\CA), \mathcal{A})
    -\ell_2^\star (\mathcal{A},\ell_2^\star( \mathcal{A},\CA))
    \big) \\
    & \hspace{4cm} + \tfrac{1}{2\cdot 3!} \,
    \big(
    \ell_2^\star(\ell_3^\star(\mathcal{A},\CA,\CA), \mathcal{A})) - \ell_2^\star(\mathcal{A}, \ell_3^\star(\mathcal{A},\CA,\CA))
    \big) \ = \ 0 \ .
\end{split}
\end{equation}

After expanding out the curvature in brackets, most terms cancel, leaving
\begin{equation}
\begin{split}
\ell_1^\star(\CF^\star_\mathcal{A}) + \tfrac{1}{2} \,
\big(\ell_2^\star (\ell_1^\star(\mathcal{A}), \mathcal{A}) - \ell_2^\star(\mathcal{A}, \ell_1^\star(\mathcal{A}))\big)
+ \tfrac{1}{6}\, \ell_1^\star\big(\ell_3^\star(\mathcal{A},\mathcal{A},\mathcal{A})\big) \ = \ 0 \ .
\end{split}
\end{equation}
This leads to the braided analogue of \eqref{eq:commutativeNoethersqed} given by
\begin{equation}
\begin{split}\label{eq:braidednoetherbsqed}
    \tfrac{1}{g}\, \partial_\mu \CF^{\star\mu}_{A} + \partial_\mu J^{\star\mu} \ = \ 0 \ ,
\end{split}
\end{equation}
where
\begin{equation}
\begin{split}\label{eq:non-covariantcurrent}
    J_\mu^\star &= \tfrac{\mathrm{i}}{2}\, \big(
    \partial_\mu \phi \star \bar\phi + \bar\phi \star \partial_\mu \phi - \phi \star \partial_\mu \bar\phi  - \partial_\mu \bar\phi \star \phi
    \big)\\
    &\quad \, + \tfrac{g}{3} \,
    \big(
    \phi \star \bar\phi \star A_\mu + \bar\phi \star \phi \star A_\mu + \phi \star A_\mu \star \bar\phi 
    + \bar\phi \star A_\mu \star \phi
    + A_\mu \star \phi \star \bar\phi + A_\mu \star \bar\phi \star \phi
    \big)\ .
\end{split}
\end{equation}
Using the covariant derivatives \eqref{eq:braidedcovderiv}, the one-form \eqref{eq:non-covariantcurrent} can be brought to the covariant form
\begin{equation}
\begin{split}\label{eq:braided_current}
    & J^{\star }_\mu :=
    \tfrac{\mathrm{i}}{4}\, \big(
    \phi \star (D^{\mbox{\tiny L}}_\mu\bar\phi + D^{ \mbox{\tiny R}}_\mu \bar\phi) - (D^{ \mbox{\tiny L}}_\mu\phi + D_\mu^{ \mbox{\tiny R}}\phi) \star \bar\phi \\
    & \hspace{2cm} 
    + (D_\mu^{ \mbox{\tiny L}}\bar\phi + D_\mu^{ \mbox{\tiny R}}\bar\phi) \star \phi - \bar\phi \star (D_\mu^{ \mbox{\tiny L}}\phi + D_\mu^{ \mbox{\tiny R}} \phi)
    \big) 
    \\
    & \hspace{3cm}
    - \tfrac{g}{12}\,
    \big(
     \phi \star \bar\phi \star A_\mu 
    + A_\mu \star \phi \star \bar\phi 
    + \bar\phi \star \phi \star A_\mu 
    +  A_\mu \star \bar\phi \star \phi \\
    & \hspace{5cm}
    - 2\, \phi \star A_\mu \star \bar\phi 
    - 2\, \bar\phi \star A_\mu \star \phi
    \big) \ .
\end{split}
\end{equation}

By putting the photon on-shell, $\CF_A^{\star\mu}=0$, this defines a new weakly conserved braided electric current $J^{\star} \in \Omega^1(\mathbbm{R}^{1,3})$ analogously to the commutative case. Again despite the manifestly non-covariant individual cubic terms proportional to $g$ in \eqref{eq:braided_current}, it is a gauge invariant quantity. This current reduces to \eqref{eq:commutativeJ} in the commutative limit $\theta=0$. In general, however, the braided $\sU(1)$ charge conservation is vastly different from the usual case, see~\cite{DimitrijevicCiric:2023hua}.

\paragraph{Maurer-Cartan Functional.}

The action functional for braided scalar electrodynamics is obtained using the formula \eqref{eq:MCaction} from Maurer-Cartan theory applied to the braided brackets in \eqref{eq:braided brackets} and the cyclic inner product in (\ref{CyclPairingBScalarQED}). It is given explicitly by
\begin{equation} \label{eq:MCbSQED}
\begin{split}
  S_\star := \CS(\CCL^\star) &= \int_{\FR^{1,3}} \, {\rm d}^4 x \ \bigg(\, \frac{1}{4}\, \big(D_\mu^{\mbox{\tiny L}}\bar\phi + D_\mu^{\mbox{\tiny R}}\bar\phi\big) \star 
    \big(D^{{\mbox{\tiny L}}\mu}\phi + D^{{\mbox{\tiny R}}\mu}\phi\big) - m^2\,\bar\phi\star\phi -\frac{1}{4}\,F_{\mu\nu}\star F^{\mu\nu}\\
    & \hspace{3cm} + \frac{g^2}{12}\,\bar\phi\star \big(A_\mu\star A^\mu\star\phi + \phi\star A_\mu\star A^\mu -2\, A^\mu\star\phi\star A_\mu \big)\bigg) \ .
\end{split}
\end{equation}
It is invariant under braided gauge transformations and reduces to \eqref{eq:scalarqedaction} in the commutative limit. 

\paragraph{Minimal Model.}

For any braided field theory~\cite{DimitrijevicCiric:2021jea,Giotopoulos:2021ieg}, the underlying cochain complex $\mathsf{Ch}(\mathscr{L}^\star)$ coincides with the cochain complex $\mathsf{Ch}(\mathscr{L})$ of the undeformed field theory, given in the present case by \eqref{eq:scalarelectrocomplex}, because its differential is unaffected by the deformation: $\ell_1^\star = \ell_1$. It follows that the minimal model ${H}^\bullet(\mathscr{L}^\star)$ of the braided theory  coincides with the minimal model ${H}^\bullet(\mathscr{L})$ of its commutative counterpart. As a result, the free field theory is unchanged. 

On the other hand, the interactions are deformed via the higher brackets $\ell_n^\star : \midwedge^n L \longrightarrow L[2-n]$ of the braided $L_\infty$-algebra $\CCL^\star$ for $n\geq2$. When pulled back to the minimal model by quasi-isomorphisms, the interacting theories differ due to non-trivial braiding in higher brackets.

\subsection{Batalin-Vilkovisky Formulation of Braided Scalar Electrodynamics} 
\label{sub:bv_action}

We study the dynamics of braided scalar electrodynamics in the braided Batalin-Vilkovisky (BV) formalism of \cite{Giotopoulos:2021ieg}.

\paragraph{Batalin-Vilkovisky Functional.}

We extend the Maurer-Cartan functional $S_\star$ to a functional $S_\BV$ on $L$, regarded as the space of fields and antifields. With the notation from \eqref{VSpace}, we consider a superfield 
\begin{align}
{\boldsymbol{\CA}} = c + \mathcal{A} + \mathcal{A}^+ + c^+ \ , 
\end{align}
viewed as a dynamical field lying in degree 1. We then compute the braided Maurer-Cartan  functional \eqref{eq:MCaction} applied to the superfield $\mbf\CA$, which defines the BV functional $S_\BV(\mbf\CA)$.

Using the non-zero brackets and pairings from \eqref{eq:braided brackets} and \eqref{CyclPairingBScalarQED}, 
the BV functional obtained in this way is given by
\begin{equation}
\begin{split}\label{eq:fullbvaction}
    S_{\BV}(\boldsymbol{\CA}) &= S_{\star}(\mathcal{A}) 
    + \tfrac{1}{2}\, \langle \mathcal{A}^+\, , \, \ell^\star_1(c) \rangle_\star
    + \tfrac{1}{2}\, \langle c\, , \, \ell^\star_1(\CA^+) \rangle_\star
    - \tfrac{1}{2}\, \langle c\, , \, \ell_2^\star(\mathcal{A,A^+}) + \ell_2^\star(\mathcal{A^+,A})\rangle_\star
    \\[4pt]
    &= S_{\star}(\mathcal{A}) 
    +\int_{\mathbbm{R}^{1,3}}\,\dd^4x \  \bigg ( 
    c \star \frac{1}{g}\, \partial_\mu A^{+\mu}\\
    & \hspace{4.5cm} + \frac{\ii}{2} \, c \star
    \Big(
     \phi^+ \star \bar\phi
    -  \bar\phi^+ \star \phi \\
    & \hspace{6.5cm} + \mathsf{R}_\alpha (\phi^+) \star \mathsf{R}^\alpha (\bar\phi\,) - 
    \mathsf{R}_\alpha (\bar\phi^+) \star \mathsf{R}^\alpha (\phi)\Big) \bigg) \ .
\end{split}
\end{equation}
Setting the antifields equal to zero in the BV functional restricts to the Becchi-Rouet-Stora-Tyutin (BRST) fields $(c,\CA)$ and recovers the original braided Maurer-Cartan functional: $S_{\BV}(\boldsymbol{\CA}) \vert_{\CA^+=c^+=0} =S_\star(\mathcal{A})$. Note that the ghost $c$ is absent from the BRST functional, because it completely decouples from the physical fields $\CA$ due to the abelian nature of the braided $\sU(1)$ gauge symmetry. In the BV functional \eqref{eq:fullbvaction}, it couples to the scalar antifields through both left and right gauge transformations of the matter fields.

\paragraph{The Space of Fields and Antifields.}

It will prove useful to have an explicit description of the fields and antifields of braided scalar electrodynamics in terms of expansions in an appropriate basis. We expand the field/antifield content in a momentum space basis using Fourier transformations.

For the scalar field/antifield pairs $(\phi,\bar\phi\,)\in L^1$ and $(\bar\phi^+,\phi^+)\in L^2$ we write
\begin{equation}
\begin{split}\label{eq:momentum bases for phiphisplus}
    \phi &= \int_k\, \widehat{\phi}(k) \  \tte_k\ , \quad 
    \bar\phi = \int_k\, \widehat{\bar\phi}(k) \ \bar{\mathtt{e}}_k \ , \quad
    \bar\phi^+ = \int_k\, \widehat{\bar\phi}{}^+(k)  \ {\mathtt{e}}^k\ , \quad
     \phi^+ = \int_k\, \widehat{\phi}{}^+(k) \ \bar{\mathtt{e}}^k \ . 
\end{split}
\end{equation}
For the scalar basis fields $(\tte_k,\bar\tte_k)\in L^1$ and $(\bar\tte^k,\tte^k)\in L^2$ we use the explicit representation as plane waves
\begin{align} \label{eq:basis elements}
 \tte_k(x) = \mathrm{e}^{-\mathrm{i}\,k \cdot x}\quad , \quad  \bar{\mathtt{e}}_k(x) = \mathrm{e}^{ -\mathrm{i}\, k \cdot x} \quad , \quad \mathtt{e}^k(x) = \mathrm{e}^{\,\mathrm{i}\,k \cdot x} \quad , \quad \bar{\mathtt{e}}^k = \mathrm{e}^{\,\mathrm{i}\, k \cdot x} \ . 
\end{align}
These bases are dual with respect to the cyclic inner product \eqref{CyclPairingBScalarQED}, in the sense that
\begin{align}
\langle \tte_k ,{\mathtt{e}}^p\rangle_\star = 
    (2 \pi)^4 \, \delta(k -p) = \langle \bar\tte_k, \bar\tte^p\rangle_\star \ .
\end{align}

For the gauge field/antifield pair $A\in L^1$ and $A^+\in L^2$, we identify the vector space $\Omega^1(\FR^{1,3})$ of gauge fields as $(\mathbbm{R}^{1,3})^* \otimes \Omega^0(\mathbbm{R}^{1,3})$, and dually the vector space of antifields as $\mathbbm{R}^{1,3} \otimes \Omega^{0}(\mathbbm{R}^{1,3})$. Given dual bases of covectors $\ttv_\mu$ and vectors $\ttv^\mu$ of $(\mathbbm{R}^{1,3})^*$ and $\mathbbm{R}^{1,3}$ respectively, we write
\begin{equation}\label{eq:momentum basis for AAplus}
    A = \int_k\, \widehat{A}_\mu(k)\  \mathtt{v}^\mu \otimes \tte_k \qquad \mbox{and} \qquad
    A^+ = \int_k\, \widehat{A}{}^+{}^\mu(k) \ \mathtt{v}_\mu\otimes \mathtt{e}^k  \ .
\end{equation}
The dual pairings are given by
\begin{equation}
\begin{split}
    \langle \mathtt{v}^\mu\otimes \tte_k, \mathtt{v}_\nu\otimes \mathtt{e}^p \rangle_\star = 
    (2 \pi)^4 \,  \delta_\nu^\mu \, \delta(k - p) \ .
\end{split}
\end{equation}
In the following we will omit the tensor product symbols, abbreviating $\mathtt{v}^\mu\otimes \tte_k$ by $\mathtt{v}^\mu\, \tte_k$ and $\mathtt{v}_\mu\otimes \mathtt{e}^k$ by $\mathtt{v}_\mu\, \mathtt{e}^k$ to simplify the notation.

Finally, we expand the ghost field/antifield pair $c\in L^0$ and $c^+\in L^3$ in a similar vein as
\begin{equation}
    c = \int_k\, \widehat{c}(k) \ \widetilde{\mathtt{e}}_k \qquad \mbox{and} \qquad c^+ = \int_k\, \widehat{c}{}^+(k) \ \widetilde{\mathtt{e}}^k \ ,
\end{equation}
where the scalar basis fields $\widetilde{\tte}_k\in L^0$ and $\widetilde{\tte}^k\in L^3$ are given by
\begin{align} \label{eq:ghbasis}
\widetilde{\tte}_k(x) = \e^{-\ii\, k\cdot x} \qquad \mbox{and} \qquad \widetilde{\tte}^k(x) = \e^{\,\ii\, k\cdot x} \ .
\end{align}
Their dual pairings are also given by
\begin{align}
\langle\widetilde{\tte}_k,\widetilde{\tte}^p\rangle_\star = (2\pi)^4 \, \delta(k-p) \ .
\end{align}

The products among momentum space basis fields, twisted by \eqref{eq:F twist}, are given by
\begin{align}\label{eq:ekstarep}
    \tte_k\star \tte_p = \e^{-\frac\ii2 \, k\cdot\theta\, p} \ \tte_{k+p} 
    = \mathrm{e}^{\,\mathrm{i}\, p \cdot \theta\, k} \ \tte_p \star \tte_k
\end{align}
where $k\cdot\theta\,p:=k_\mu\,\theta^{\mu\nu}\,p_\nu=-p\cdot\theta\, k$. The action of the inverse $R$-matrix on tensor products of basis fields is given by
\begin{equation}
\begin{split}\label{eq:Rekotimesep}
\RR^{-1}(\tte_k\otimes \tte_p) &= \sfR_\alpha(\tte_k)\otimes\sfR^\alpha(\tte_p) = \e^{\,\ii\,k\cdot\theta\,p} \ \tte_k\otimes \tte_p \ .
\end{split}
\end{equation}
Similar formulas hold among combinations of the other basis fields from \eqref{eq:basis elements} and \eqref{eq:ghbasis}.

\section{Braided Scalar Quantum Electrodynamics} 
\label{sec:braided_scalar_qed}

In this paper we compute correlation functions of polynomial observables for braided scalar quantum electrodynamics using the algebraic framework of braided BV quantization. These observables live in the braided symmetric algebra $\Sym_\star L[2]=\bigoplus_{n\geq0} \, (\Sym_\star L[2])^n$ over $\FR$, defined by twisting the usual symmetric tensor product $\odot$ to the braided symmetric tensor product $\odot_\star$ and graded by polynomial degree. We denote  the identity element by $\mathbbm{1} \in (\mathrm{Sym}_\star L[2])^0$. In this section we lay the foundations for all subsequent calculations and analysis in this paper.

\subsection{Extended Batalin-Vilkovisky Formalism} 
\label{sub:perturbative_calculations_finally1}

To move from the description in terms of the braided $L_\infty$-algebra $\CCL^\star = (L,\ell_1^\star,\ell_2^\star,\ell_3^\star,\langle-,-\rangle_\star)$ to the braided symmetric algebra, we introduce \textit{contracted coordinate functions}~\cite{Jurco:2018sby} in the space $\mathrm{Sym}_\star (L[2]) \otimes L$.
The cyclic braided $L_\infty$-algebra of braided scalar electrodynamics from \cref{sub:braided_qed} is extended to the space of contracted coordinate functions similarly to \cite{DimitrijevicCiric:2023hua}, and denoted 
\begin{align}
\Sym_\star(L[2])\otimes\CCL^\star = \big(\mathrm{Sym}_\star (L[2]) \otimes L\,,\,\mbf\ell^\star_1\,,\,\mbf\ell^\star_2\,,\,\mbf\ell^\star_3\,,\,\langle\!\langle-,-\rangle\!\rangle_\star\big) \ .
\end{align}

In this extended framework, the BV functional $S_\BV(\mbf\CA)\in\FR$ from \eqref{eq:fullbvaction} is lifted to a translation invariant functional ${\mathcal{S}}_\BV(\mbf\xi) \in (\mathrm{Sym}_{\star} L[2])^0$, called the extended BV functional. It is defined by the Maurer-Cartan functional $\CS(\Sym_\star(L[2])\otimes\CCL^\star)$ applied to the translation invariant contracted coordinate function $\boldsymbol{\xi} \in (\mathrm{Sym}_{\star} (L[2]) \otimes L )^1 $ of symmetric tensor and total degree~1. In terms of the decompositions of \cref{sub:bv_action}, this is given by
\begin{align}
\begin{split}\label{eq:superfield}
    \boldsymbol{\xi} &= \int_k\, \mathtt{v}_\mu\, \mathtt{e}^k \otimes \mathtt{v}^\mu\, \tte_k + \mathtt{e}^k \otimes \tte_k + \bar{\mathtt{e}}^{k} \otimes \bar{\mathtt{e}}_{k} \\
    & \qquad \, + \int_k\, \widetilde{\mathtt{e}}^k \otimes \widetilde{\mathtt{e}}_k + \widetilde{\mathtt{e}}_k \otimes \widetilde{\mathtt{e}}^k
    + \mathtt{v}^\mu\, \tte_k \otimes \mathtt{v}_\mu\, \mathtt{e}^k
    + \tte_k \otimes \mathtt{e}^k + \bar{\mathtt{e}}_k \otimes \bar{\mathtt{e}}^k \ .
\end{split}
\end{align}

The first line of \eqref{eq:superfield} represents the \textit{physical} part of the extended superfield for the extended theory, corresponding to pairs $\CA^+\otimes\CA$ of antifields and fields in $L[2]^0\otimes L^1$, which is the part relevant to the computation of correlation functions of physical fields $\CA$. The second line incorporates braided gauge symmetries: the first term corresponds to the ghost antifield/field pairs $c^+ \otimes c$ and $c\otimes c^+$,  living in $L[2]^1 \otimes L^0$ and $L[2]^{-2}\otimes L^3$ respectively, while the remaining terms correspond to $\mathcal{A} \otimes \mathcal{A}^+$ in $L[2]^{-1} \otimes L^2$.

The extended BV functional is a 0-cochain given in terms of the extended cyclic inner product and extended brackets by
\begin{align} \label{eq:BVactionext}
\CS_\BV = \tfrac{1}{2!}\, \langle \!\langle \boldsymbol{\xi}, \boldsymbol{\ell}_1^\star(\boldsymbol{\xi}) \rangle\! \rangle_\star -\tfrac{1}{3!}\, \langle\! \langle \boldsymbol{\xi}, \boldsymbol{\ell}^\star_2(\boldsymbol{\xi}, \boldsymbol{\xi}) \rangle \!\rangle_{\star}
    -\tfrac{1}{4!}\, \langle\! \langle \boldsymbol{\xi}, \boldsymbol{\ell}^\star_3(\boldsymbol{\xi}, \boldsymbol{\xi}, \boldsymbol{\xi}) \rangle \!\rangle_{\star} \ .
\end{align}
The first term contains the free part $\CS_0$. The remaining terms define the interaction part of the extended BV functional which we write as a sum $\CS_{\rm int}+\CS_{\BRST}$, where
\begin{align}
\CS_{\rm int} = \CS_{\rm int}^\swthree + \CS_{\rm int}^\swfour 
\end{align}
represents respectively the trivalent and quadrivalent interactions among the physical fields $\CA$, while 
\begin{align}
\CS_{\BRST} = \CS_{\BRST}^\swtwo + \CS_{\BRST}^\swthree
\end{align}
represents respectively the bivalent and trivalent interactions of the ghost $c$ with the physical fields $\CA$ and their antifields $\CA^+$. We proceed to analyse these four interaction terms in turn.

\paragraph{$\mbf{\phi\bar\phi A}$ Vertex.}

Using strict cyclicity of the inner product to reduce the number of terms to two, together with translation invariance of the contracted coordinate functions, we compute the trivalent interaction from the physical part of the extended superfield \eqref{eq:superfield} as
\begin{equation}
\begin{split}\label{eq:S3calcu}
    \mathcal{S}^\swthree_{\rm int} &= -\tfrac{1}{3!}\, \langle\! \langle \boldsymbol{\xi}, \boldsymbol{\ell}^\star_2(\boldsymbol{\xi}, \boldsymbol{\xi}) \rangle \!\rangle_{\star}\\[4pt]
    & =-\int_{k_1,k_2,k_3}\,
    \langle \!\langle \mathtt{v}_\mu \, \mathtt{e}^{k_1}  \otimes \mathtt{v}^\mu\, \tte_{k_1}
    \, , \,
    \boldsymbol{\ell}_2^{\star}(\mathtt{e}^{k_2} \otimes \tte_{k_2}
    ,
    \bar{\mathtt{e}}^{k_3} \otimes \bar{\mathtt{e}}_{k_3})
    \rangle\! \rangle_\star\\[4pt]
    & = -\int_{k_1,k_2,k_3}\,
    \langle \!\langle \mathtt{v}_\mu\, \mathtt{e}^{k_1}  \otimes \mathtt{v}^\mu\, \tte_{k_1}
    \, , \, 
    (\mathtt{e}^{k_2} \odot_\star \mathsf{R}_\alpha (\bar{\mathtt{e}}^{k_3})) \otimes
    \ell_2^\star( \mathsf{R}^\alpha(\tte_{k_2})
    ,
    \bar{\mathtt{e}}_{k_3})
    \rangle\! \rangle_\star \\[4pt]
    & = - \int_{k_1,k_2,k_3} \,
    \mathrm{e}^{\,\mathrm{i}\,k_2\cdot \theta \, k_3}
    \ \mathtt{v}_\mu\, \mathtt{e}^{k_1} \odot_\star \mathsf{R}_\alpha(\mathtt{e}^{k_2} \odot_\star  \bar{\mathtt{e}}^{k_3} ) \
    \langle 
    \mathsf{R}^\alpha(\mathtt{v}^\mu\, \tte_{k_1}) \, , \, 
    \ell_2^\star( \tte_{k_2} , \bar{\mathtt{e}}_{k_3})
    \rangle_\star\\[4pt]
    & = -\int_{k_1,k_2,k_3}\, \mathrm{e}^{\,\mathrm{i}\,\sum\limits_{i<j}\, k_i \cdot\theta\, k_j} 
    \ \mathtt{v}_\mu\, \mathtt{e}^{k_1} \odot_\star \mathtt{e}^{k_2} \odot_\star  \bar{\mathtt{e}}^{k_3}  \
    \langle 
    \mathtt{v}^\mu\, \tte_{k_1} \, , \, \ell_2^\star( \tte_{k_2} , \bar{\mathtt{e}}_{k_3})
    \rangle_\star \ .
\end{split}
\end{equation}

Using the last entry of \eqref{eq:2-bracket BQED} to compute
\begin{equation}
\begin{split}
    \langle 
    \mathtt{v}^\mu\, \tte_{k_1} \,,\,
    \ell_2^\star( \tte_{k_2} , \bar{\mathtt{e}}_{k_3})
    \rangle_\star &= 
    \mathrm{i} \,g\,\langle \mathtt{v}^\mu \, \tte_{k_1},
    \tte_{k_2} \star \ttv_\nu\,\partial^\nu \bar{\mathtt{e}}_{k_3} - \ttv_\nu\,\partial^\nu \tte_{k_2} \star \bar{\mathtt{e}}_{k_3}
    \rangle_\star \\[4pt]
    &=  g\,(  k_3-k_2)^\mu \ \e^{-\frac\ii2\,\sum\limits_{i<j} \, k_i\cdot\theta\, k_j} \ (2\pi)^4 \, \delta(k_1+k_2+k_3) \ , 
\end{split}
\end{equation}
this evaluates to
\begin{equation}
\begin{split}\label{eq:S3int}
    \mathcal{S}^\swthree_{\rm int} = \int_{k_1,k_2,k_3}\, V^\mu_3(k_1, k_2,k_3) \
    \mathtt{v}_\mu\, \mathtt{e}^{k_1} \odot_\star \mathtt{e}^{k_2} \odot_\star  \bar{\mathtt{e}}^{k_3} \ ,
\end{split}
\end{equation}
where the scalar-scalar-photon vertex is given by
\begin{equation}
\begin{split}\label{eq:3-vertex diagram}
    V^\mu_3(k_1, k_2,k_3) =  g \ \e^{\frac{\mathrm{i}}{2}\, \sum\limits_{i<j}\, k_i \cdot \theta \,k_j} \, 
    (k_2 - k_3)^\mu \
    (2  \pi)^4 \,
    \delta(k_1 + k_2 + k_3) \ .
\end{split}
\end{equation}

Labelling momenta as incoming, this is represented graphically by 
\begin{equation}
\begin{split}
\normalsize
\begin{tikzpicture}[scale = 0.8, line width=0.2 pt]
{\scriptsize
    \draw[scalplus] (-140:1.8)node[below]{}--(0,0);
    \draw[Aplus] (140:1.8)node[left]{$\mu$}--(0,0);
    \draw[scalplusreverse] (0:1.8)node[right]{} --(0,0);
    \draw[->, shift={(-0.25,0.6)}] (140:.5) node[above]{$k_1$} -- (0,0);
    \draw[->, shift={(-0.25,-0.6)}] (-140:.5) node[below]{$k_3$} -- (0,0);
    \draw[->, shift={(0.29,0.3)}] (0:.5) node[above]{$k_2$} -- (0,0);
    } \normalsize
\end{tikzpicture}
\normalsize
\end{split}
\end{equation}
where the photon field $A$ is represented by a wavy line and the scalar fields $(\phi,\bar\phi\,)$ by directed straight lines.

\paragraph{$\mbf{\phi\bar\phi AA}$ Vertex.}

The four-point interaction is given by
\begin{equation}
    \mathcal{S}^\swfour_{\rm int} = 
    - \tfrac{1}{4!}\,
    \langle \!\langle 
    \boldsymbol{\xi}, \boldsymbol{\ell}_3^\star(\boldsymbol{\xi},\boldsymbol{\xi},\boldsymbol{\xi})
    \rangle\! \rangle_\star \ .
\end{equation}
Since the extended bracket and cyclic structure are compatible, we consider the cyclically inequivalent terms, of which there are three.
We then use the translation invariance of contracted coordinate functions $\boldsymbol{\xi}$, and strict cyclicity of the inner product to identify all such terms.
This gives 
\begin{equation}
\begin{split}
    \mathcal{S}^\swfour_{\rm int} 
    & = - \frac12\,\int_{k_1, k_2, k_3, k_4}\, \langle \!\langle 
    \mathtt{v}_\mu\, \mathtt{e}^{k_1} \otimes \mathtt{v}^\mu\, \tte_{k_1} \,,\, \boldsymbol{\ell}_3^\star( \mathtt{e}^{k_2} \otimes  \tte_{k_2}, \bar{\mathtt{e}}^{k_3} \otimes \bar{\mathtt{e}}_{k_3}, \mathtt{v}_\nu{\mathtt{e}}^{k_4} \otimes \mathtt{v}^\nu{\mathtt{e}}_{k_4})
    \rangle\! \rangle_\star \\[4pt]
    &= - \frac12\,\int_{k_1, k_2, k_3, k_4}\, 
    \mathtt{v}_\mu \, \mathtt{e}^{k_1} \odot_\star \mathsf{R}_\delta\big( \mathtt{e}^{k_2} \odot_\star  \mathsf{R}_\alpha(\bar{\tte}^{k_3}) \odot_\star  \mathsf{R}_\beta\, \mathsf{R}_\gamma(\mathtt{v}_\nu\,\tte^{k_4})\big)  \\
    &\hspace{4cm} \times \langle \mathsf{R}^\delta(\mathtt{v}^\mu\, \tte_{k_1}) \,,\, \ell_3^\star(\mathsf{R}^\beta\, \mathsf{R}^\alpha( \tte_{k_2}), \mathsf{R}^\gamma (\bar{\tte}_{k_3}), \mathtt{v}^\nu\,\tte_{k_4}) \rangle_\star \\[4pt]
    &=
    - \frac12\,\int_{k_1, k_2, k_3, k_4}\,  \mathrm{e}^{\,\mathrm{i}\, \sum\limits_{i<j}\, k_i \cdot \theta\, k_j} \
    \mathtt{v}_\mu\, \mathtt{e}^{k_1} \odot_\star \mathtt{e}^{k_2} \odot_\star \bar{\tte}^{k_3} \odot_\star \mathtt{v}_\nu\, \tte^{k_4} \\
    & \hspace{6cm} \times  \langle \mathtt{v}^\mu\, \tte_{k_1} \, , \, \ell_3^\star( \tte_{k_2}, \bar{\tte}_{k_3}, \mathtt{v}^\nu\, \tte_{k_4}) \rangle_\star \ .
\end{split}
\end{equation}

Using the last entry of \eqref{eq:l_3} we compute
\begin{equation}
\begin{split}
    \langle \mathtt{v}^\mu\, \tte_{k_1} \, , \, \ell_3^\star( \tte_{k_2}, \bar{\tte}_{k_3},\mathtt{v}^\nu\, \tte_{k_4}) \rangle_\star &= 
     -2\, g^2\, \langle \ttv^\mu\,\tte_{k_1} , \tte_{k_2} \star  \bar{\tte}_{k_3} \star \ttv^\nu\,\tte_{k_4}\rangle_\star \\[4pt]
    &=-2\, g^2\, \eta^{\mu \nu}\, \mathrm{e}^{-\frac{\mathrm{i}}{2}\,\sum\limits_{i<j}\, k_i \cdot \theta\, k_j}
    \ (2\pi)^4 \, \delta(k_1+k_2+k_3+k_4) \ .
\end{split}
\end{equation}
This gives the braided symmetric four-point interaction 
\begin{equation}\label{eq:one term}
\begin{split}
    \mathcal{S}^\swfour_{\rm int} = 
    \int_{k_1, k_2, k_3, k_4}\,  
    V_4^{\mu \nu}(k_1,k_2,k_3,k_4) \
    \mathtt{v}_\mu \,
    \mathtt{e}^{k_1} \odot_\star \mathtt{e}^{k_2} \odot_\star \bar{\mathtt{e}}^{k_3} \odot_\star \mathtt{v}_\nu\,\mathtt{e}^{k_4}
    \ ,
\end{split}
\end{equation}
where the four-vertex function is given by
\begin{equation}
\begin{split}\label{eq:4vertex}
    V_4^{\mu \nu}(k_1,k_2,k_3,k_4) = g^2\, \eta^{\mu \nu}\,
    \e^{\,\frac{\ii}{2} \, \sum\limits_{i<j }\,k_i \cdot \theta\, k_j} \ 
    (2 \pi)^4\, \delta(k_1 + k_2 + k_3 + k_4) \ .
\end{split}
\end{equation}

This is represented graphically by 
\begin{equation}
\begin{split}
\begin{tikzpicture}[scale = 0.8, line width=0.3 pt]
    {\scriptsize
    \draw[Aplus] (135:1.8)node[left]{$\mu$}--(0,0);
    \draw[scalplus] (-135:1.8)node[below]{}--(0,0);
    \draw[scalplusreverse] (-45:1.8)node[right]{} --(0,0);
    \draw[Aplus] (45:1.8)node[right]{$\nu$} --(0,0);
    \draw[->, shift={(-0.35, 0.75)}] (135:.5)  node[above]{$k_4$} -- (0,0);
    \draw[->, shift={(-0.35,-0.75)}] (-135:.5) node[below]{$k_3$} -- (0,0);
    \draw[->, shift={( 0.35,-0.75)}] (-45:.5)  node[below]{$k_2$} -- (0,0);
    \draw[->, shift={( 0.35, 0.75)}] (45:.5)   node[above]{$k_1$} -- (0,0);
    } \normalsize
    \end{tikzpicture}
    \end{split}
\end{equation}

\paragraph{$\mbf{cA^+}$ Vertex.}

Using \eqref{eq:1-bracketBQED} together with cyclicity of the inner product, we find that the lift of the term $ \langle \mathcal{A}^+, \ell^\star_1(c) \rangle_\star
$ in the full BV functional \eqref{eq:fullbvaction} is given by
\begin{equation} \label{eq:Sbrst2}
\begin{split}
    \CS_{\BRST}^\swtwo &=  \int_{k_1,k_2} \,
    \langle \!\langle \mathtt{v}^\mu\, \mathtt{e}_{k_1} \otimes \mathtt{v}_\mu\, \mathtt{e}^{k_1} \,,\,
    \boldsymbol{\ell}_1^{\star}(\mathtt{\widetilde e}^{k_2} \otimes \mathtt{\widetilde e}_{k_2})  \rangle\! \rangle_\star = \int_{k_1,k_2}\, V_{2\, \mu}^\BRST(k_1, k_2) \ \mathtt{v}^{\mu}\, \mathtt{e}_{k_1} \odot_\star \mathtt{\widetilde e}^{k_2} \ ,
\end{split}
\end{equation}
where the linear interaction vertex is given by
\begin{align}
V_{2\,\mu}^\BRST(k_1,k_2) = \tfrac{\mathrm{i}}{g} \, k_{1\,\mu} \ (2 \pi)^4\, \delta(k_1 - k_2) \ . 
\end{align}
 This corresponds to braided gauge transformations of the photon field.
Note the sign appearing from $\boldsymbol{\ell}_1^\star (\mathtt{\widetilde e}^{k_2} \otimes \mathtt{\widetilde e}_{k_2})
    = - \mathtt{\widetilde e}^{k_2} \otimes \ell_1^\star(\mathtt{\widetilde e}_{k_2})$ due to the grading of the ghosts, see \cite{Jurco:2018sby}. 
    
We represent this graphically by the diagram
\begin{equation}
    \vcenter{\hbox{\scriptsize{\begin{tikzpicture}[scale=0.5]
    \node  at (-0.5, 0.4) { $\mu$};
    \node  at (4.5, -0.4) {};
    \draw[A] (0,0) -- (2,0);
    \draw[dashed] (2,0) -- (4,0);
    \draw[<-, shift={(0.5,0.7)}] (0:0.8) node[above]{$k_1$} -- (0,0);
   \draw[->, shift={(2.5,0.7)}] (0:0.8) node[above]{$k_2$} -- (0,0);
    \end{tikzpicture}} \normalsize}}
\end{equation}
where the photon antifield $A^+$ is depicted with the dotted pattern and the ghost field $c$ with the dashed pattern.

\paragraph{$\mbf{c\bar\phi\phi^+ {+} c\phi\bar\phi^+}$ Vertex.}

The three-point BRST vertex corresponding to braided gauge transformations of the scalar fields is calculated from the lift of the terms $-\frac{1}{2}\, \langle c\, , \, \ell_2^\star(\mathcal{A,A^+}) + \ell_2^\star(\mathcal{A}^+, \mathcal{A}) \rangle_\star$ in the full BV functional \eqref{eq:fullbvaction}. 
Using the brackets \eqref{eq:l2AA+} and performing completely analogous calculations to those above, we find that this contributes
\begin{equation}
\begin{split}\label{eq:S3BRST}
    \CS_\BRST^\swthree &= -\int_{k_1,k_2,k_3} \, 
    \langle \!\langle
    \mathtt{\widetilde{e}}^{k_3} \otimes \mathtt{\widetilde{e}}_{k_3} 
    \,,\,
    \boldsymbol{\ell}_2^{\star}(
    \mathtt{e}^{k_1} \otimes \mathtt{e}_{k_1}
    +
    \bar{\mathtt{e}}^{k_1} \otimes \bar{\mathtt{e}}_{k_1}
    \ , \
    {\mathtt{e}}_{k_2} \otimes {\mathtt{e}}^{k_2}
    +
    \bar{\mathtt{e}}_{k_2} \otimes \bar{\mathtt{e}}^{k_2}
    )
    \rangle\! \rangle_\star \\[4pt]
    &=  \int_{k_1,k_2,k_3} \, V_3^{\BRST }(k_1,k_2,k_3) \ 
    \big(
    -
    \widetilde{\mathtt{e}}^{k_3}
    \odot_\star
    \mathtt{e}_{k_2} 
    \odot_\star
    \mathtt{e}^{k_1} 
    \ + \
    \mathtt{\widetilde{e}}^{k_3} 
    \odot_\star
    \bar{\mathtt{e}}_{k_2} 
    \odot_\star
    \bar{\mathtt{e}}^{k_1} 
    \big)
    \end{split}
\end{equation}
to the extended BV functional, where
\begin{align}
V_3^\BRST(k_1,k_2,k_3) = \mathrm{i} \
    \mathrm{e}^{\,\frac{\mathrm{i}}{2}\,\sum\limits_{i<j}\, k_i \cdot \theta\, k_j} \ (2 \pi)^4 \, \delta(k_1 - k_2 + k_3) 
    \ .
\end{align}

This is represented by the diagrams 
\begin{equation}
\begin{split}
    \vcenter{\hbox{{\scriptsize
    \begin{tikzpicture}[scale = 0.8, line width=0.2 pt]
    \draw[dashed] (140:1.8)node[above]{}--(0,0);
    \draw[scalplusreverse] (0:1.8)node[right]{} --(0,0);
    \draw[scalreverse] (-140:1.8)node[below]{}--(0,0);
    \draw[->, shift={(-0.25,0.6)}] (140:.5) node[above]{$k_2$} -- (0,0);
    \draw[->, shift={(-0.25,-0.6)}] (-140:.5) node[below]{$k_3$} -- (0,0);
    \draw[->, shift={(0.29,0.3)}] (0:.5) node[above]{$k_1$} -- (0,0);
    \end{tikzpicture} 
    }}} \normalsize
    \ \ 
    + \ \ 
    \vcenter{\hbox{{\scriptsize
    \begin{tikzpicture}[scale = 0.8, line width=0.2 pt]
    \draw[dashed] (140:1.8)node[above]{}--(0,0);
    \draw[scalplus] (0:1.8)node[right]{} --(0,0);
    \draw[scal] (-140:1.8)node[below]{}--(0,0);
    \draw[->, shift={(-0.25,0.6)}] (140:.5) node[above]{$k_2$} -- (0,0);
    \draw[->, shift={(-0.25,-0.6)}] (-140:.5) node[below]{$k_3$} -- (0,0);
    \draw[->, shift={(0.29,0.3)}] (0:.5) node[above]{$k_1$} -- (0,0);
    \end{tikzpicture}
    }}}  \normalsize
\end{split}
\end{equation}
where the scalar antifields $(\bar\phi{}^+,\phi^+)$ are depicted with the directed dotted lines.

\subsection{Homological Perturbation Theory} 
\label{sub:perturbative_calculations_finally}

To study correlation functions of braided scalar quantum electrodynamics, we need to extend the deformation retract of the original theory from \cref{sub:minimal_model_of_sqed} to the braided symmetric algebra $\Sym_\star L[2]$.
We use the ``symmetric tensor trick'' (reviewed in \cite{Berglund2014}) to extend the retract to a contraction of symmetric algebras, generalized to the symmetric monoidal category with non-trivial braiding isomorphism in which our $L_\infty$-algebra $\CCL^\star$ lives.

We represent our extended deformation retract by a diagram of ``thick'' maps given by
\begin{equation} 
    \begin{tikzcd}\label{eq:HEdatasym}
        \arrow[loop left, distance=3em, start anchor={[yshift=-1ex]west}, end anchor={[yshift=1ex]west}]{}{\mathsf{H}} 
        (\mathrm{Sym}_\star L[2], \boldsymbol\ell^\star_1  ) 
        \arrow["\mathsf{P}_0 "]{r} 
        & ( \mathrm{Sym}_\star H^\bullet( L[2]), 0) \arrow[bend left, " \mathsf{I}_0"]{l}
    \end{tikzcd}\ .
\end{equation}
Here we used translation invariance of the differential $\ell_1^\star$ of the braided $L_\infty$-algebra $\CCL^\star$ to extend it to a differential $\boldsymbol\ell_1^\star : \mathrm{Sym}_\star L[2] \longrightarrow (\mathrm{Sym}_\star L[2])[1]$ of symmetric degree $0$ as a strict graded derivation.
The maps in \eqref{eq:HEdatasym} are all of symmetric degree $0$ while their respective total degrees are inherited from the minimal model of \cref{sub:minimal_model_of_sqed}.

The homotopy equivalence is captured by the relation
\begin{equation}\label{eq:HEdata}
    \boldsymbol\ell^\star_1 \circ\mathsf{H} + \mathsf{H} \circ \boldsymbol\ell^\star_1 + \mathrm{id}_{\mathrm{Sym}_\star L[2]} = \mathsf{I}_0 \circ \mathsf{P}_0 \ .
\end{equation}
By the isomorphism $L[2] \simeq L[1]^*$ induced by the cyclic inner product, the degree 0 cochain maps $\iota$ and $ p$ extend as algebra homomorphisms with $\mathsf{I}_0 = {}^{\rm t}p$ and $ \mathsf{P}_0 = {}^{\rm t}\iota$ in symmetric degree~$1$. Since any deformation retract can be refined to a strong deformation retract, we can assume without loss of generality the relations
\begin{equation}
\begin{split}\label{eq:SDR}
        \mathsf{P}_0 \circ \mathsf{I}_0 = \mathrm{id}_{\mathrm{Sym}_\star H^\bullet( L[2])}\quad , \quad 
        \mathsf{P}_0 \circ \mathsf{H} = 0\quad , \quad 
        \mathsf{H} \circ \mathsf{I}_0 = 0\quad , \quad 
        \mathsf{H}^2 = 0
\end{split}\ .
\end{equation}

Vacuum correlation functions are computed by projecting to the trivial degree~$0$ subspace  $\mathbbm{R} \subset \Sym_\star H^\bullet(L[2])$~\cite{Masuda:2020tfa,Okawa:2022sjf}; this is done using the map $\sP_0:\Sym_\star L[2]\longrightarrow (\Sym_\star H^\bullet(L[2]))^0$ defined by
\begin{align}\label{eq:trivial project}
    \sP_0(\mathbbm{1})=1 \qquad \mbox{and} \qquad \sP_0(\varphi_1\odot_\star\cdots\odot_\star\varphi_n) = 0 \ ,
\end{align}
for all $\varphi_i\in L[2]$, with $i=1,\dots,n$. Then $\mathsf{I}_0:\Sym_\star H^\bullet(L[2])\longrightarrow (\Sym_\star L[2])^0$ is the canonical inclusion $\FR\subset\Sym_\star L[2]$ sending all higher degree elements to zero. The associated contracting homotopy is given by the map $\sH:\Sym_\star (L[2])\longrightarrow\Sym_\star (L[2])[-1]$ of symmetric degree 0 defined as
\begin{align}\label{eq:sfH}
\begin{split}
    \sH(\mathbbm{1})&=0 \ , \\[4pt]
    \sH(\varphi_1\odot_\star\cdots\odot_\star\varphi_n) &= \frac1n \, \sum_{i=1}^n \, \pm \ \varphi_1\odot_\star\cdots\odot_\star\varphi_{i-1}\odot_\star\sfh(\varphi_i) \odot_\star\varphi_{i+1} \odot_\star\cdots\odot_\star \varphi_n \ ,
\end{split}
\end{align}
where we used translation invariance of $\sfh$ in \eqref{eq:sfH} which trivializes the actions of $R$-matrices; here and in the following all explicit Koszul sign factors $\pm$ can be found in~\cite{Nguyen:2021rsa}. 

The braided extension of the homological perturbation lemma~\cite{Nguyen:2021rsa} allows one to deform this homotopy equivalence data with differential $\boldsymbol\ell^\star_1 + \boldsymbol{\delta}$ for ``small'' translation invariant perturbation map $\boldsymbol{\delta}$, i.e. for which the map $\mathrm{id}_{\mathrm{Sym}_\star L[2]} - \boldsymbol{\delta}\, \mathsf{H}$ is invertible, while preserving the strong deformation retract structure \eqref{eq:SDR}.
The deformation retract \eqref{eq:HEdatasym} is then perturbed to
\begin{equation}\label{eq:perturbed data of BV}
    \begin{tikzcd}
        \arrow[loop left, distance=3em, start anchor={[yshift=-1ex]west}, end anchor={[yshift=1ex]west}]{}{\mathsf{H} + \mathsf{H_{\mbf\delta}}} 
        (\mathrm{Sym}_\star L[2], \boldsymbol\ell^\star_1  + \boldsymbol{\delta}) 
        \arrow["\mathsf{P}_0 + \mathsf{P}_{\boldsymbol{\delta}}"]{rr} 
        & & ( \mathrm{Sym}_\star H^\bullet( L[2]), {\boldsymbol{\delta}}') \arrow[bend left, "\mathsf{I}_0 + \mathsf{I_{\mbf\delta}}"]{ll}
    \end{tikzcd}
\end{equation}
where the projection operator is modified by the map $\mathsf{P}_{\boldsymbol{\delta}} : \Sym_\star L[2]\longrightarrow \mathbbm{R} = (\Sym_\star H^\bullet(L[2]))^0$ given by
\begin{equation}
    \begin{split}\label{eq:perturbedmaps}
    \mathsf{P}_{\mbf\delta} &= \mathsf{P}_0 \,\big(\mathrm{id}_{\mathrm{Sym}_\star L[2]} - \boldsymbol{\delta}\, \mathsf{H}\big)^{-1}\, \boldsymbol{\delta} \, \mathsf{H} 
    =\mathsf{P}_0 \, \sum_{k=1}^\infty\, ( \boldsymbol{\delta}\, \mathsf{H})^k \ .
    \end{split}
\end{equation}
The explicit forms of the remaining perturbed maps in \eqref{eq:perturbed data of BV} can be found in~\cite{Nguyen:2021rsa}.

For a perturbation datum $\mbf\delta$, we denote the perturbed differential by
\begin{align}
Q_{\mbf\delta} := \boldsymbol\ell^\star_1 + {\mbf\delta} \ .
\end{align}

\paragraph{Free Theory.} 

The free theory is constructed from the perturbed differential
\begin{align}
Q_{\hbar} := \boldsymbol\ell_1^\star + \ii\,\hbar\,\BVL \ ,
\end{align}
which is an example of a quantum differential, see e.g.~\cite{Fuster:2005eg}.
The equivariant cochain map 
\begin{align}
\mathsf{\Delta}_{\textrm{\tiny BV}}:\Sym_\star L[2]\longrightarrow (\Sym_\star L[2])[1]
\end{align}
of symmetric degree $-2$ is called the BV Laplacian, since it reduces the symmetric degree by $2$. It satisfies $(\BVL)^2=0$ and $\boldsymbol\ell_1^\star\circ\BVL=-\BVL\circ\boldsymbol\ell_1^\star$, which together guarantee $(Q_\hbar)^2=0$. Explicitly, it is defined by extending the differential
\begin{align}\label{eq:BVL}
\begin{split}
\mathsf{\Delta}_{\textrm{\tiny BV}}(\mathbbm{1})=0 & 
\qquad , \qquad \mathsf{\Delta}_{\textrm{\tiny BV}}(\varphi_1)=0 
\qquad , \qquad  \mathsf{\Delta}_{\textrm{\tiny BV}}(\varphi_1\odot_\star\varphi_2) = \langle\varphi_1,\varphi_2\rangle_\star\, \mathbbm{1} 
\end{split}
\end{align}
on the symmetric algebra to 
\begin{equation}
\begin{split}
\mathsf{\Delta}_{\textrm{\tiny BV}}(\varphi_1\odot_\star\cdots\odot_\star\varphi_n) 
& = \sum_{i<j} \, \pm \, \langle \varphi_i,\sfR_{\alpha_{i+1}}\cdots\sfR_{\alpha_{j-1}}(\varphi_j)\rangle_\star 
\ \varphi_1\odot_\star\cdots\odot_\star\varphi_{i-1}\\ 
& \hspace{1.2cm} \odot_\star\sfR^{\alpha_{i+1}}(\varphi_{i+1})\odot_\star\cdots\odot_\star\sfR^{\alpha_{j-1}}(\varphi_{j-1})\odot_\star \varphi_{j+1}\odot_\star\cdots\odot_\star\varphi_n \ ,
\end{split}
\end{equation}
for all $\varphi_1,\dots,\varphi_n\in L[2]$. 

This yields the strong deformation retract
\begin{equation}\label{eq:Delta_perturbed}
    \begin{tikzcd}
        \arrow[loop left, distance=3em, start anchor={[yshift=-1ex]west}, end anchor={[yshift=1ex]west}]{}{\mathsf{H} + \mathsf{H_\hbar}} 
        (\mathrm{Sym}_\star L[2], \boldsymbol\ell_1^\star + \mathrm{i}\, \hbar\, \mathsf{\Delta}_{\textrm{\tiny BV}}) 
        \arrow["\mathsf{P}_0 + \mathsf{P}_\hbar"]{rr} &
        & ( \mathrm{Sym}_\star H^\bullet( L[2]), \mathrm{i}\, \hbar\, \mathsf{\Delta}_{\textrm{\tiny BV}}') 
        \arrow[bend left, "\mathsf{I}_0 +\mathsf{I}_\hbar"]{ll}
    \end{tikzcd} \ .
\end{equation}
Since $\mathsf{H}\, \mathsf{\Delta}_{\textrm{\tiny BV}}\, \mathsf{I}_0 = 0 $, 
 the induced BV differential on $\mathrm{Sym}_\star H^\bullet( L[2])$ is given by~\cite{Doubek:2017naz} $\mathsf{\Delta}_{\textrm{\tiny BV}}' = \mathsf{P}_0\, \mathsf{\Delta}_{\textrm{\tiny BV}}\, \mathsf{I}_0$.

\paragraph{Interacting Theory.} 

The interacting quantum theory is constructed by further deforming the BV Laplacian with the interactions $\mathcal{S}_{\mathrm{int}}$ from \cref{sub:perturbative_calculations_finally1}, giving the degree 1 small perturbation
\begin{equation} \label{eq:deltahbarint}
    \boldsymbol{\delta}_{\hbar,{\rm int}} =  \mathrm{i}\, \hbar\, {\mathsf \Delta}_\BV +  \left\{\mathcal{S}_{\rm int} ,  - \right\}_\star \ ,
\end{equation}
and the quantum differential $Q_{\hbar,{\rm int}} := \boldsymbol\ell_1^\star + \boldsymbol{\delta}_{\hbar,{\rm int}}$.
The BV antibracket 
\begin{align}
\{-,-\}_\star:\Sym_\star L[2]\otimes\Sym_\star L[2] \longrightarrow (\Sym_\star L[2])[1]
\end{align}
is the braided $(-1)$-shifted Poisson structure 
defined on generators $\varphi_1,\varphi_2\in L[2]$ using the cyclic inner product $\langle-,-\rangle_\star$ of the braided $L_\infty$-algebra $\CCL^\star$ as
\begin{align}
\{\varphi_1,\varphi_2\}_\star 
    = \langle\varphi_1 , \varphi_2\rangle_\star \, \mathbbm{1} 
    =\pm\,
    \,\{\sfR_\alpha(\varphi_2),\sfR^\alpha(\varphi_1)\}_\star \ ,
\end{align}
and extended as a braided graded derivation in each of its slots. 

Because of cyclicity of the inner product $\langle-,-\rangle_\star$, the differential~${\boldsymbol\ell}^\star_1$ is compatible with the antibracket, making $\big( \mathrm{Sym}_\star L[2], {\boldsymbol\ell}^\star_1, \{- ,-\}_\star)$ into a braided $P_0$-algebra~\cite{Nguyen:2021rsa}. The braided $L_\infty$-structure implies that the full classical theory with BV functional $\mathcal{S}_\BV$ from \eqref{eq:BVactionext} satisfies the classical master equation $\{\CS_\BV,\CS_\BV\}_\star=0$.
On the other hand, the failure of the BV Laplacian to be a strict graded derivation of the braided commutative product is measured by 
\begin{align}\label{eq:BVLantibracket}
    \mathsf{\Delta}_{\textrm{\tiny BV}}(a_1\odot_\star a_2) = \mathsf{\Delta}_{\textrm{\tiny BV}}(a_1)\odot_\star a_2 
    \pm
    a_1\odot_\star \mathsf{\Delta}_{\textrm{\tiny BV}}(a_2) +
    \{a_1,a_2\}_\star \ ,
\end{align}
for all $a_1,a_2\in\Sym_\star L[2]$, making $\big( \mathrm{Sym}_{\star}L[2], {\boldsymbol\ell}^\star_1, \left\{- , -\right\}_\star, \mathsf{\Delta}_{\textrm{\tiny BV}}\big)$ into a differential graded braided BV algebra. This implies that the BV differential act as a strict graded derivation of the Poisson structure~\cite{Nguyen:2021rsa}:
\begin{equation}
\begin{split}\label{eq:BV derivation poisson}
    \mathsf{\Delta}_{\textrm{\tiny BV}} (\left\{ a_1, a_2 \right\}_\star) = 
    \{\mathsf{\Delta}_{\textrm{\tiny BV}} (a_1), a_2\}_\star \pm \{a_1, \mathsf{\Delta}_{\textrm{\tiny BV}} (a_2) \}_\star \ .
\end{split}
\end{equation}

This perturbs the free strong deformation retract to
\begin{equation}\label{eq:sdr_full}
    \begin{tikzcd}
        \arrow[loop left, distance=3em, start anchor={[yshift=-1ex]west}, end anchor={[yshift=1ex]west}]{}{\mathsf{H} + \mathsf{H}_{\hbar,{\rm int}}} 
        \left(\mathrm{Sym}_\star L[2], \boldsymbol\ell_1^\star + \mathrm{i}\, \hbar\, \mathsf{\Delta}_{\textrm{\tiny BV}} + \{\mathcal{S}_{\rm int}, -\}_\star \right) 
        \arrow["\mathsf{P}_0 +\mathsf{P}_{\hbar,{\rm int}}"]{rr} &
        & ( \mathrm{Sym}_\star H^\bullet( L[2]), \boldsymbol{{\delta}}_{\hbar,{\rm int}}') 
        \arrow[bend left, "\mathsf{I}_0 + \mathsf{I}_{\hbar,{\rm int}}"]{ll}
    \end{tikzcd} \ .
\end{equation}
The nilpotency condition for the deformed differential $ {\boldsymbol\ell}^\star_1 + \boldsymbol{\delta}_{\hbar,{\rm int}}$ on $\mathrm{Sym}_{\star} L[2]$ is equivalent to the quantum master equation
\begin{equation}\label{eq:qme}
    \boldsymbol\ell_1^\star(\CS_{\rm int}) + \mathrm{i}\, \hbar \, {\mathsf{\Delta}}_\BV(\mathcal{S}_{\rm int}) + \tfrac{1}{2}\, \{ \mathcal{S}_{\rm int}, \mathcal{S}_{\rm int} \}_\star = 0 
\end{equation}
for the degree~0 cochain $\CS_{\rm int}$, which follows trivially from the classical master equation for $\CS_\BV$ together with $\BVL(\CS_\BV)=0$~\cite{Nguyen:2021rsa,Giotopoulos:2021ieg, DimitrijevicCiric:2023hua}.

\paragraph{Using the Homological Perturbation Lemma.} 

In this paper we will compute various one-loop correlation functions of physical fields in braided scalar quantum electrodynamics. Correlation functions are computed via the homological perturbation lemma by applying the perturbed projector \eqref{eq:perturbedmaps} with the deformation retract \eqref{eq:sdr_full} to insertions of \textit{antifields} on external states at positions $x_i\in\FR^{1,3}$. 

For this, we use the momentum space decompositions \eqref{eq:momentum basis for AAplus} and \eqref{eq:momentum bases for phiphisplus} for the field and antifield content of braided scalar electrodynamics. 
The external antifield states are related to a position space basis by Fourier transformation
\begin{equation}\label{eq:postion space FT}
    \delta^{A_\mu}_{x_i} =\int_{p_i}\, \mathtt{v}_\mu\, \mathtt{e}^{p_i} \ \mathrm{e}^{-\mathrm{i}\,p_i \cdot x_i} \quad , \quad
    \delta^{\bar\phi}_{x_i} =\int_{p_i}\, \bar{\mathtt{e}}^{p_i} \ \mathrm{e}^{-\mathrm{i}\,p_i \cdot x_i}
    \quad , \quad
    \delta^\phi_{x_i} = \int_{p_i}\, \mathtt{e}^{p_i} \ \mathrm{e}^{-\mathrm{i}\,p_i \cdot x_i}\ .
\end{equation}

Contractions of states are achieved through application of the operator $\BVL\,\sH$.
By contracting external states with each other using the cyclic structure and contracting homotopy, we recover the respective free propagators in position space: for the scalar fields
\begin{equation}\label{eq:contraction_positionsp}
    \langle \delta_{x_i}^\phi , \mathsf{h} \, \delta_{x_j}^{\bar\phi} \rangle_\star
    = -\mathsf{G}(x_i - x_j) = - \int_k \,\tilde{\mathsf{G}}(k) \ \mathrm{e}^{-\mathrm{i}\, k\cdot (x_i - x_j)} \ ,
\end{equation}
and similarly for the photon contractions.
For the scalar fields, contractions between internal and external states yield
\begin{equation}
    \langle \mathtt{e}^{k_i}\,,\, \mathsf{h}\, \delta_{x_j}^{\bar\phi} \rangle_\star 
    = -\tilde{\mathsf{G}}(k_i) \, \mathrm{e}^{\,\mathrm{i}\, k_i \cdot x_j} 
    = \langle \mathsf{h}\, \mathtt{e}^{k_i}\,,\, \delta_{x_j}^{\bar\phi} \rangle_\star \ ,
\end{equation}
and similarly for the photon internal/external state contractions.

In this paper we mostly use the momentum space representation, electing to Fourier transform the correlation function at the source.
Internal contractions of two fields using $\Delta_{\BV}\, \mathsf{H}$ in momentum space are described by \eqref{eq:momentum basis for AAplus}, \eqref{eq:momentum bases for phiphisplus} and \eqref{eq:basis elements} together with \eqref{eq:contracting hom on coordfunc}. They satisfy
\begin{equation}\label{eq:cylicgivesinternalprop1}
    \langle \mathtt{v}_\mu\, \mathtt{e}^{k_i}, \mathsf{h}\, \mathtt{v}_\nu\, \mathtt{e}^{k_j} \rangle_\star 
    = - \mathsf{\tilde{D}}(k_i) \, \mathsf{\tilde\Pi}_{\mu \nu}(k_i) \ (2 \pi)^4 \, \delta(k_i + k_j) 
\end{equation}
for the photon field, which we represent graphically by the wavy line
\begin{equation}
\begin{split}\label{eq:propag graph1}
\vcenter{\hbox{\begin{tikzpicture}[scale=0.5]{\scriptsize
\node  at (-0.3, 0.4) { $\mu$};
\node  at (4.3, 0.4) { $\nu$};
\draw[Aplus] (0,0) -- (4,0);
\draw[<-, shift={(1.5,0.5)}] (0:0.8) node[above]{$k_i$} -- (0,0); } \normalsize
\end{tikzpicture}}}
\end{split}
\end{equation}
and
\begin{equation}\label{eq:cylicgivesinternalprop2}
\langle \mathtt{e}^{k_i}, \mathsf{h}\, \bar{\mathtt{e}}^{k_j} \rangle_\star 
    = - \mathsf{\tilde{G}}(k_i) \ (2 \pi)^4 \, \delta(k_i + k_j) 
\end{equation}
for the scalar fields, which we represent graphically by the directed straight line
\begin{equation}
\begin{split}\label{eq:propag graph2}
\vcenter{\hbox{\begin{tikzpicture}[scale=0.5]{\scriptsize
\draw[scalplusreverse] (0,0) -- (4,0);
\draw[<-, shift={(1.5,0.5)}] (0:0.8) node[above]{$k_i$} -- (0,0); } \normalsize
\end{tikzpicture}}} 
\end{split}
\end{equation}

Finally, the interaction vertices are generated by the BV antibrackets with the translation invariant degree~$0$ extended action functionals \eqref{eq:S3int} and \eqref{eq:one term} from \cref{sub:perturbative_calculations_finally1}:
\begin{equation}
\begin{split}\label{eq:vertices}
    \{ \mathcal{S}^\swthree_{\rm int}, - \}_\star 
    & = \int_{k_1,k_2,k_3} \, V^{\mu}_3(k_1,k_2,k_3) \
    \big\{  \mathtt{v}_{\mu} \, \mathtt{e}^{k_1} \odot_{\star} \mathtt{e}^{k_2}\odot_\star \bar{\mathtt{e}}^{k_3},   -  \big\}_\star  \ , \\[4pt]
    \{\mathcal{S}_{\rm int}^{\swfour} , - \}_\star
    &= \int_{k_1, k_2, k_3,k_4} \,
    V_4^{\mu\nu}(k_1, k_2, k_3, k_4) \
    \big\{
    \mathtt{v}_{\mu}\,\mathtt{e}^{k_1} 
    \odot_\star 
    \mathtt{e}^{k_2} 
    \odot_\star 
    \bar{\mathtt{e}}^{k_3} 
    \odot_\star 
    \mathtt{v}_{\nu}\,\mathtt{e}^{k_4} 
    , - \big\}_\star \ .
\end{split}
\end{equation}
Thanks to momentum conservation across vertices, the expansions \eqref{eq:vertices} in terms of a basis of antifields can always be cyclically rotated to better suit calculations; we will always act from the left. For the same reason, the interacting action functional acts as a \textit{strict} graded derivation on $\mathrm{Sym}_\star L[2]$, that is, the $R$-matrices trivialise when moving the vertex across a braided symmetric product; equivalently, $\mathcal{S}_\mathrm{int}$ is translation invariant. We will only consider degree 0 elements of $L[2]$, so this further restricts to a derivation:
    \begin{equation}
    \begin{split}
        \left\{ \mathcal{S}_{\rm int}, \varphi_1 \odot_\star \varphi_2 \right\}_\star
        = \left\{ \mathcal{S}_{\rm int}, \varphi_1 \right\}_\star \odot_\star \varphi_2
        + 
        \varphi_1 \odot_\star \left\{ \mathcal{S}_{\rm int}, \varphi_2  \right\}_\star
    \end{split}\ ,
    \end{equation}
    for $\varphi_1, \varphi_2 \in L[2]^0$.
    
\section{Correlation Functions at One-Loop} 
\label{sec:correlation_functions_in_braided_scalar_qed}

In this paper we are interested in the $n$-point correlation functions of braided scalar quantum electrodynamics by perturbed projection maps \eqref{eq:perturbedmaps} acting on antifield insertions of Dirac distributions supported at spacetime points $x_i$.
Our basis of test Dirac distributions is given by the set \smash{$\big\{ \delta_{x_i}^{\phi}, \delta_{x_i}^{\bar\phi}, \delta_{x_i}^{A_\mu}  \big\} \subset L[2]^0$} described in \cref{sub:perturbative_calculations_finally}, of which a generic element is denoted $\delta^{\mathcal{A}_i}_{x_i}$.

We define correlation functions in this setting as
\begin{equation}
    \begin{split}\label{eq:BQFTnpoint}
    \mathsf{G}^{\mbf\delta}_{\CA_1,\dots,\CA_n }({x_1}, \ldots, {x_n}) 
    &=
    \mathsf{P}_{\mbf\delta}
    \big(\delta^{\mathcal{A}_1}_{x_1} \odot_\star \cdots \odot_\star \delta^{\mathcal{A}_n}_{x_n}\big) =
    \sum_{k=1}^\infty \, \sP_0\,({\mbf\delta} \,\sH)^k\,
    \big(\delta^{\mathcal{A}_1}_{x_1} \odot_\star \cdots \odot_\star \delta^{\mathcal{A}_n}_{x_n}\big) \ .
    \end{split}
\end{equation}
Because only $\sP_0(\mathbbm{1})=1$ is non-zero, the correlation function is a map 
\begin{align}
\sG_n^{\mbf\delta} : (\mathrm{Sym}_{\star}L[2] )^0 \longrightarrow \Omega^0\big((\FR^{1,3})^{\times n}\big) \ , 
\end{align}
understood in a distributional sense. For the same reason only correlators involving the same number of scalar antifields $\delta_{x_i}^{\phi}$ and their conjugates \smash{$\delta_{x_i}^{\bar\phi}$} are non-vanishing, as only such pairs can lead to non-vanishing inner products; this is the statement of charge conjugation symmetry in braided scalar QED. We will generically work with the strong deformation retract \eqref{eq:sdr_full} for which 
\smash{$\sG_n^{\hbar,{\rm int}}:=\sG_n^{\mbf\delta_{\hbar,{\rm int}}}$} is a formal power series in $\hbar$ and $g$; we write \smash{$\sG_n^\hbar$} for the correlators of the free field theory associated to the deformation retract \eqref{eq:Delta_perturbed}.

Alternatively, we can decompose a general antifield in terms of a basis \smash{$\mathtt{e}^{p_i}_{\mathcal{A}_i}$} of plane waves dual to a basis \smash{$ \mathtt{e}_{p_i}^{\mathcal{A}_i} \in L[2]^0$} as
\begin{equation}
\begin{split}
    \delta^{\mathcal{A}_i}_{x_i} =  \int_{p_i} \, \mathtt{e}_{\mathcal{A}_i}^{p_i} \ \mathrm{e}^{-\mathrm{i}\, p_i \cdot x_i} \ .
\end{split}
\end{equation}
This defines the $n$-point correlation function in momentum space as
\begin{equation}
\begin{split}\label{eq:bqftnpointmom}
    \mathsf{\tilde{G}}^{\mbf\delta}_{\mathcal{A}_1, \ldots, \mathcal{A}_n}(p_1, \ldots, p_n) &= 
   \sum_{k=1}^\infty \, \mathsf{P}_0\,(\boldsymbol{\delta} \,\mathsf{H})^k \,
    \big(
    \mathtt{e}^{p_1}_{\mathcal{A}_1} 
    \odot_\star \cdots \odot_\star
    \mathtt{e}^{p_n}_{\mathcal{A}_n}
    \big) \ .
\end{split}
\end{equation}
This prescription computes both connected and disconnected diagrams with $n$ external propagators. In this section we detail the calculation of several correlators up to one-loop order.

These are vacuum correlation functions in the terminology of \cite{DimitrijevicCiric:2023hua}. This is unlike the computation of scattering amplitudes done in \cite{Gomez2021,Szabo:2023cmv}, where the cyclic $L_\infty$-structure is pulled back to the minimal model, which yields infinitely many higher brackets encoding the $n$-point amplitudes.
In the present setting, the analogue of the Lehmann-Symanzik-Zimmermann (LSZ) reduction formula is obtained by considering a basis of fields in $L[2]^0$ that are $\ell^\star_1$-exact.

Throughout this paper we will mostly leave aside two important analytic issues to avoid clouding the main messages of this paper. Firstly, many of our formal manipulations can only be made precise when one uses wavepackets for the external particles, rather than delta-distributions. Secondly, we do not explicitly spell out regularisations of loop integrals, which are also needed to make sense of our algebraic manipulations. Both of these technical problems can be handled using standard quantum field theory techniques.

\subsection{One-Point Functions} 
\label{sub:one_point_functions}

As a warm-up, let us start by computing the interacting one-point correlation functions using the homological perturbation lemma with the deformation retract \eqref{eq:sdr_full}, confirming that all tadpole diagrams vanish.

Using \eqref{eq:cylicgivesinternalprop1} and \eqref{eq:cylicgivesinternalprop2}, the one-point photon correlator at one-loop in momentum space is calculated as
\begin{equation}
\begin{split}\label{eq:1pointphoton}
     \mathsf{\tilde G}^{\hbar,{\rm int}}_{A_{\mu}}(p)^\swone
    & =  \mathrm{i}\, \hbar\, \BVL\, \mathsf{H}\,
    \big\{\mathcal{S}_{\rm int}^{\swthree}\,,\,\mathsf{h}\,\mathtt{v}_\mu\mathtt{e}^p \big\}_\star \\[4pt]
    & = \mathrm{i}\, \hbar\, \BVL\, \mathsf{H}\, \int_{k_1,k_2, k_3}\, V_3^{\nu} (k_1, k_2, k_3) \
    \big\{ \mathtt{v}_\nu\, \mathtt{e}^{k_1}\odot_{\star} \mathtt{e}^{k_2}\odot_{\star} \bar{\mathtt{e}}^{k_3} \,,\, \mathsf{h}\, \mathtt{v}_\mu\,\mathtt{e}^p \big\}_\star \\[4pt]
    & = -\mathrm{i}\, \hbar\, \BVL\, \mathsf{H}\, \int_{k_1,k_2, k_3}\, V_3^{\nu} (k_1, k_2, k_3) \ 
    \mathrm{e}^{\,\mathrm{i}\,(k_2+k_3)\cdot \theta\, k_1} \,  (2\pi)^4\,\delta(p+k_1)\\
    & \quad \, \hspace{8cm} \times \tilde{\mathsf{D}}(p)\, \tilde{\mathsf{\Pi}}_{\nu\mu}(p) \ \mathtt{e}^{k_2}\odot_{\star} \bar{\mathtt{e}}^{k_3} \\[4pt]
& = \mathrm{i}\, \hbar\, \int_{k_1,k_2, k_3}\, V_3^{\nu} (k_1, k_2, k_3) \ (2\pi)^4\,\delta(k_2+k_3)\, \tilde{\mathsf{G}}(k_2) \,
(2\pi)^4\,\delta(p+k_1)\, \tilde{\mathsf{D}}(p)\, \tilde{\mathsf{\Pi}}_{\nu\mu}(p) \\[4pt]
& = 2\, \mathrm{i}\, \hbar\, g\, \int_k\, k^{\nu}\, \tilde{\mathsf{G}}(k)\, (2\pi)^4\,\delta(p)\,\tilde{\mathsf{D}}(p)\, \tilde{\mathsf{\Pi}}_{\nu\mu}(p) = 0 \ .
 \end{split}
\end{equation}
In the last line we formally set $\int_k\, k^\nu\, \tilde{\mathsf{G}}(k) = 0$, after reflection symmetric cutoff regularisation, because the integrand is an odd function of the loop
momentum. This vanishing contribution is represented graphically by the tadpole diagram
\begin{equation}
\begin{split}
\vcenter{\hbox{\begin{tikzpicture}[scale=0.4]
{\scriptsize
    \draw[Aplus] (0,-1) -- (0,0); 
    \draw[scalplus] (0,0) arc (-90:270:0.8); 
    } \normalsize
    \end{tikzpicture} }} \ = \ 0
\end{split}
\end{equation}

For the tadpoles of charged scalars at one-loop we find
\begin{equation}
\begin{split}
    \mathsf{\tilde G}^{\hbar,{\rm int}}_{\phi}(p)^\swone & =  \mathrm{i}\, \hbar\, \BVL\, \mathsf{H}\,
    \big\{\mathcal{S}_{\rm int}^{\swthree}\,,\,\mathsf{h}\,\mathtt{e}^p \big\}_\star \\[4pt]
    & = \mathrm{i}\, \hbar\, \BVL\, \mathsf{H}\, \int_{k_1,k_2, k_3}\, V_3^{\nu} (k_1, k_2, k_3) \
    \big\{ \mathtt{v}_\nu \, \mathtt{e}^{k_1}\odot_{\star} \mathtt{e}^{k_2}\odot_{\star} \bar{\mathtt{e}}^{k_3}\,,\, \mathsf{h}\, \mathtt{e}^p \big\}_\star \\[4pt]
    & = -\mathrm{i}\, \hbar\, \BVL\, \mathsf{H}\, \int_{k_1,k_2, k_3}\, V_3^{\nu} (k_1, k_2, k_3) \
    (2\pi)^4\,\delta(p+k_3)\,\tilde{\mathsf{G}}(p) \ \mathtt{v}_\nu\, \mathtt{e}^{k_1}\odot_{\star} \mathtt{e}^{k_2}  = 0 \ ,
    \end{split}
\end{equation}
where the last line vanishes because there are no contractions between photons and scalars: 
\begin{align}
\mathsf{\Delta}_{\textrm{\tiny BV}}\, \mathsf{H}\,\big(\mathtt{v}_\nu\, \mathtt{e}^{k_1}\odot_{\star} \mathtt{e}^{k_2}\big) = 0 \ .
\end{align}
Similarly, \smash{$ \mathsf{\tilde G}^{\hbar,{\rm int}}_{\bar\phi}(p)^\swone  =  \mathrm{i}\, \hbar\, \BVL\, \mathsf{H}\,
 \big\{\mathcal{S}_{\rm int}^{\swthree}\,,\,\mathsf{h}\,\bar{\mathtt{e}}^{p} \big\}_\star = 0$}
by the same argument.


%
\subsection{Photon Vacuum Polarisation} 
\label{sub:photon_polarisation}

Using the definition (\ref{eq:bqftnpointmom}) of the $n$-point correlation function in momentum space,  we now proceed to calculate the photon two-point function. 

The tree-level photon two-point function is given by the leading order contribution in the homological perturbation lemma and coincides with the bare photon propagator:
\begin{equation}
\begin{split}\label{eq:barephotonpropbv}
    \mathsf{\tilde{G}}^{\hbar}_{A_\mu,A_\nu}(p_1,p_2)^\swzero &= 
     \mathrm{i}\,\hbar\,\mathsf{\Delta}_{\textrm{\tiny BV}}\, \mathsf{H}\,
    ( \mathtt{v}_\mu \,\mathtt{e}^{p_1} \odot_\star \mathtt{v}_\nu\, \mathtt{e}^{p_2}) \\[4pt]
    &=  - \mathrm{i}\, \hbar \, (2 \pi)^4\, \delta(p_1 + p_2)\, \mathsf{\tilde{D}}(p_1)\, \mathsf{\tilde\Pi}_{\mu \nu}(p_1) \\[4pt]
    &=   \mathrm{i}\, \hbar \, (2 \pi)^4\, \delta(p_1 + p_2) \,\frac{1}{p_1^2}\, \left(\eta_{\mu \nu} - \frac{p_{1\, \mu}\, p_{1\, \nu}}{p_1^2}\right) \ .
\end{split}
\end{equation}

At order $\hbar^2$, there are three terms that contribute to the two-point  function. They are given by
\begin{equation}
\begin{split}\label{eq:2-point photon polar}
    \mathsf{\tilde{G}}^{\hbar,{\rm int}}_{A_\mu,A_\nu}(p_1,p_2)^\swone
    &=  
    (\mathrm{i}\,\hbar\,\mathsf{\Delta}_{\textrm{\tiny BV}}\, \mathsf{H} )^2
    \left\{ \mathcal{S}^\swthree_{\rm int},
    \mathsf{H}\left\{ \mathcal{S}^\swthree_{\rm int}, 
    \mathsf{H}\, ( \mathtt{v}_\mu \,\mathtt{e}^{p_1} \odot_\star \mathtt{v}_\nu\, \mathtt{e}^{p_2}) \right\}_\star \right\}_\star \\
    & \quad \, +
    \mathrm{i}\,\hbar\,\mathsf{\Delta}_{\textrm{\tiny BV}}\, \mathsf{H} 
    \left\{ \mathcal{S}^\swthree_{\rm int},
    \mathsf{H}\, (\mathrm{i}\,\hbar\,\mathsf{\Delta}_{\textrm{\tiny BV}}\, \mathsf{H})
    \left\{ \mathcal{S}^\swthree_{\rm int}, 
    \mathsf{H}\, ( \mathtt{v}_\mu\, \mathtt{e}^{p_1} \odot_\star \mathtt{v}_\nu\, \mathtt{e}^{p_2}) \right\}_\star \right\}_\star \\
    & \quad \, +    
    (\mathrm{i}\,\hbar\, \mathsf{\Delta}_{\textrm{\tiny BV}}\, \mathsf{H})^2
    \left\{ \mathcal{S}_{\rm int}^\swfour,
    \mathsf{H}\, (\mathtt{v}_\mu\, \mathtt{e}^{p_1} \odot_\star \mathtt{v}_\nu\, \mathtt{e}^{p_2})
    \right\}_\star\\[4pt]
    &=: \mathsf{\tilde{G}}_{\mu\nu}(p_1,p_2)_{\tt A} + \mathsf{\tilde{G}}_{\mu\nu}(p_1,p_2)_{\tt B} + \mathsf{\tilde{G}}_{\mu\nu}(p_1,p_2)_{\tt C} \ .
\end{split}
\end{equation}
Using the same techniques as for the calculation of tadpole diagrams in \cref{sub:one_point_functions}, as well as in the analogue calculations for braided spinor QED detailed in~\cite{DimitrijevicCiric:2023hua}, we now calculate each of these three terms separately. 

Compared to the one-point functions, the calculations for higher multiplicity correlators become increasingly more lengthy and cumbersome. They are simplified with the help of a diagrammatic calculus, similarly to those developed for (braided) scalar field theories in~\cite{Nguyen:2021rsa,Bogdanovic:2024jnf}, as well as in~\cite{Gaunt2022} for matrix models with gauge symmetries, which we employ below using the diagrammatic rules discussed in \cref{sub:perturbative_calculations_finally}. Throughout this paper, diagrams should be read from the bottom up, with input at the bottom and output at the top.
We explicitly add momentum labels in the initial instances and in final results, but otherwise mostly suppress them.

\paragraph{Evaluation of $\mbf{\mathsf{\tilde{G}}_{\mu\nu}(p_1,p_2)_{\tt A}}$.}

Denote the momentum space representation $\mathtt{v}_\mu\, \mathtt{e}^{p_1} \odot_\star \mathtt{v}_\nu\, \mathtt{e}^{p_2}$ by two vertical photon lines.
Apply the three-vertex operator $\big\{ \mathcal{S}^\swthree_{\rm int}, - \big\}_\star\,\mathsf{H} : \mathrm{Sym}_{\star}L[2] \longrightarrow \mathrm{Sym}_{\star}L[2]$ which is  a  strict derivation of total degree~0 by the discussion of \cref{sub:perturbative_calculations_finally}; however, it raises the symmetric degree by 1. This results in the non-zero terms
\begin{equation}
\begin{split}\label{eq:photonS3}
    \bigg\{ \mathcal{S}^\swthree_{\rm int}\,,\,
    \mathsf{H}\, \bigg(
    \,
    \vcenter{\hbox{\begin{tikzpicture}[scale=0.4]
    { \scriptsize
    \draw[Aplus] (0,-0.25) -- (0,2.25) node[above] { $ p_1$};
    \draw[Aplus] (1,-0.25) -- (1,2.25) node[above] { $ p_2$};
    } \normalsize
    \end{tikzpicture}}}
    \, \bigg)
    \bigg\}_\star
    &= \tfrac12\,\bigg(
    \vcenter{\hbox{
    \begin{tikzpicture}[scale=0.4]
    { \scriptsize
    \draw[Aplus] (0,-0.25) -- (0,1.25);
    \draw[scalplusreverse] (0,1.25) -- (-1,2.25) node[above] {$k_2$};
    \draw[scalplus] (0,1.25) -- (1,2.25) node[above] {$k_3$};
    \draw[Aplus] (2,-0.25) -- (2,2.25) node[above] {$p_2$};
    } \normalsize
    \end{tikzpicture}}}
    \
    +
    \
    \vcenter{\hbox{
    \begin{tikzpicture}[scale=0.4]
    { \scriptsize
    \draw[Aplus] (0.5,-0.25) -- (0.5,2.25) node[above] {$p_1$};
    \draw[Aplus] (2.75,-0.25) -- (2.75,1.25);
    \draw[scalplusreverse] (2.75,1.25) -- (1.75,2.25)  node[above] {$ k_2 $};
    \draw[scalplus] (2.75,1.25) -- (3.75,2.25)  node[above] {$ k_3$};
    } \normalsize
    \end{tikzpicture}}} \, \bigg)
    \end{split}
\end{equation}
Recall that we can also rotate the basis of antifields in the three-vertex \eqref{eq:S3int}.

Further application of the three-vertex operator $\big\{ \mathcal{S}^3_{\rm int}, - \big\}_\star\,\mathsf{H}$ results in six more non-zero terms. For the first term on the right-hand side of \eqref{eq:photonS3} this gives
\begin{equation}
\begin{split}
\bigg\{ \mathcal{S}^\swthree_{\rm int}\,,\,
    \mathsf{H}\, \bigg(
    \,
        \vcenter{\hbox{
    \begin{tikzpicture}[scale=0.4]
    { \scriptsize
    \draw[Aplus] (0,-0.25) -- (0,1.25);
    \draw[scalplusreverse] (0,1.25) -- (-1,2.25) node[above] {$k_2$};
    \draw[scalplus] (0,1.25) -- (1,2.25) node[above] {$k_3$};
    \draw[Aplus] (2,-0.25) -- (2,2.25) node[above] {$p_2$};
    } \normalsize
    \end{tikzpicture}}}
     \, \bigg)
    \bigg\}_\star
    &= \tfrac13\,\Bigg(\,
    \vcenter{\hbox{
    \begin{tikzpicture}[scale=0.4]
    {\scriptsize
    \draw[Aplus] (0,-0.25) -- (0,0.75);
    \draw[scalplusreverse] (0.0,0.75) -- (-0.75,1.5);
    \draw[Aplus] (-0.75,1.5) -- (-1.5, 2.25) node[above] {$k_4$};
    \draw[scalplusreverse] (-0.75,1.5) -- (0,2.25) node[above] {$k_5$};
    \draw[scalplus] (0.0,0.75) -- (1.5,2.25) node[above]{$k_3$};
    \draw[Aplus] (2.25, -0.25) -- (2.25,2.25) node[above]{$p_2$};
    } \normalsize
    \end{tikzpicture}
    }}
    \ + \
    \vcenter{\hbox{
    \begin{tikzpicture}[scale=0.4]
    {\scriptsize
    \draw[Aplus] (0,-0.25) -- (0,0.75);
    \draw[scalplus] (0.0,0.75) -- (0.75,1.5);
    \draw[Aplus]  (0.75,1.5) -- (1.5, 2.25)  node[above] {$k_4$};
    \draw[scalplus] (0.75,1.5) -- (0,2.25) node[above] {$k_6$} ;
    \draw[scalplusreverse] (0.0,0.75) -- (-1.5,2.25) node[above] {$k_2$};
    \draw[Aplus] (2.25, -0.25) -- (2.25,2.25) node[above] {$p_2$};
      } \normalsize
    \end{tikzpicture}}}
    \ + \
    \vcenter{\hbox{
    \begin{tikzpicture}[scale=0.4]
    {\scriptsize
    \draw[Aplus] (0,-0.25) -- (0,1.25);
    \draw[scalplusreverse] (0,1.25) -- (-1,2.25) node[above] {$k_2$};
    \draw[scalplus] (0,1.25) -- (1,2.25) node[above] {$k_3$};
    \draw[Aplus] (2.75,-0.25) -- (2.75,1.25);
    \draw[scalplusreverse] (2.75,1.25) -- (1.75,2.25) node[above] {$k_5$};
    \draw[scalplus] (2.75,1.25) -- (3.75,2.25) node[above] {$k_6$};
      } \normalsize
    \end{tikzpicture}}}
    \Bigg) \ ,
\end{split}
\end{equation}
and similarly for the second term
\begin{equation}
\begin{split}
\bigg\{ \mathcal{S}^\swthree_{\rm int}\,,\,
    \mathsf{H}\, \bigg(
    \,
        \vcenter{\hbox{
    \begin{tikzpicture}[scale=0.4]
    { \scriptsize
    \draw[Aplus] (0.5,-0.25) -- (0.5,2.25) node[above] {$p_1$};
    \draw[Aplus] (2.75,-0.25) -- (2.75,1.25);
    \draw[scalplusreverse] (2.75,1.25) -- (1.75,2.25)  node[above] {$ k_2 $};
    \draw[scalplus] (2.75,1.25) -- (3.75,2.25)  node[above] {$ k_3$};
    } \normalsize
    \end{tikzpicture}}}
     \, \bigg)
    \bigg\}_\star
    &= \tfrac13\,\Bigg(\,
    \vcenter{\hbox{
    \begin{tikzpicture}[scale=0.4]
    {\scriptsize
    \draw[Aplus] (0,-0.25) -- (0,1.25);
    \draw[scalplusreverse] (0,1.25) -- (-1,2.25)  node[above] {$ k_5$};
    \draw[scalplus] (0,1.25) -- (1,2.25)  node[above] {$ k_6 $};
    \draw[Aplus] (2.75,-0.25) -- (2.75,1.25);
    \draw[scalplusreverse] (2.75,1.25) -- (1.75,2.25)  node[above] {$ k_2 $};
    \draw[scalplus] (2.75,1.25) -- (3.75,2.25)  node[above] {$ k_3$};
     } \normalsize
    \end{tikzpicture}}}
    \ + \
    \vcenter{\hbox{
    \begin{tikzpicture}[scale=0.4]
    {\scriptsize
    \draw[Aplus] (0,-0.25) -- (0,0.75);
    \draw[scalplusreverse] (0.0,0.75) -- (-0.75,1.5);
    \draw[Aplus] (-0.75,1.5) -- (-1.5, 2.25) node[above] {$ k_4$};
    \draw[scalplusreverse] (-0.75,1.5) -- (0,2.25) node[above] {$ k_5$} ;
    \draw[scalplus] (0.0,0.75) -- (1.5,2.25) node[above] {$ k_3$};
    \draw[Aplus] (-2.25, -0.25) -- (-2.25,2.25) node[above] {$p_1$};
    } \normalsize
    \end{tikzpicture}}}
    \ + \
    \vcenter{\hbox{
    \begin{tikzpicture}[scale=0.4]
    {\scriptsize
    \draw[Aplus] (0,-0.25) -- (0,0.75);
    \draw[scalplus] (0.0,0.75) -- (0.75,1.5);
    \draw[Aplus] (0.75,1.5) -- (1.5, 2.25) node[above] {$ k_4$};
    \draw[scalplus] (0.75,1.5) -- (0,2.25) node[above] {$ k_6$}  ;
    \draw[scalplusreverse] (0.0,0.75) -- (-1.5,2.25) node[above] {$ k_2$};
    \draw[Aplus] (-2.25, -0.25) -- (-2.25,2.25) node[above] {$ p_1$};
      } \normalsize
    \end{tikzpicture}}}
    \Bigg) \ .
\end{split}
\end{equation}
Note that any extra noncommutative phase factors can be removed by braiding the basis of antifields in the symmetric algebra. Contracting with $(\mathrm{i}\,\hbar\,\mathsf{\Delta}_{\textrm{\tiny BV}}\, \mathsf{H} )^2$ discerns where the connected and disconnected diagrams originate.

Altogether one finds that all phase factors disappear and the first contribution to \eqref{eq:2-point photon polar} results in
\begin{equation}
\begin{split}
\mathsf{\tilde{G}}_{\mu\nu}(p_1,p_2)_{\tt A} &= -\hbar^2\, g^2\, \int_k\, (2\pi)^4\, \delta(p_1 + p_2)\, (p_1 + 2k)^{\lambda}\, (p_1 + 2k)^{\rho}\\
& \hspace{4cm} \times \tilde{\sf D}(p_1)\, \tilde{\sf \Pi}_{\lambda \mu}(p_1)\, \tilde{\sf D}(p_2)\, \tilde{\sf \Pi}_{\rho \nu}(p_2)\, \tilde{\sf G}(k)\, \tilde{\sf G}(k+p_1) \\
& \quad \, - \frac{4\,\hbar^2 \,g^2}{3} \,\int_{k, q}\, (2\pi)^8\, \delta(p_1)\, \delta(p_2)\,
k^{\lambda}\, q^{\rho}\\
& \hspace{4cm} \times \tilde{\sf D}(p_1)\, \tilde{\sf \Pi}_{\lambda \mu}(p_1)\, \tilde{\sf D}(p_2)\, \tilde{\sf \Pi}_{\rho \nu}(p_2)\, \tilde{\sf G}(k) \,\tilde{\sf G}(q) \ .
\end{split} 
\end{equation}
The first integral corresponds to the connected diagram 
\begin{equation}
\begin{split}
\begin{tikzpicture}[scale=0.5]
{ \scriptsize
    \draw[<-, shift={(0,0.5)}] (0:0.6) node[above]{$p_1$} -- (0,0);
    \draw[<-, shift={(1.2,-1.3)}] (-30:0.8) node[below]{${k{+}p_1}$} -- (0,0);
    \node at (-0.5,0) [below]{$\mu$}; 
    \node at (4.9,0) [below]{$\nu$};
    \draw[Aplus] (-0.5,0) -- (1,0);
    \draw[scalplusreverse] (1,0) arc (-180:180:1.2); 
    \draw[Aplus] (3.4,0) -- (4.9,0);
    } \normalsize
    \end{tikzpicture}
    \end{split}
\end{equation}
and it makes a non-trivial contribution to the the photon two-point function. The second integral corresponds to the disconnected diagram 
\begin{equation}\label{fig:VPa}
\begin{split}
\begin{tikzpicture}[scale=0.4]
{\scriptsize
    \draw[Aplus] (0,-1) -- (0,0); 
    \draw[scalplus] (0,0) arc (-90:270:0.8); 
    \draw[Aplus] (2,-1) -- (2,0); 
    \draw[scalplus] (2,0) arc (-90:270:0.8); 
    } \normalsize
    \end{tikzpicture}
    \end{split}
\end{equation}
with vanishing contribution to the vacuum polarisation by \cref{sub:one_point_functions}. 

\paragraph{Evaluation of $\mbf{\mathsf{\tilde{G}}_{\mu\nu}(p_1,p_2)_{\tt B}}$.}

Using our graphical calculus, we apply the contraction $\mathrm{i}\,\hbar\,\mathsf{\Delta}_{\textrm{\tiny BV}}\, \mathsf{H}$ to \eqref{eq:photonS3} and find
\begin{equation}
\begin{split}\label{eq:2nd term lead to tadpole}
    \mathrm{i}\, \hbar\, \mathsf{\Delta}_{\textrm{\tiny BV}}\, \mathsf{H}\,
    \bigg\{ \mathcal{S}^\swthree_{\rm int}\,,\,
    \mathsf{H}\, \bigg(
    \,
    \vcenter{\hbox{\begin{tikzpicture}[scale=0.5]
    \draw[Aplus] (0,0) -- (0,1.5);
    \draw[Aplus] (1,0) -- (1,1.5);
    \node at (0,2) {\scriptsize $ p_1$};
    \node at (1,2) {\scriptsize $ p_2$};
    \end{tikzpicture}}}
    \, \bigg)
    \bigg\}_\star
    = \tfrac13\,\Bigg( \
    \vcenter{\hbox{\begin{tikzpicture}[scale=0.4]
    \draw[Aplus] (0,-1) -- (0,0); 
    \node at (0,-1) [below]{\scriptsize $\mu$};
    \draw[scalplus] (0,0) arc (-90:270:0.8); 
    \draw[Aplus] (2,-1) -- (2,1.5); 
    \node at (2,-1) [below]{\scriptsize $\nu$};
    \end{tikzpicture}}}
    \
    +
    \
    \vcenter{\hbox{\begin{tikzpicture}[scale=0.4]
    { \scriptsize
    \draw[Aplus] (2,-1) -- (2,0); 
    \node at (2,-1) [below]{$\nu$};
    \draw[scalplus] (2,0) arc (-90:270:0.8); 
    \draw[Aplus] (0,-1) -- (0,1.5); 
    \node at (0,-1) [below]{$\mu$};
    } \normalsize
    \end{tikzpicture}}} \ \Bigg)
    \end{split}
\end{equation}
Further applying the operator $\mathrm{i}\, \hbar\, \BVL\, \mathsf{H}\, \big\{ \mathcal{S}^\swthree_{\rm int},-\big\}\, \mathsf{H} $ gives another tadpole. In fact, as noted in \cite{DimitrijevicCiric:2023hua}, this factorisation is also given in terms of one-point functions at one-loop from \cref{sub:one_point_functions}. 

Altogether the second contribution to \eqref{eq:2-point photon polar} results in the same disconnected diagram \eqref{fig:VPa}, and hence has no contribution:
\smash{$
\mathsf{\tilde{G}}_{\mu\nu}(p_1,p_2)_{\tt B} =0 $}.

\paragraph{Evaluation of $\mbf{\mathsf{\tilde{G}}_{\mu\nu}(p_1,p_2)_{\tt C}}$.}

Finally, the third contribution to \eqref{eq:2-point photon polar} involves the four-vertex (\ref{eq:4vertex}). Applying the strict derivation $\big\{ \mathcal{S}^\swfour_{\rm int}, - \big\}_\star\, \mathsf{H} : \mathrm{Sym}_{\star}L[2] \longrightarrow \mathrm{Sym}_{\star}L[2]$ of symmetric degree~2 to our basis of photon fields has the graphical representation
\begin{equation}
\begin{split}
  \bigg\{ \mathcal{S}^\swfour_{\rm int}\,,\,
    \mathsf{H}\, \bigg(
    \,
    \vcenter{\hbox{\begin{tikzpicture}[scale=0.5]
    { \scriptsize
    \draw[Aplus] (0,0) -- (0,1.5) node[above] {$p_1$};
    \draw[Aplus] (1,0) -- (1,1.5) node[above] {$p_2$};
    } \normalsize
    \end{tikzpicture}}}
    \, \bigg)
    \bigg\}_\star
    = \tfrac12\,\Bigg( \,
    \vcenter{\hbox{\begin{tikzpicture}[scale=0.3]
    {\scriptsize
    \draw[Aplus] (0,-0.5) -- (0,0.75);
    \draw[Aplus] (0,0.75) -- (120:3.5) node[above]{$k_1$}; 
    \draw[scalplus] (0, 0.75) -- (90:3) node[above]{$k_2$}; 
    \draw[scalplus] (0,0.75) -- (60:3.5) node[above]{$k_3$}; 
    \draw[Aplus] (2.8,-0.5) -- (2.8,3)node[above]{$p_2$};
    }\normalsize
    \end{tikzpicture}}}
    \ + \
    \vcenter{\hbox{\begin{tikzpicture}[scale=0.3]
    {\scriptsize
    \draw[Aplus] (0,-0.5) -- (0,0.75);
    \draw[scalplusreverse] (0,0.75) -- (120:3.5) node[above]{\scriptsize $k_2$}; 
    \draw[scalplusreverse] (0, 0.75) -- (90:3) node[above]{$k_3$}; 
    \draw[Aplus] (0,0.75) -- (60:3.5) node[above]{ $k_4$}; 
    \draw[Aplus] (2.8,-0.5) -- (2.8,3)node[above]{ $p_2$};
    }\normalsize
    \end{tikzpicture}}}
    \ + \
    \vcenter{\hbox{\begin{tikzpicture}[scale=0.3]
    {\scriptsize
    \draw[Aplus] (-2.8,-0.5) -- (-2.8,3) node[above]{$p_1$};
    \draw[Aplus] (0,-0.5) -- (0,0.75);
    \draw[Aplus] (0,0.75) -- (120:3.5) node[above]{$k_1$}; 
    \draw[scalplus] (0,0.75) -- (90:3) node[above]{\scriptsize $k_2$}; 
    \draw[scalplus] (0,0.75) -- (60:3.5) node[above]{\scriptsize $k_3$}; 
    }\normalsize
    \end{tikzpicture}}}
    \ + \
    \vcenter{\hbox{\begin{tikzpicture}[scale=0.3]
    {\scriptsize
    \draw[Aplus] (-2.8,-0.5) -- (-2.8,3) node[above]{$p_1$};
    \draw[Aplus] (0,-0.5) -- (0,0.75);
    \draw[scalplusreverse] (0,0.75) -- (120:3.5) node[above]{$k_2$}; 
    \draw[scalplusreverse] (0,0.75) -- (90:3) node[above]{$k_3$}; 
    \draw[Aplus] (0,0.75) -- (60:3.5) node[above]{$k_4$}; 
    }\normalsize
    \end{tikzpicture}}} \, \Bigg)
\end{split}
\end{equation}
The contractions with $(\mathrm{i}\,\hbar\,\mathsf{\Delta}_{\textrm{\tiny BV}}\, \mathsf{H} )^2$ are unambiguous and one can always jump across fields at the price of $R$-matrices, which in this instance end up trivialising due to momentum conservation. 

Altogether the final result is represented graphically by the connected tadpole diagram
\begin{equation} \label{fig:VPc}
\begin{split}
\begin{tikzpicture}[scale=0.6]
{\scriptsize
    \draw[Aplus] (-0.5,0) -- (1,0);
    \node at (-0.5,0) [above]{$\mu$};
    \node at (2.5,0) [above]{$\nu$};
    \draw[<-, shift={(-0.5,-0.5)}] (0:0.6) node[below]{$p_1$} -- (0,0);
    \draw[scalplus] (1,0) to [out=50,in=0] (1,2) to [out=180, in=130] (1,0);
    \draw[<-, shift={(0.5,0.4)}] (120:0.6) node[left]{$k$} -- (0,0);
    \draw[Aplus] (1,0) -- (2.5,0);
    } \normalsize
    \end{tikzpicture}
\end{split}
\end{equation}
It has a non-trivial contribution to the vacuum polarisation given by
\begin{equation}
\mathsf{\tilde{G}}_{\mu\nu}(p_1,p_2)_{\tt C} = 2\, \hbar^2\,g^2\, (2\pi)^4\, \delta(p_1+p_2)\, \tilde{\sf D}(p_1)\,\tilde{\sf \Pi}_{\lambda\mu}(p_1)\, \tilde{\sf D}(p_2)\,\tilde{\sf \Pi}^{\lambda}_{\ \nu}(p_2) \, \int_{k}\,  \tilde{\sf G}(k)  \ .
\end{equation}

\paragraph{Photon Self-Energy.}

The photon two-point function at one-loop receives contributions from only bubble diagrams, and adding the contributions gives 
\begin{equation}
\begin{split}
    \mathsf{\tilde{G}}^{\hbar, {\rm int}}_{A_\mu,A_\nu}(p_1, p_2)^\swone
    &=(\mathrm{i}\, \hbar\,g)^2\,
    (2 \pi)^4\,  \delta(p_1 + p_2)\,
    \tilde{\mathsf{\Pi}}_{\mu \lambda}(p_1)  \,
    \tilde{\mathsf{D}}(p_1)\,
    \tilde{\mathsf{\Pi}}_{\nu \rho}(p_2)\,
    \tilde{\mathsf{D}}(p_2)
     \\
    & \quad \, \times
    \int_{k} \,
    \left(
    (2k + p_1)^\lambda\,
    (2k + p_1)^\rho\,
    \tilde{\mathsf{G}}(k+p_1)\,  \tilde{\mathsf{G}}(k)
    -
    2\,  \eta^{\lambda \rho}\,
    \tilde{\mathsf{G}}(k)
    \right) \ .
\end{split}
\end{equation}
Equivalently, one can dress the photon propagator of \eqref{eq:barephotonpropbv} in terms of the one-loop photon vacuum polarisation tensor ${\Pi}_\star(p)^\swone$ by setting 
\begin{equation}
    \mathsf{\tilde{G}}^{\hbar}_{A_\mu,A_\nu}(p_1, p_2)^\swzero
    + \mathsf{\tilde{G}}^{\hbar,{\rm int}}_{A_\mu,A_\nu}(p_1, p_2)^\swone =  \mathrm{i}\, \hbar\,
    (2 \pi)^4\, \delta(p_1 + p_2)\,
    \left(\frac{1}{p_1^2\, \mathsf{\tilde\Pi}^{-1}(p_1) - {\Pi}_{\star}(p_1)^\swone }\right)_{\mu \nu} \ ,
\end{equation}
where the transverse projector $\mathsf{\tilde\Pi}(k)$ is given by \eqref{eq:projector}. This coincides with the commutative result
\begin{equation} \label{eq:poltensor}
\begin{split}
    -\frac{\mathrm{i}}{\hbar}\,{\Pi}_{\star}^{\mu\nu}(p)^\swone
    &= \
    \vcenter{\hbox{\begin{tikzpicture}[scale=0.3]
    \draw[scalplusreverse] (1,0) arc (-180:180:1.2);
    \end{tikzpicture}}}
     \ + \
    \vcenter{\hbox{\begin{tikzpicture}[scale=0.3]
    \draw[scalplus] (1,0) to [out=50,in=0] (1,2) to [out=180, in=130] (1,0);
    \end{tikzpicture}}} \\[4pt]
    &= g^2\,
    \int_{k} \,
   \frac{
    (2k + p)^\mu\,
    (2k + p)^\nu}{\big((k+p)^2-m^2\big)\,\big(k^2-m^2\big)}
    - 2\, g^2\,
    \int_{k} \,
    \frac{\eta^{\mu\nu}}{(k+p)^2-m^2} \ .
\end{split}
\end{equation}

Since the loop diagrams here reproduce the commutative results, this implies that there is no UV/IR mixing in the photon vacuum polarisation at the one-loop level. This raises hopes that our braided field theory is perturbatively renormalisable and so has a good quantum theory. In particular, like the free photon propagator, after dimensional regularisation the one-loop polarisation tensor is transverse: 
\begin{align}
p_\mu\,\Pi_\star^{\mu\nu}(p)^\swone = 0 \ .
\end{align}
An easy but formal way to argue this is by using the identity
\begin{align}
p\cdot(2k+p) = \big((k+p)^2-m^2\big) - \big(k^2-m^2\big)
\end{align}
and shifting integration variable $k\longmapsto k-p$ in two of the three resulting integrals from \eqref{eq:poltensor}.

\subsection{Scalar Vacuum Polarisation}
\label{sub:scalar_self_energy}

The tree-level two-point function for the charged scalar field follows from (\ref{eq:bqftnpointmom}) and  is given by
\begin{equation}
\begin{split}\label{eq:bare scal prop}
    \mathsf{\tilde{G}}^{\hbar}_{\phi, \bar\phi}(p_1,p_2)^\swzero &= 
     \mathrm{i}\,\hbar\,\mathsf{\Delta}_{\textrm{\tiny BV}}\, \mathsf{H}\,
    ( \mathtt{e}^{p_1} \odot_\star \bar{\mathtt{e}}^{p_2}) \\[4pt]
    &= - \mathrm{i}\, \hbar \,(2 \pi)^4 \,\delta(p_1 + p_2)\, \mathsf{\tilde{G}}(p_1)\\[4pt]
    &= - \mathrm{i}\, \hbar\, (2 \pi)^4 \,\delta(p_1 + p_2) \ \frac{1}{p_1^2 - m^2}\ .
\end{split}
\end{equation}

Similarly to the photon two-point function at one-loop, the homological perturbation lemma gives three terms contributing the scalar two-point function at one-loop. These are given by
\begin{equation}
\begin{split}\label{eq:scalarself1loop}
\mathsf{\tilde{G}}^{\hbar,{\rm int}}_{\phi,\bar\phi}(p_1, p_2)^\swone &=  (\mathrm{i}\, \hbar\, \BVL\, \mathsf{H})^2 \left\{ \mathcal{S}_{\rm int}^{\swthree}, \mathsf{H}\left\{\mathcal{S}_{\rm int}^{\swthree}, \mathsf{H}\,(\mathtt{e}^{p_1} \odot_\star \bar{\mathtt{e}}^{p_2})\right\}_\star\right\}_\star\\
& \quad \, + \mathrm{i}\, \hbar\, \BVL\, \mathsf{H} \left\{\mathcal{S}_{\rm int}^{\swthree}, \mathsf{H}\,(\mathrm{i} \,\hbar\, \BVL\, \mathsf{H}) \left\{ \mathcal{S}_{\rm int}^{\swthree}, \mathsf{H}\,(\mathtt{e}^{p_1} \odot_\star \bar{\mathtt{e}}^{p_2}) \right\}_\star\right\}_\star \\
&\quad \, + (\mathrm{i}\, \hbar\, \BVL\, \mathsf{H})^2 
\left\{\mathcal{S}_{\rm int}^{\swfour},\mathsf{H}\,( \mathtt{e}^{p_1} \odot_\star \bar{\mathtt{e}}^{p_2})\right\}_\star \\[4pt]
&=: \mathsf{\tilde{G}}(p_1, p_2)_{\tt A} + \mathsf{\tilde{G}}(p_1, p_2)_{\tt B} + \mathsf{\tilde{G}}(p_1, p_2)_{\tt C}\ .
\end{split}
\end{equation}

\paragraph{Evaluation of $\mbf{\mathsf{\tilde{G}}(p_1, p_2)_{\tt A}}$.}

Our basis of fields $\mathtt{e}^{p_1} \odot_\star \mathtt{\bar{e}}^{p_2}$ is now represented by a pair of vertical charged scalar lines with opposite orientation. The three-vertex operator from \eqref{eq:vertices} acts on this basis as a strict derivation and results in
\begin{equation}
\begin{split}\label{eq:poissonS3scalr}
    \bigg\{ \mathcal{S}^\swthree_{\rm int}\,,\,
    \mathsf{H}\, \bigg(\,
    \vcenter{\hbox{\begin{tikzpicture}[scale=0.4]
    { \scriptsize
    \draw[scalplusreverse] (0,-0.25) -- (0,2.25) node[above] { $ p_1$};
    \draw[scalplus] (1,-0.25) -- (1,2.25) node[above] { $ p_2$};
    } \normalsize
    \end{tikzpicture}}}
    \,
    \bigg)
    \bigg\}_\star
    = \tfrac12\,\bigg(
    \vcenter{\hbox{
    \begin{tikzpicture}[scale=0.4]
    { \scriptsize
\draw[scalplusreverse] (0,-0.25) -- (0,1.25);
    \draw[Aplus] (0,1.25) -- (-1,2.25) node[above] {$k_1$};
    \draw[scalplusreverse] (0,1.25) -- (1,2.25) node[above] {$k_2$};
    \draw[scalplus] (2,-0.25) -- (2,2.25) node[above] {$p_2$};
    } \normalsize
    \end{tikzpicture}}}
    \ + \
    \vcenter{\hbox{
    \begin{tikzpicture}[scale=0.4]
    { \scriptsize
    \draw[scalplusreverse] (0.5,-0.25) -- (0.5,2.25) node[above] {$p_1$};
    \draw[scalplus] (2.75,-0.25) -- (2.75,1.25);
    \draw[scalplus] (2.75,1.25) -- (1.75,2.25)  node[above] {$ k_3 $};
    \draw[Aplus] (2.75,1.25) -- (3.75,2.25)  node[above] {$ k_1$};
    } \normalsize
    \end{tikzpicture}}} \, \bigg)
    \end{split}
\end{equation}
Acting again with the three-vertex operator on the first term in the right-hand side of \eqref{eq:poissonS3scalr} has the graphical representation
\begin{equation}
\begin{split}\label{eq:graphs3s3poissonscal}
    \bigg\{ \mathcal{S}^\swthree_{\rm int}\,,\,
    \mathsf{H}\, \bigg(\,
    \vcenter{\hbox{\begin{tikzpicture}[scale=0.4]
    { \scriptsize
\draw[scalplusreverse] (0,-0.25) -- (0,1.25);
    \draw[Aplus] (0,1.25) -- (-1,2.25) node[above] {$k_1$};
    \draw[scalplusreverse] (0,1.25) -- (1,2.25) node[above] {$k_2$};
    \draw[scalplus] (2,-0.25) -- (2,2.25) node[above] {$p_2$};
    } \normalsize
    \end{tikzpicture}}}
    \,
    \bigg)
    \bigg\}_\star
    = \tfrac13\,\Bigg( \,
    \vcenter{\hbox{
    \begin{tikzpicture}[scale=0.4]
    {\scriptsize
    \draw[scalplusreverse] (0,-0.25) -- (0,0.75);
    \draw[Aplus] (0.0,0.75) -- (-0.75,1.5);
    \draw[scalplusreverse] (-0.75,1.5) -- (-1.5, 2.25) node[above] {$k_5$};
    \draw[scalplus] (-0.75,1.5) -- (0,2.25) node[above] {$k_6$};
    \draw[scalplusreverse] (0.0,0.75) -- (1.5,2.25) node[above]{$k_2$};
    \draw[scalplus] (2.25, -0.25) -- (2.25,2.25) node[above]{$p_2$};
    } \normalsize
    \end{tikzpicture}}}
    \
    +
    \
    \vcenter{\hbox{
    \begin{tikzpicture}[scale=0.4]
    {\scriptsize
    \draw[scalplusreverse] (0,-0.25) -- (0,0.75);
    \draw[scalplusreverse] (0.0,0.75) -- (0.75,1.5);
    \draw[scalplusreverse]  (0.75,1.5) -- (1.5, 2.25)  node[above] {$k_5$};
    \draw[Aplus] (0.75,1.5) -- (0,2.25) node[above] {$k_4$} ;
    \draw[Aplus] (0.0,0.75) -- (-1.5,2.25) node[above] {$k_1$};
    \draw[scalplus] (2.25, -0.25) -- (2.25,2.25) node[above] {$p_2$};
      } \normalsize
    \end{tikzpicture}}}
    \
    +
    \
    \vcenter{\hbox{
   \begin{tikzpicture}[scale=0.4]
    {\scriptsize
    \draw[scalplusreverse] (0,-0.25) -- (0,1.25);
    \draw[Aplus] (0,1.25) -- (-1,2.25) node[above] {$k_1$};
    \draw[scalplusreverse] (0,1.25) -- (1,2.25) node[above] {$k_2$};
    \draw[scalplus] (2.75,-0.25) -- (2.75,1.25);
    \draw[scalplus] (2.75,1.25) -- (1.75,2.25) node[above] {$k_6$};
    \draw[Aplus] (2.75,1.25) -- (3.75,2.25) node[above] {$k_4$};
      } \normalsize
    \end{tikzpicture}}} \ \Bigg)
\end{split}
\end{equation}
and similarly for the second term.

Contracting with $(\mathrm{i}\, \hbar\,\mathsf{\Delta}_{\textrm{\tiny BV}}\, \mathsf{H})^2$ then shows that the first contribution to \eqref{eq:scalarself1loop} results in the two connected diagrams
\begin{equation}
\begin{split}
    \begin{tikzpicture}[scale=0.6]
    {\scriptsize
    \draw[scalplusreverse] (-0.5,0) -- (1,0);
    \draw[<-, shift={(1.5,-0.5)}] (0:0.6) node[below]{$k{+}p_1$} -- (0,0);
    \draw[<-, shift={(-0.5,-0.5)}] (0:0.6) node[below]{$p_1$} -- (0,0);
    \draw[Aplus] (1,0) to [out=90,in=180] (2,1) to [out=0, in=90] (3,0);
    \draw[scalplusreverse] (1,0) -- (3,0);
    \draw[scalplusreverse] (3,0) -- (4.5,0);
    } \normalsize
    \end{tikzpicture}
    \end{split}
\end{equation}
and
\begin{equation}\label{fig:SSEa}
\begin{split}
\begin{tikzpicture}[scale=0.6]
{\scriptsize
    \draw[Aplus] (0,-1) -- (0,0); 
    \draw[scalplus] (0,0) arc (-90:270:0.8); 
    \draw[scalplusreverse] (-2,-1) -- (0,-1);
    \draw[scalplusreverse] (0,-1) -- (2,-1);
     } \normalsize
    \end{tikzpicture} 
    \end{split}
    \end{equation}
They respectively sum to
\begin{equation}
\begin{split}
\mathsf{\tilde{G}}(p_1, p_2)_{\tt A} &= - \hbar^2\,  g^2  \, 
    (2 \pi)^4\,  \delta(p_1 + p_2)\, 
    \mathsf{\tilde{G}}(p_1) \, 
    \mathsf{\tilde{G}}(p_2) \\
    & \hspace{5cm} \times
    \int_k \,  (k+2p_1)^\mu\,  (k+2p_1)^\nu\, 
    \tilde{\mathsf{G}}(k+p_1)\,  \tilde{\mathsf{D}}(k)\,  \tilde{\mathsf{\Pi}}_{\mu \nu}(k) \\
& \quad \,  -  \frac{4\,\hbar^2\,  g^2}{3}\,  \int_{k,q}\,  (2\pi)^8\,  \delta(p_1+p_2)\,  \delta(k)\,  \tilde{\mathsf{G}}(p_1)\,  \tilde{\mathsf{G}}(p_2)\,  \tilde{\mathsf{G}}(q)\,  \tilde{\mathsf{D}}(k)\,  \tilde{\mathsf{\Pi}}_{\mu\nu}(k)\,  p_1^{\mu}\, q^{\nu} \ .
\end{split}
\end{equation}
The second integral is a tadpole contribution which vanishes by \cref{sub:one_point_functions}, leaving only the contribution from the first integral. The noncommutative phase coming from the vertex in \eqref{eq:3-vertex diagram} vanishes due to momentum conservation.

\paragraph{Evaluation of $\mbf{\mathsf{\tilde{G}}(p_1, p_2)_{\tt B}}$.}

The second contribution to \eqref{eq:scalarself1loop} is the same scalar tadpole diagram \eqref{fig:SSEa}. This is most easily seen by following  the symmetric degree inside the calculation. The application of $\mathrm{i}\, \hbar\, \BVL\, \mathsf{H}$ to \eqref{eq:poissonS3scalr} has symmetric degree $1$, involving the field $\mathtt{v}_\mu\, \mathtt{e}^{k_1}$, so further applying $\mathrm{i}\, \hbar\, \BVL\, \mathsf{H}\, \big\{ \mathcal{S}^\swthree_{\rm int}\ , - \big\}\,\sH$ closes a tadpole, thereby giving a vanishing contribution to the scalar two-point function at one-loop:
\smash{$
\mathsf{\tilde{G}}(p_1, p_2)_{\tt B} = 0 $}.

\paragraph{Evaluation of $\mbf{\mathsf{\tilde{G}}(p_1, p_2)_{\tt C}}$.}

Similarly, the four-vertex operator yields the graphical representation
\begin{equation}\label{eq:graphs4poissonscal}
\begin{split}
  \bigg\{ \mathcal{S}^\swfour_{\rm int}\,,\,
    \mathsf{H}\, \bigg(
    \,
    \vcenter{\hbox{\begin{tikzpicture}[scale=0.4]
    { \scriptsize
    \draw[scalplusreverse] (0,-0.5) -- (0,1.75) node[above] {$p_1$};
    \draw[scalplus] (1,-0.5) -- (1,1.75) node[above] {$p_2$};
    } \normalsize
    \end{tikzpicture}}}
    \, \bigg)
    \bigg\}_\star
    = \tfrac12\,\Bigg( \,
    \vcenter{\hbox{\begin{tikzpicture}[scale=0.3]
    {\scriptsize
    \draw[scalplusreverse] (0,-0.5) -- (0,0.75);
    \draw[Aplus] (0,0.75) -- (120:3.5)node[above]{$k_4$}; 
    \draw[Aplus] (0, 0.75) -- (90:3) node[above]{$k_1$}; 
    \draw[scalplusreverse] (0,0.75) -- (60:3.5) node[above]{$k_2$}; 
    \draw[scalplus] (2.8,-0.5) -- (2.8,3.0)node[above]{$p_2$};
    }\normalsize
    \end{tikzpicture}}}
    \ + \
    \vcenter{\hbox{\begin{tikzpicture}[scale=0.3]
    {\scriptsize
    \draw[scalplus] (0,-0.5) -- (0,0.75);
    \draw[scalplus] (0,0.75) -- (120:3.5)node[above]{\scriptsize $k_3$}; 
    \draw[Aplus] (0, 0.75) -- (90:3) node[above]{$k_4$}; 
    \draw[Aplus] (0,0.75) -- (60:3.5) node[above]{ $k_1$}; 
    \draw[scalplusreverse] (-2.8,-0.5) -- (-2.8,3)node[above]{ $p_1$};
    }\normalsize
    \end{tikzpicture}}}
     \, \Bigg)
\end{split}
\end{equation}
Contracting again with $(\ii\,\hbar\,\BVL\,\sH)^2$, there is no need to switch fields so no $R$-matrices appear. We find that the noncommutative phase factor arising from \eqref{eq:4vertex} disappears as before because of momentum conservation.

Finally, the third contribution to \eqref{eq:scalarself1loop} is the connected diagram
\begin{equation}\label{fig:SSEc}
\begin{split}
\begin{tikzpicture}[scale=0.6]
{\scriptsize
    \draw[scalplusreverse] (-0.5,0) -- (1,0);
    \draw[<-, shift={(-0.5,-0.5)}] (0:0.6) node[below]{$p_1$} -- (0,0);
    \draw[Aplus] (1,0) to [out=50,in=0] (1,2) to [out=180, in=130] (1,0);
    \draw[<-, shift={(0.5,0.4)}] (120:0.6) node[left]{$k$} -- (0,0);
    \draw[scalplusreverse] (1,0) -- (2.5,0);
    } \normalsize
    \end{tikzpicture}
    \end{split}
\end{equation}
The explicit result is
\begin{equation}
\mathsf{\tilde{G}}(p_1, p_2)_{\tt C} = -3\,   \hbar^2\,  g^2  \, 
    (2 \pi)^4\,  \delta(p_1 + p_2)\, 
    \mathsf{\tilde{G}}(p_1) \, 
    \mathsf{\tilde{G}}(p_2) \, 
    \int_k\,   \tilde\sD(k) \ . 
\end{equation}

\paragraph{Scalar Self-Energy.}

Altogether we find that the charged scalar field two-point function at one-loop is given by
\begin{align}
\begin{split}
    \tilde{\mathsf{G}}^{\hbar,{\rm int}}_{\phi, \bar\phi}(p_1, p_2)^\swone &= 
    (\mathrm{i}\, \hbar\, g)^2  \,
    (2 \pi)^4\, \delta(p_1 + p_2)\,
    \mathsf{\tilde{G}}(p_1) \,
    \mathsf{\tilde{G}}(p_2) \\
    & \quad \, \times 
    \int_k\,
    \left(
    (k+2p_1)^\mu \,(k+2p_1)^\nu\,
    \tilde{\mathsf{G}}(k+p_1)\, \tilde{\mathsf{D}}(k)\, \tilde{\mathsf{\Pi}}_{\mu \nu}(k)
    - 3\, \mathsf{\tilde{D}}(k)
    \right)\ .
\end{split}
\end{align}
The dressed scalar propagator at one-loop is defined by
\begin{equation}\label{eq:dressed scal prop}
    \tilde{\mathsf{G}}^{\hbar}_{\phi, \bar\phi}(p_1, p_2)^\swzero +
    \tilde{\mathsf{G}}^{\hbar,{\rm int}}_{\phi, \bar\phi}(p_1, p_2)^\swone = - \mathrm{i}\, \hbar \,
    (2 \pi)^4\, \delta(p_1 + p_2) \
    \frac{1}{p_1^2 - m^2 - {\Sigma}_\star(p_1)^\swone} \ ,
\end{equation}
where the self-energy $\Sigma_\star(p)^\swone$ is obtained by removing external scalar propagators $\ii\,\hbar\,\tilde{\mathsf{G}}(p)$ and is the sum of amputated diagrams
\begin{equation}
\begin{split}\label{eq:scal_Self_energy}
    \frac{\mathrm{i}}{\hbar}\,{\Sigma}_\star(p)^\swone &= \ 
    \vcenter{\hbox{\begin{tikzpicture}[scale=0.6]
    \draw[Aplus] (1,0) to [out=90,in=180] (2,1) to [out=0, in=90] (3,0);
    \draw[scalplusreverse] (1,0) -- (3,0);
    \end{tikzpicture}}}
     \ + \ 
    \vcenter{\hbox{\begin{tikzpicture}[scale=0.6]
    \draw[Aplus] (1,0) to [out=50,in=0] (1,2) to [out=180, in=130] (1,0);
    \end{tikzpicture}}} \\[4pt]
    &= -g^2\,
    \int_k\,  \frac{(k+2p)^\mu\, (k+2p)^\nu}{k^2\,\big((k+p)^2-m^2\big)} \, \left(\eta_{\mu\nu}-\frac{k_\mu\,k_\nu}{k^2}\right)
    - 3\, g^2\, \int_k\, \frac{1}{k^2} \ .
\end{split}
\end{equation}

In summary, we recover the commutative result for the charged scalar field self-energy and no UV/IR mixing.

\subsection{Three-Vertex Function} 
\label{app:3_point_vertex_correction}

The three-vertex function  at tree-level is
\begin{equation}
\begin{split}\label{eq:treelevelcurrent2}
\mathsf{\tilde{G}}^{\hbar,{\rm int}}_{\phi,\bar\phi,A_\mu}(p_1,p_2,p_3)^\swzero &=
    (\mathrm{i}\,\hbar\, \mathsf{\Delta}_{\textrm{\tiny BV}}\, \mathsf{H})^2\,
    \left\{ \mathcal{S}^\swthree_{\mathrm{int}}\,,\, \mathsf{H}\,(
    \mathtt{e}^{p_1} 
    \odot_\star 
    \bar{\mathtt{e}}^{p_2} 
    \odot_\star 
    \mathtt{v}_{\mu}\,\mathtt{e}^{p_3}
    ) \right\}_\star
    \\[4pt]
    &  =
    - (\mathrm{i} \,\hbar)^2 \,  
    V^\lambda_3(p_3,-p_2,-p_1) \, \mathsf{\tilde G}(p_1) \,
    \mathsf{\tilde{G}}(p_2)\,
    \mathsf{\tilde{D}}(p_3)\, \mathsf{\tilde\Pi}_{\lambda \mu }(p_3) \\
    & \qquad \, + \tfrac{1}{3}\,
    \mathsf{\tilde{G}}^{\hbar,{\rm int}}_{A_{\mu}}(p_3)^\swone  \,
    \mathsf{\tilde{G}}^{\hbar}_{\phi, \bar\phi}(p_1,p_2)^\swzero \ .
\end{split}
\end{equation}
The last line is proportional to the photon one-point function at one-loop, which we found to vanish in \eqref{eq:1pointphoton}. Hence
\begin{align} \label{eq:3vertextree}
\begin{split}
\mathsf{\tilde{G}}^{\hbar,{\rm int}}_{\phi,\bar\phi,A_\mu}(p_1,p_2,p_3)^\swzero &= - \hbar^2\,g \, (2\pi)^4\,\delta(p_3-p_1-p_2) \\
& \qquad \, \times  \e^{-\frac\ii2\,\sum\limits_{i<j}\,p_i\cdot\theta\,p_j}  \, \frac{(p_1-p_2)^\lambda}{\big(p_1^2-m^2\big)\,\big(p_2^2-m^2\big)} \, \frac1{p_3^2}\,\left(\eta_{\lambda\mu}-\frac{p_{3\,\lambda}\,p_{3\,\mu}}{p_3^2}\right) \ ,
\end{split}
\end{align}
which differs from the commutative result by the phase factor depending on the external momenta.

We shall now detail the calculation of the three-point vertex at one-loop order.
For this, one needs to calculate the combinations of interactions and contractions, omitting obvious tadpole terms which vanish.
We recall the degrees of relevant operators, which are useful in identifying contributions that are obviously zero:
\begin{equation}
    \begin{tabular}{|c|c|}
        \hline
        Operator & Symmetric Degree \\
        \hline \hline
        $\big\{ \mathcal{S}^{\swthree}_\mathrm{int},- \big\}_\star\,\sH$
        & $+1$ \\[0.5em]
        $\big\{ \mathcal{S}^{\swfour}_\mathrm{int}, - \big\}_\star\,\sH$
        & $+2$ \\[0.5em]
        $\mathrm{i}\,\hbar\, \mathsf{\Delta}_{\textrm{\tiny BV}}\, \mathsf{H}$
        & $-2$
        \\
        \hline
    \end{tabular}
\end{equation}

The term
\begin{equation}
\begin{split}
    \mathrm{i}\, \hbar\, \mathsf{\Delta}_{\textrm{\tiny BV}}\, \mathsf{H}\,
    \left\{ \CS^\swthree_{\rm int}\,,\, \mathsf{H}\,
    (\mathrm{i}\, \hbar\, \mathsf{\Delta}_{\textrm{\tiny BV}}\, \mathsf{H})^2\,
    \left\{ \CS^\swthree_{\rm int}\,,\, \mathsf{H}\,
    \left\{ \CS^\swthree_{\rm int}\ ,\ \mathsf{H}\, (\mathtt{e}^{p_1} \odot_\star \bar{\mathtt{e}}^{p_2} \odot_\star \mathtt{v}_\mu\, \mathtt{e}^{p_3}) \right\}_\star  \right\}_\star \right\}_\star = 0
\end{split}
\end{equation}
vanishes by symmetric power counting:
After the application of $(\mathrm{i}\, \hbar \,\mathsf{\Delta}_{\textrm{\tiny BV}}\, \mathsf{H})^2$ we end up with a monomial in the symmetric algebra, which then leads to a vanishing tadpole diagram by \cref{sub:one_point_functions}.
Several other terms vanish similarly.
Note that any operator landing in symmetric degree 0, for example
\begin{equation}
    (\mathrm{i} \, \hbar\,  \mathsf{\Delta}_{\textrm{\tiny BV}} \, \mathsf{H})^2\, 
    \left\{ \CS^\swthree_{\rm int}\ ,\ \mathsf{H}\,  (\mathtt{e}^{p_1} \odot_\star \bar{\mathtt{e}}^{p_2} \odot_\star \mathtt{v}_\nu\,  \mathtt{e}^{p_3}) \right\}_\star  \ \in \ \big(\mathrm{Sym}_\star L[2]\big)^0 \ ,
\end{equation}
yields a vanishing contribution.
The upshot is that terms where the contraction operator $\mathrm{i}\, \hbar\, \mathsf{\Delta}_{\textrm{\tiny BV}}\, \mathsf{H}$ acts early enough vanish, and we omit them from now on.

The manifestly non-zero $\mathsf{P}_0$-projections that contribute to the one-loop three-point function 
\begin{align}
\mathsf{\tilde{G}}^{\hbar,{\rm int}}_{\phi,\bar\phi,A_\mu}(p_1,p_2,p_3)^\swone = {\text{\small{\tt(I)}}} + {\text{\small{\tt(II)}}} + {\text{\small{\tt(III)}}} + {\text{\small{\tt(IV)}}} + {\text{\small{\tt(V)}}} + {\text{\small{\tt(VI)}}} + {\text{\small{\tt(VII)}}} 
\end{align}
are given by
\begin{subequations}\label{eqs:pains}
\begin{align}
    \label{eq:pain_I}
& \hspace{0cm}  {\text{\small{\tt(I)}}} 
    = 
    (\mathrm{i}\, \hbar\, \mathsf{\Delta}_{\textrm{\tiny BV}}\, \mathsf{H})^2\,
    \left\{ \CS^\swthree_{\rm int}\,,\, \mathsf{H}\,
    \left\{ \CS^\swthree_{\rm int}\,,\, \mathsf{H}\,
    (\mathrm{i}\, \hbar\, \mathsf{\Delta}_{\textrm{\tiny BV}}\, \mathsf{H})\,
    \left\{ \CS^\swthree_{\rm int}\ ,\ \mathsf{H}\, (\mathtt{e}^{p_1} \odot_\star \bar{\mathtt{e}}^{p_2} \odot_\star \mathtt{v}_\mu\, \mathtt{e}^{p_3}) \right\}_\star  \right\}_\star \right\}_\star \ , \\[4pt]
    \label{eq:pain_II}
 & \hspace{-0.2cm}  {\text{\small{\tt(II)}}} 
    =
    (\mathrm{i}\, \hbar\, \mathsf{\Delta}_{\textrm{\tiny BV}}\, \mathsf{H})^2\,
    \left\{ \CS^\swthree_{\rm int}\,,\, \mathsf{H}\,(
    \mathrm{i}\, \hbar\, \mathsf{\Delta}_{\textrm{\tiny BV}}\, \mathsf{H})\,
    \left\{ \CS^\swthree_{\rm int}\,,\, \mathsf{H}\,
    \left\{ \CS^\swthree_{\rm int}\ ,\ \mathsf{H}\, (\mathtt{e}^{p_1} \odot_\star \bar{\mathtt{e}}^{p_2} \odot_\star \mathtt{v}_\mu\, \mathtt{e}^{p_3}) \right\}_\star  \right\}_\star \right\}_\star \ , \\[4pt]
    \label{eq:pain_III}
& \hspace{-0.4cm}  {\text{\small{\tt(III)}}} 
    =
    (\mathrm{i} \,\hbar\, \mathsf{\Delta}_{\textrm{\tiny BV}}\, \mathsf{H})^3\,
    \left\{ \CS^\swthree_{\rm int}\,,\, \mathsf{H}\,
    \left\{ \CS^\swthree_{\rm int}\,,\, \mathsf{H}\,
    \left\{ \CS^\swthree_{\rm int}\ ,\ \mathsf{H}\, (\mathtt{e}^{p_1} \odot_\star \bar{\mathtt{e}}^{p_2} \odot_\star \mathtt{v}_\mu \,\mathtt{e}^{p_3}) \right\}_\star  \right\}_\star \right\}_\star \ , \\[4pt]
    \label{eq:pain_IV}
 & \hspace{-0.2cm}  {\text{\small{\tt(IV)}}} 
    =
    (\mathrm{i}\, \hbar\, \mathsf{\Delta}_{\textrm{\tiny BV}}\, \mathsf{H})^2\,
    \left\{ \CS^\swfour_{\rm int}\,,\,  
    \mathsf{H}\,(\mathrm{i}\, \hbar\, \mathsf{\Delta}_{\textrm{\tiny BV}}\, \mathsf{H})\,
    \left\{ \CS^\swthree_{\rm int}\ ,\ \mathsf{H}\, (\mathtt{e}^{p_1} \odot_\star \bar{\mathtt{e}}^{p_2} \odot_\star \mathtt{v}_\mu\, \mathtt{e}^{p_3}) \right\}_\star \right\}_\star \ , \\[4pt]
    \label{eq:pain_V}
 & \hspace{0cm} {\text{\small{\tt(V)}}}  
    =
    (\mathrm{i}\, \hbar\, \mathsf{\Delta}_{\textrm{\tiny BV}}\, \mathsf{H})^3\,
    \left\{ \CS^\swthree_{\rm int}\,,\, \mathsf{H} \,
    \left\{ \CS^\swfour_{\rm int}\ ,\ \mathsf{H}\, (\mathtt{e}^{p_1} \odot_\star \bar{\mathtt{e}}^{p_2} \odot_\star \mathtt{v}_\mu\, \mathtt{e}^{p_3}) \right\}_\star
    \right\}_\star \ , \\[4pt]
    \label{eq:pain_VI}
  & \hspace{-0.2cm}  {\text{\small{\tt(VI)}}}  
    =
    (\mathrm{i}\, \hbar\, \mathsf{\Delta}_{\textrm{\tiny BV}}\, \mathsf{H})^3\,
    \left\{ \CS^\swfour_{\rm int}\,, \,\mathsf{H} \,
    \left\{ \CS^\swthree_{\rm int}\ ,\ \mathsf{H}\, (\mathtt{e}^{p_1} \odot_\star \bar{\mathtt{e}}^{p_2} \odot_\star \mathtt{v}_\mu\, \mathtt{e}^{p_3}) \right\}_\star
    \right\}_\star \ , \\[4pt]
    \label{eq:pain_VII}
& \hspace{-0.4cm}    {\text{\small{\tt(VII)}}}
    =
    (\mathrm{i}\, \hbar\, \mathsf{\Delta}_{\textrm{\tiny BV}}\, \mathsf{H})^2\,
    \left\{ \CS^\swthree_{\rm int}\,,\,\mathsf{H}\,( \mathrm{i}\, \hbar\, \mathsf{\Delta}_{\textrm{\tiny BV}}\, \mathsf{H})\,
    \left\{ \CS^\swfour_{\rm int}\ ,\ \mathsf{H}\, (\mathtt{e}^{p_1} \odot_\star \bar{\mathtt{e}}^{p_2} \odot_\star \mathtt{v}_\mu\, \mathtt{e}^{p_3}) \right\}_\star
    \right\}_\star \ .
\end{align}
\end{subequations}
In the remainder of this section we summarise the results of these computations.

\paragraph{Calculation of ${\mbf{\text{\small{\tt(I)}}}}$.}

Consider the contribution ${\text{\small{\tt(I)}}}$ from \eqref{eq:pain_I}  to the correlation function.
We present this calculation in some detail using the graphical calculus, reading diagrams from bottom up, starting from
\begin{equation}
\begin{split}
    \hspace{-2cm}
    &\mathrm{i}\, \hbar\, \mathsf{\Delta}_{\textrm{\tiny BV}}\, \mathsf{H}
    \bigg\{ \mathcal{S}^\swthree_{\rm int}\,,\,
    \mathsf{H}\, \bigg(\,
    \vcenter{\hbox{\begin{tikzpicture}[scale=0.4]
    { \scriptsize
    \draw[scalplusreverse] (0,-0.25) -- (0,2.25) node[above] { $ p_1$};
    \draw[scalplus] (1,-0.25) -- (1,2.25) node[above] { $ p_2$};
    \draw[Aplus] (2,-0.25) -- (2,2.25) node[above] { $ p_3$};
    } \normalsize
    \end{tikzpicture}}}
    \,
    \bigg)
    \bigg\}_\star \\[4pt]
    & \hspace{3cm}
    = \frac{\mathrm{i}\,\hbar}{3}\,\bigg( \mathrm{e}^{\,\mathrm{i}\, p_3\cdot \theta\, (k_2 + p_2)}\, \times \,
    \vcenter{\hbox{
    \begin{tikzpicture}[scale=0.4]
    { \scriptsize
\draw[scalplusreverse] (0,-0.25) -- (0,1) node[above] {$k_2$};
    \draw[Aplus] (0,-0.25) -- (-120:2);
    \draw[scalplus] (0,-0.25) -- (-60:2);
    \draw[scalplus] (2,-1.75) -- (2,1) node[above] {$p_2$};
    } \normalsize
    \end{tikzpicture}}}
    \ + \
    \vcenter{\hbox{
    \begin{tikzpicture}[scale=0.4]
    { \scriptsize
\draw[Aplus] (0,-0.25) -- (0,1) node[above] {$k_1$};
    \draw[scalplus] (0,-0.25) -- (-120:2);
    \draw[scalplusreverse] (0,-0.25) -- (-60:2);
    \draw[Aplus] (2,-1.75) -- (2,1) node[above] {$p_3$};
    } \normalsize
    \end{tikzpicture}}}
    \ + \
    \vcenter{\hbox{
    \begin{tikzpicture}[scale=0.4]
    { \scriptsize
\draw[scalplus] (0,-0.25) -- (0,1) node[above] {$k_3$};
    \draw[Aplus] (0,-0.25) -- (-120:2);
    \draw[scalplusreverse] (0,-0.25) -- (-60:2);
    \draw[scalplusreverse] (-2,-1.75) -- (-2,1) node[above] {$p_1$};
    } \normalsize
    \end{tikzpicture}}} \\[5pt]
    & \hspace{6cm}
    \ + \
    \frac{1}{2} \times
    \vcenter{\hbox{
    \begin{tikzpicture}[scale=0.4]
    { \scriptsize
\draw[Aplus] (0,0.25) -- (0,-1);
    \draw[scalplusreverse] (0,0.25) -- (120:2) node[above] {$k_2$};
    \draw[scalplus] (0,0.25) -- (60:2) node[above] {$k_3$};
    \draw[scalplusreverse] (-1,-1.5) -- (1,-1.5);
    } \normalsize
    \end{tikzpicture}}}
    \ + \ \frac{1}{2} \times
    \vcenter{\hbox{
    \begin{tikzpicture}[scale=0.4]
    { \scriptsize
\draw[scalplusreverse] (-3,-1) -- (-3,1.75) node[above] {$p_1$};
\draw[scalplus] (-2,-1) -- (-2,1.75) node[above] {$p_2$};
    \draw[Aplus] (0,0.5) -- (0,-1);
    \draw[scalplus] (0,0.5) arc (-90:270:0.75);
 } \normalsize
\end{tikzpicture}}} \, \bigg)  \ .
\end{split}
\end{equation}
Of the five diagrams generated, only the first three are relevant because tadpoles vanish by \cref{sub:one_point_functions}, while disconnected diagrams do not contribute in the end.

Next we apply $\big\{ \mathcal{S}_\mathrm{int}^\swthree\,,\, - \big\}_\star\,\sH$, resulting in the six diagrams
\begin{equation}
\begin{split}
    \hspace{-2cm}
    &\bigg\{ \mathcal{S}^\swthree_{\rm int}\,,\,
    \mathsf{H}\,(\mathrm{i}\, \hbar\, \mathsf{\Delta}_{\textrm{\tiny BV}}\, \mathsf{H})
    \bigg\{ \mathcal{S}^\swthree_{\rm int}\,,\,
    \mathsf{H}\, \bigg(\,
    \vcenter{\hbox{\begin{tikzpicture}[scale=0.4]
    { \scriptsize
    \draw[scalplusreverse] (0,-0.25) -- (0,2.25) node[above] { $ p_1$};
    \draw[scalplus] (1,-0.25) -- (1,2.25) node[above] { $ p_2$};
    \draw[Aplus] (2,-0.25) -- (2,2.25) node[above] { $ p_3$};
    } \normalsize
    \end{tikzpicture}}}
    \,
    \bigg)
    \bigg\}_\star \bigg\}_\star \\[4pt]
    & \hspace{2cm} = \frac{\mathrm{i}\,\hbar}{6}\,\bigg(
     \vcenter{\hbox{
    \begin{tikzpicture}[scale=0.4]
    { \scriptsize
\draw[scalplus] (2,0) -- (2,1.5) node[above] {$k_3$};
    \draw[scalplusreverse] (2,0) -- +(-120:1.75);
    \draw[Aplus] (2,0) -- +(-60:1.75);
    \draw[scalplusreverse] (0,-1.5) -- (0,0);
    \draw[Aplus] (0,0) -- +(120:1.75) node[above] {$k_4$};
    \draw[scalplusreverse] (0,0) -- +(60:1.75) node[above] {$k_5$};
    } \normalsize
    \end{tikzpicture}}}
    \ + \
    \vcenter{\hbox{
    \begin{tikzpicture}[scale=0.4]
    { \scriptsize
\draw[scalplus] (2,-0.5) -- (2,0.5);
    \draw[scalplus] (2,0.5) -- +(120:1.5) node[above]{$k_6$};
    \draw[Aplus] (2,0.5) -- +(60:1.5) node[above]{$k_4$};
    \draw[Aplus] (2,-0.5) -- +(-120:1.5);
    \draw[scalplusreverse] (2,-0.5) -- +(-60:1.5);
    \draw[scalplusreverse] (0,-1.75) -- (0,1.75) node[above] {$p_1$};
    } \normalsize
    \end{tikzpicture}}} 
    \ + \
    \mathrm{e}^{\,\mathrm{i}\, p_3\cdot \theta\, (k_2 + p_2)}\, \times \,
    \vcenter{\hbox{
    \begin{tikzpicture}[scale=0.4]
    { \scriptsize
\draw[scalplusreverse] (0,-0.5) -- (0,0.5);
    \draw[Aplus] (0,0.5) -- +(120:1.5) node[above]{$k_4$};
    \draw[scalplusreverse] (0,0.5) -- +(60:1.5) node[above]{$k_5$};
    \draw[Aplus] (0,-0.5) -- +(-120:1.5);
    \draw[scalplus] (0,-0.5) -- +(-60:1.5);
    \draw[scalplus] (2,-1.75) -- (2,1.75) node[above] {$p_2$};
    } \normalsize
    \end{tikzpicture}}}
    \\ &
    \hspace{4cm}
    \ + \
    \mathrm{e}^{\,\mathrm{i}\, p_3\cdot \theta\, (k_2 + p_2)}\, \times \,
    \vcenter{\hbox{
    \begin{tikzpicture}[scale=0.4]
    { \scriptsize
\draw[scalplusreverse] (0,0) -- (0,1.5) node[above] {$k_2$};
    \draw[Aplus] (0,0) -- +(-120:1.75);
    \draw[scalplus] (0,0) -- +(-60:1.75);
    \draw[scalplus] (2,-1.5) -- (2,0);
    \draw[scalplus] (2,0) -- +(120:1.75) node[above] {$k_6$};
    \draw[Aplus] (2,0) -- +(60:1.75) node[above] {$k_4$};
    } \normalsize
    \end{tikzpicture}}}
    \ + \
    \vcenter{\hbox{
    \begin{tikzpicture}[scale=0.4]
    { \scriptsize
\draw[Aplus] (0,-0.5) -- (0,0.5);
    \draw[scalplusreverse] (0,0.5) -- +(120:1.5) node[above]{$k_5$};
    \draw[scalplus] (0,0.5) -- +(60:1.5) node[above]{$k_6$};
    \draw[scalplus] (0,-0.5) -- +(-120:1.5);
    \draw[scalplusreverse] (0,-0.5) -- +(-60:1.5);
    \draw[Aplus] (2,-1.75) -- (2,1.75) node[above] {$p_3$};
    } \normalsize
    \end{tikzpicture}}}
    \ + \
    \vcenter{\hbox{
    \begin{tikzpicture}[scale=0.4]
    { \scriptsize
\draw[Aplus] (0,0) -- (0,1.5) node[above] {$k_1$};
    \draw[scalplus] (0,0) -- +(-120:1.75);
    \draw[scalplusreverse] (0,0) -- +(-60:1.75);
    \draw[scalplus] (2,-1.5) -- (2,0);
    \draw[scalplusreverse] (2,0) -- +(120:1.75) node[above] {$k_5$};
    \draw[scalplus] (2,0) -- +(60:1.75) node[above] {$k_6$};
    } \normalsize
    \end{tikzpicture}}}
     \, \bigg)  \ .
\end{split}
\end{equation}

The next vertex insertion yields 18 diagrams.
By reading the diagrams from bottom up, the introduction of vertices \eqref{eq:3-vertex diagram} is ordered; the bottom most vertex comes with ingoing momenta $\{k_1,k_2,k_3\}$, while the second and third vertices have momenta $\{l_1,l_2,l_3\}$ and $\{q_1,q_2,q_3\}$, respectively, with labelling convention of momenta following the definition of the three-vertex. This gives
\begin{equation}
\begin{split}
 \label{eq:varThetadiagrams}
    \hspace{-2cm}
    &\bigg\{ \mathcal{S}^\swthree_{\rm int}\,,\,
    \mathsf{H}\, \bigg\{ \mathcal{S}^\swthree_{\rm int}\,,\,
    \mathsf{H}\,(\mathrm{i}\, \hbar\, \mathsf{\Delta}_{\textrm{\tiny BV}}\, \mathsf{H})
    \bigg\{ \mathcal{S}^\swthree_{\rm int}\,,\,
    \mathsf{H}\, \bigg(\,
    \vcenter{\hbox{\begin{tikzpicture}[scale=0.4]
    { \scriptsize
    \draw[scalplusreverse] (0,-0.25) -- (0,2.25) node[above] { $ p_1$};
    \draw[scalplus] (1,-0.25) -- (1,2.25) node[above] { $ p_2$};
    \draw[Aplus] (2,-0.25) -- (2,2.25) node[above] { $ p_3$};
    } \normalsize
    \end{tikzpicture}}}
    \,
    \bigg)
    \bigg\}_\star \bigg\}_\star \bigg\}_\star \\[4pt]
    & \hspace{1cm} = \frac{\mathrm{i}\,\hbar}{18}\,\Bigg( \mathrm{e}^{\,\mathrm{i}\, p_3\cdot \theta\, (k_2 + p_2)}\, \times\, \bigg(
    \vcenter{\hbox{
    \begin{tikzpicture}[scale=0.4]
    { \scriptsize
\draw[scalplusreverse] (0,-0.5) -- (0,0.5);
    \draw[Aplus] (0,0.5) -- +(-0.5,1);
    \draw[scalplusreverse] (-0.5,1.5) -- +(120:1) node[above]{$q_2$};
    \draw[scalplus] (-0.5,1.5) -- +(60:1) node[above]{$q_3$};
     \draw[scalplusreverse] (0,0.5) -- +(60:2) node[above]{$l_2$};
    \draw[Aplus] (0,-0.5) -- +(-120:1.5);
    \draw[scalplus] (0,-0.5) -- +(-60:1.5);
    \draw[scalplus] (2,-1.75) -- (2,2.25) node[above] {$p_2$};
    } \normalsize
    \end{tikzpicture}}}
    \ + \
    \vcenter{\hbox{
    \begin{tikzpicture}[scale=0.4]
    { \scriptsize
\draw[scalplusreverse] (0,-0.5) -- (0,0.5);
    \draw[Aplus] (0,0.5) -- +(120:2) node[above]{$l_1$};
    \draw[scalplusreverse] (0,0.5) -- +(0.5,1);
    \draw[Aplus] (0.5,1.5) -- +(120:1) node[above]{$q_1$};
    \draw[scalplusreverse] (0.5,1.5) -- +(60:1) node[above]{$q_2$};
    \draw[Aplus] (0,-0.5) -- +(-120:1.5);
    \draw[scalplus] (0,-0.5) -- +(-60:1.5);
    \draw[scalplus] (2,-1.75) -- (2,2.25) node[above] {$p_2$};
    } \normalsize
    \end{tikzpicture}}}
    \ + \
    2\, \times\vcenter{\hbox{
    \begin{tikzpicture}[scale=0.4]
    { \scriptsize
\draw[scalplusreverse] (0,-0.5) -- (0,0.5);
    \draw[Aplus] (0,0.5) -- +(120:1.5) node[above]{$l_1$};
    \draw[scalplusreverse] (0,0.5) -- +(60:1.5) node[above]{$l_2$};
    \draw[Aplus] (0,-0.5) -- +(-120:1.5);
    \draw[scalplus] (0,-0.5) -- +(-60:1.5);
    \draw[scalplus] (2.5,-1.75) -- (2.5,0);
    \draw[scalplus] (2.5,0) -- +(120:2) node[above]{$q_3$};
    \draw[Aplus] (2.5,0) -- +(60:2) node[above]{$q_1$};
    } \normalsize
    \end{tikzpicture}}} \\
    & \hspace{8cm}
    \ + \
    \vcenter{\hbox{
    \begin{tikzpicture}[scale=0.4]
    { \scriptsize
\draw[scalplusreverse] (0,0.5) -- (0,1.75) node[above] {$k_2$};
    \draw[Aplus] (0,0.5) -- +(-120:2);
    \draw[scalplus] (0,0.5) -- +(-60:2);
    \draw[scalplus] (2,-1.25) -- (2,0);
    \draw[scalplus] (2,0) -- +(-0.5,1);
    \draw[scalplus] (1.5,1) -- +(120:1) node[above]{$q_3$};
    \draw[Aplus] (1.5,1) -- +(60:1) node[above]{$q_1$};
    \draw[Aplus] (2,0) -- +(60:2.25) node[above] {$l_1$};
    } \normalsize
    \end{tikzpicture}}}
    \ + \
    \vcenter{\hbox{
    \begin{tikzpicture}[scale=0.4]
    { \scriptsize
\draw[scalplusreverse] (0,0.5) -- (0,1.75) node[above] {$k_2$};
    \draw[Aplus] (0,0.5) -- +(-120:2);
    \draw[scalplus] (0,0.5) -- +(-60:2);
    \draw[scalplus] (2,-1.25) -- (2,0);
    \draw[Aplus] (2,0) -- +(0.5,1);
    \draw[scalplusreverse] (2.5,1) -- +(120:1) node[above]{$q_2$};
    \draw[scalplus] (2.5,1) -- +(60:1) node[above]{$q_3$};
    \draw[scalplus] (2,0) -- +(120:2.25) node[above] {$l_3$};
    } \normalsize
    \end{tikzpicture}}} \bigg) \\
    & \hspace{3cm}
    \ + \
    \vcenter{\hbox{
    \begin{tikzpicture}[scale=0.4]
    { \scriptsize
\draw[Aplus] (0,-0.5) -- (0,0.5);
    \draw[scalplusreverse] (0,0.5) -- +(-0.5,1);
    \draw[Aplus] (-0.5,1.5) -- +(120:1) node[above]{$q_1$};
    \draw[scalplusreverse] (-0.5,1.5) -- +(60:1) node[above]{$q_2$};
     \draw[scalplus] (0,0.5) -- +(60:2) node[above]{$l_3$};
    \draw[scalplus] (0,-0.5) -- +(-120:1.5);
    \draw[scalplusreverse] (0,-0.5) -- +(-60:1.5);
    \draw[Aplus] (2,-1.75) -- (2,2.25) node[above] {$p_3$};
    } \normalsize
    \end{tikzpicture}}}
    \ + \
    \vcenter{\hbox{
    \begin{tikzpicture}[scale=0.4]
    { \scriptsize
\draw[Aplus] (0,-0.5) -- (0,0.5);
    \draw[scalplusreverse] (0,0.5) -- +(120:2) node[above]{$l_2$};
    \draw[scalplus] (0,0.5) -- +(0.5,1);
    \draw[scalplus] (0.5,1.5) -- +(120:1) node[above]{$q_3$};
    \draw[Aplus] (0.5,1.5) -- +(60:1) node[above]{$q_1$};
    \draw[scalplus] (0,-0.5) -- +(-120:1.5);
    \draw[scalplusreverse] (0,-0.5) -- +(-60:1.5);
    \draw[Aplus] (2,-1.75) -- (2,2.25) node[above] {$p_3$};
    } \normalsize
    \end{tikzpicture}}}
    \ + \
    2\, \times\vcenter{\hbox{
    \begin{tikzpicture}[scale=0.4]
    { \scriptsize
\draw[Aplus] (0,-0.5) -- (0,0.5);
    \draw[scalplusreverse] (0,0.5) -- +(120:1.5) node[above]{$l_2$};
    \draw[scalplus] (0,0.5) -- +(60:1.5) node[above]{$l_3$};
    \draw[scalplus] (0,-0.5) -- +(-120:1.5);
    \draw[scalplusreverse] (0,-0.5) -- +(-60:1.5);
    \draw[Aplus] (2.5,-1.75) -- (2.5,0);
    \draw[scalplusreverse] (2.5,0) -- +(120:2) node[above]{$q_2$};
    \draw[scalplus] (2.5,0) -- +(60:2) node[above]{$q_3$};
    } \normalsize
    \end{tikzpicture}}}
    \ + \
    \vcenter{\hbox{
    \begin{tikzpicture}[scale=0.4]
    { \scriptsize
\draw[Aplus] (0,0.5) -- (0,1.75) node[above] {$k_1$};
    \draw[scalplus] (0,0.5) -- +(-120:2);
    \draw[scalplusreverse] (0,0.5) -- +(-60:2);
    \draw[Aplus] (2,-1.25) -- (2,0);
    \draw[scalplusreverse] (2,0) -- +(-0.5,1);
    \draw[Aplus] (1.5,1) -- +(120:1) node[above]{$q_1$};
    \draw[scalplusreverse] (1.5,1) -- +(60:1) node[above]{$q_2$};
    \draw[scalplus] (2,0) -- +(60:2.25) node[above] {$l_3$};
    } \normalsize
    \end{tikzpicture}}} \\
    & \hspace{3cm}
    \ + \
    \vcenter{\hbox{
    \begin{tikzpicture}[scale=0.4]
    { \scriptsize
\draw[Aplus] (0,0.5) -- (0,1.75) node[above] {$k_1$};
    \draw[scalplus] (0,0.5) -- +(-120:2);
    \draw[scalplusreverse] (0,0.5) -- +(-60:2);
    \draw[Aplus] (2,-1.25) -- (2,0);
    \draw[scalplus] (2,0) -- +(0.5,1);
    \draw[scalplus] (2.5,1) -- +(120:1) node[above]{$q_3$};
    \draw[Aplus] (2.5,1) -- +(60:1) node[above]{$q_1$};
    \draw[scalplusreverse] (2,0) -- +(120:2.25) node[above] {$l_2$};
    } \normalsize
    \end{tikzpicture}}}
    \ + \
    \vcenter{\hbox{
    \begin{tikzpicture}[scale=0.4]
    { \scriptsize
\draw[scalplus] (2,0.5) -- (2,1.75) node[above] {$k_3$};
    \draw[Aplus] (2,0.5) -- +(-120:2);
    \draw[scalplusreverse] (2,0.5) -- +(-60:2);
    \draw[scalplusreverse] (0,-1.25) -- (0,0);
    \draw[Aplus] (0,0) -- +(-0.5,1);
    \draw[scalplusreverse] (-0.5,1) -- +(120:1) node[above]{$q_2$};
    \draw[scalplus] (-0.5,1) -- +(60:1) node[above]{$q_3$};
    \draw[scalplusreverse] (0,0) -- +(60:2.25) node[above] {$l_2$};
    } \normalsize
    \end{tikzpicture}}}
    \ + \
    \vcenter{\hbox{
    \begin{tikzpicture}[scale=0.4]
    { \scriptsize
\draw[scalplus] (2,0.5) -- (2,1.75) node[above] {$k_3$};
    \draw[Aplus] (2,0.5) -- +(-120:2);
    \draw[scalplusreverse] (2,0.5) -- +(-60:2);
    \draw[scalplusreverse] (0,-1.25) -- (0,0);
    \draw[scalplusreverse] (0,0) -- +(0.5,1);
    \draw[Aplus] (0.5,1) -- +(120:1) node[above]{$q_1$};
    \draw[scalplusreverse] (0.5,1) -- +(60:1) node[above]{$q_2$};
    \draw[Aplus] (0,0) -- +(120:2.25) node[above] {$l_1$};
    } \normalsize
    \end{tikzpicture}}}
    \ + \
    2\, \times\vcenter{\hbox{
    \begin{tikzpicture}[scale=0.4]
    { \scriptsize
\draw[scalplus] (2.5,-0.5) -- (2.5,0.5);
    \draw[scalplus] (2.5,0.5) -- +(120:1.5) node[above]{$q_3$};
    \draw[Aplus] (2.5,0.5) -- +(60:1.5) node[above]{$q_1$};
    \draw[Aplus] (2.5,-0.5) -- +(-120:1.5);
    \draw[scalplusreverse] (2.5,-0.5) -- +(-60:1.5);
    \draw[scalplusreverse] (0,-1.75) -- (0,0);
    \draw[Aplus] (0,0) -- +(120:2) node[above]{$l_1$};
    \draw[scalplusreverse] (0,0) -- +(60:2) node[above]{$l_2$};
    } \normalsize
    \end{tikzpicture}}} \\
    & \hspace{8cm}
    \ + \
    \vcenter{\hbox{
    \begin{tikzpicture}[scale=0.4]
    { \scriptsize
\draw[scalplus] (2,-0.5) -- (2,0.5);
    \draw[scalplus] (2,0.5) -- +(-0.5,1);
    \draw[scalplus] (1.5,1.5) -- +(120:1) node[above]{$q_3$};
    \draw[Aplus] (1.5,1.5) -- +(60:1) node[above]{$q_1$};
     \draw[Aplus] (2,0.5) -- +(60:2) node[above]{$l_1$};
    \draw[Aplus] (2,-0.5) -- +(-120:1.5);
    \draw[scalplusreverse] (2,-0.5) -- +(-60:1.5);
    \draw[scalplusreverse] (0,-1.75) -- (0,2.25) node[above] {$p_1$};
    } \normalsize
    \end{tikzpicture}}}
    \ + \
    \vcenter{\hbox{
    \begin{tikzpicture}[scale=0.4]
    { \scriptsize
\draw[scalplus] (0,-0.5) -- (0,0.5);
    \draw[scalplus] (0,0.5) -- +(120:2) node[above]{$l_3$};
    \draw[Aplus] (0,0.5) -- +(0.5,1);
    \draw[scalplusreverse] (0.5,1.5) -- +(120:1) node[above]{$q_2$};
    \draw[scalplus] (0.5,1.5) -- +(60:1) node[above]{$q_3$};
    \draw[Aplus] (0,-0.5) -- +(-120:1.5);
    \draw[scalplus] (0,-0.5) -- +(-60:1.5);
    \draw[scalplusreverse] (-2,-1.75) -- (-2,2.25) node[above] {$p_1$};
    } \normalsize
    \end{tikzpicture}}}
  \,\hskip 0.5cm \Bigg)  \ .
\end{split}
\end{equation}

From the diagrammatic expression \eqref{eq:varThetadiagrams} it is straightforward to see that all diagrams are two-particle irreducible (2PI) upon applying the operator $(\mathrm{i}\, \hbar\, \mathsf{\Delta}_{\textrm{\tiny BV}}\, \mathsf{H})^2$.
The resulting connected diagrams contributing to \eqref{eq:pain_I} are
\begin{equation}\label{eq:not1PIvertex}
 \begin{split}
 {\text{\small{\tt(I)}}} \ = \
		-\frac{ \mathrm{i}\, \hbar^3}{3}\,\left(
    \quad
    \vcenter{\hbox{\scriptsize{\begin{tikzpicture}[scale=0.6]
    \draw[scalplusreverse] (0,-1) -- (2,1);
    \draw[scalplus] (4,-1) -- (2,1);
    \draw[Aplus] (2,1) -- (2,3);
    \draw[Aplus] (2.5,0.5) to [out=45,in=135] (3.25,0.5) to [out=-45, in=45] (3.5,-0.5);
       \node at (2, 3.3) { $ \mu, p_3 $ };
    \node at (0, -1.3) { $ p_1 $};
    \node at (4, -1.3) { $ p_2 $};
    \end{tikzpicture}}}} \normalsize \quad
    + \quad
    \vcenter{\hbox{\scriptsize{\begin{tikzpicture}[scale=0.6]
    \draw[scalplusreverse] (0,-1) -- (2,1);
    \draw[scalplus] (4,-1) -- (2,1);
    \draw[Aplus] (2,1) -- (2,3);
    \draw[Aplus] (1.5,0.5) to [out=135,in=45] (0.75,0.5) to [out=-135, in=135] (0.5,-0.5);
    \node at (2, 3.3) { $ \mu, p_3 $ };
    \node at (0, -1.3) { $ p_1 $};
    \node at (4, -1.3) { $ p_2 $};
    \end{tikzpicture}}}} \normalsize \quad
    + \quad
    \vcenter{\hbox{\scriptsize{\begin{tikzpicture}[scale=0.6]
    \draw[scalplusreverse] (0,-1) -- (2,1);
    \draw[scalplus] (4,-1) -- (2,1);
    \draw[Aplus] (2,1) -- (2,2);
    \draw[scalplus] (2,2) arc (-90:270:0.5);
    \draw[Aplus] (2,3) -- (2,4);
    \node at (2, 4.3) { $ \mu, p_3 $ };
    \node at (0, -1.3) { $ p_1 $};
    \node at (4, -1.3) { $ p_2 $};
    \end{tikzpicture}}}} \normalsize \quad
    \right)
 \end{split}
\end{equation}
where each external leg is corrected at one-loop.
Of course the photon and scalar legs only have contributions from the trivalent vertices in this term, while the corrections of the form \eqref{fig:VPc} and \eqref{fig:SSEc} will come from a later term.
The explicit analytic expressions for these and other diagrams are summarised at the end of this section.

\paragraph{Calculation of ${\mbf{\text{\small{\tt(II)}}}}$.}

In the rest of this section we will not draw all diagrams; one can find the relevant contributions following the procedure outlined above.
 The second term ${\text{\small{\tt(II)}}}$ from \eqref{eq:pain_II} contributes to the one-particle irreducible (1PI) part of the correlation function.
After simple relabellings of internal momentum variables, we find that all contributions can be combined and yield
\begin{equation}\label{eq:triangle1}
 \begin{split}
 {\text{\small{\tt(II)}}} \ = \  \frac45 \, \times \, {\text{\small{\tt(I)}}}\quad
   - \quad \frac{2\, \mathrm{i}\, \hbar^3}{5}\, \ \mathrm{e}^{\,\mathrm{i}\,k_1\cdot \theta\, p_3} \times
    \vcenter{\hbox{\scriptsize{\begin{tikzpicture}[scale=0.6]
    \draw[scalplusreverse] (0, -1) -- (1, 0);
    \draw[scalplusreverse] (1, 0) -- (2, 1);
    \draw[scalplus] (4, -1) -- (3, 0);
    \draw[scalplus] (3, 0) -- (2, 1);
    \draw[Aplus] (2, 1) -- (2, 3);
    \draw[Aplus] (1, 0) -- (3, 0);
    \node at (2, 3.3) { $ \mu, p_3 $ };
    \node at (0, -1.3) { $ p_1 $};
    \node at (4, -1.3) { $ p_2 $};
    \node at (1.4, -0.5) {  $ \rho, k_1 $};
    \end{tikzpicture}}}} \normalsize
 \end{split}
\end{equation}
With the vertex function \eqref{eq:3-vertex diagram}, we notice that the phase factor appearing in \eqref{eq:triangle1} combines with the phase factors appearing in the vertex function to factor out of the loop integral.
This is nothing but the corresponding commutative diagram dressed by an external phase factor.

\paragraph{Calculation of ${\mbf{\text{\small{\tt(III)}}}}$.}

The 1PI part of the third term ${\text{\small{\tt(III)}}}$ from \eqref{eq:pain_III} is of the same form as the triangle diagram contribution to \eqref{eq:triangle1}.
Altogether, its contributions sum up to
\begin{equation}
 \begin{split}
 {\text{\small{\tt(III)}}} \ = \ \frac32\,\times\, {\text{\small{\tt(II)}}}
\end{split}
\end{equation}
which is just the commutative result times the external phase factor. 

\paragraph{Calculation of ${\mbf{\text{\small{\tt(IV)}}}}$.}

For the fourth correction term ${\text{\small{\tt(IV)}}}$, we find that all terms contributing to \eqref{eq:pain_IV} are  of the 2PI form 
\begin{equation}
\begin{split}\label{eq:2PIwithv4v3}
{\text{\small{\tt(IV)}}} \ &= \ - \frac{\mathrm{i}\, \hbar^3}{3}\,\left(
    \quad
    \vcenter{\hbox{\scriptsize{\begin{tikzpicture}[scale=0.6]
    \draw[scalplusreverse] (0,-1) -- (1,0);
    \draw[scalplusreverse] (1,0) -- (2,1);
    \draw[scalplus] (4,-1) -- (2,1);
    \draw[Aplus] (2,1) -- (2,3);
    \draw[Aplus] (1,0) to [out=75,in=30] (-0.3,1.5) to [out=-135, in=-165] (1,0);
    \node at (2, 3.3) { $ \mu, p_3 $ };
    \node at (0, -1.3) { $ p_1 $};
    \node at (4, -1.3) { $ p_2 $};
    \end{tikzpicture}}}} \normalsize \quad
    + \quad
    \vcenter{\hbox{\scriptsize{\begin{tikzpicture}[scale=0.6]
    \draw[scalplusreverse] (0,-1) -- (2,1);
    \draw[scalplus] (4,-1) -- (3,0);
    \draw[scalplus] (3,0) -- (2,1);
    \draw[Aplus] (2,1) -- (2,3);
    \draw[Aplus] (3,0) to [out=-15,in=-40] (4.3,1.5) to [out=130, in=105] (3,0);
    \node at (2, 3.3) { $ \mu, p_3 $ };
    \node at (0, -1.3) { $ p_1 $};
    \node at (4, -1.3) { $ p_2 $};
    \end{tikzpicture}}}} \normalsize \right. \\
    & \hspace{8cm} \left. + \quad 2 \times
    \vcenter{\hbox{\scriptsize{\begin{tikzpicture}[scale=0.6]
    \draw[scalplusreverse] (0,-1) -- (2,1);
    \draw[scalplus] (4,-1) -- (2,1);
    \draw[Aplus] (2,1) -- (2,2);
    \draw[Aplus] (2,2) -- (2,3);
    \draw[scalplus] (2,2) to [out=120,in=90] (0,2) to [out=-90, in=-120] (2,2);
    \node at (2, 3.3) { $ \mu, p_3 $ };
    \node at (0, -1.3) { $ p_1 $};
    \node at (4, -1.3) { $ p_2 $};
    \end{tikzpicture}}}} \normalsize \quad
    \right)
 \end{split}
\end{equation}
where the correction happens at the level of the propagators.
Of course, the sum over all 2PI diagrams can be recovered by dressing the external propagators with their one-loop correction (modulo higher order graphs), which we calculated in \cref{sub:photon_polarisation} and \cref{sub:scalar_self_energy}.

\paragraph{Calculation of ${\mbf{\text{\small{\tt(V)}--\small{\tt(VII)}}}}$.}

The last three terms of the contributions to \eqref{eqs:pains} turn out to be proportional to each other.
They all carry a four-point vertex and a three-point vertex.
Then \eqref{eq:pain_V}, \eqref{eq:pain_VI} and \eqref{eq:pain_VII} contribute
\begin{equation}
 \begin{split}
   {\text{\small{\tt(V)}}} \ = \ -\frac{\mathrm{4\,i}\, \hbar^3}{5}\,\left(
    \quad 
    \vcenter{\hbox{\scriptsize{\begin{tikzpicture}[scale=0.6]
    { \scriptsize
    \draw[scalplusreverse] (0,-1) -- (2,1);
    \draw[scalplus] (4,-1) -- (2,1);
    \draw[Aplus] (2,1) -- (2,3);
    \draw[Aplus] (2,1) to [out=45,in=135] (3.25,0.75) to [out=-45, in=45] (3.5,-0.5);
    \node at (2, 3.3) { $ \mu, p_3 $ };
    \node at (1.3, 1.4) { $ \lambda, k_1 $ };
    \node at (2.9, 1.6) { $ \rho, k_2 $ };
    \node at (0, -1.3) { $ p_1 $};
    \node at (4, -1.3) { $ p_2 $};
    } \normalsize
    \end{tikzpicture}}}} \normalsize \quad
    + \quad 
    \vcenter{\hbox{\scriptsize{\begin{tikzpicture}[scale=0.6]
    { \scriptsize
    \draw[scalplusreverse] (0,-1) -- (2,1);
    \draw[scalplus] (4,-1) -- (2,1);
    \draw[Aplus] (2,1) -- (2,3);
    \draw[Aplus] (2,1) to [out=135,in=45] (0.75,0.75) to [out=-135, in=135] (0.5,-0.5);
    \node at (2, 3.3) { $ \mu, p_3 $ };
    \node at (1.1, 1.6) { $ \lambda, k_1 $ };
    \node at (2.7, 1.4) { $ \rho, k_2 $ };
    \node at (0, -1.3) { $ p_1 $};
    \node at (4, -1.3) { $ p_2 $};
    } \normalsize
    \end{tikzpicture}}}} \normalsize
    \quad \right) \quad + \quad \frac35\,\times\,{\text{\small{\tt(IV)}}}
 \end{split}
\end{equation}
together with
\begin{align}
{\text{\small{\tt(VI)}}} \ = \ \frac56\,\times\,{\text{\small{\tt(V)}}} \qquad \mbox{and} \qquad {\text{\small{\tt(VII)}}} \ = \ \frac23\,\times\,{\text{\small{\tt(V)}}} \ .
\end{align}
This is the commutative result up to the external noncommutative phase factor.

\paragraph{One-Loop Corrections to the Three-Vertex.}

The three-vertex correction corresponds to the contributions from 1PI diagrams.
Combining the above results, we obtain the 1PI three-point correlation function
\begin{equation}
 \begin{split}
 \mathsf{\tilde G}^{\hbar,{\rm int}}_{\phi, \phi^{\dagger}, A_{\mu}}(p_1, p_2, p_3)_{\text{\tiny 1PI}}^\swone \ &= \ -\mathrm{i}\,\hbar^3\, \left(
  \quad 2 \times
    \vcenter{\hbox{\begin{tikzpicture}[scale=0.6]
    { \scriptsize
    \draw[scalplusreverse] (0,-1) -- (2,1);
    \draw[scalplus] (4,-1) -- (2,1);
    \draw[Aplus] (2,1) -- (2,3);
    \draw[Aplus] (2,1) to [out=45,in=135] (3.25,0.75) to [out=-45, in=45] (3.5,-0.5);
    \node at (2, 3.3) { $ \mu, p_3 $ };
    \node at (1.3, 1.4) { $ \lambda, k_1 $ };
    \node at (2.9, 1.6) { $ \rho, k_2 $ };
    \node at (0, -1.3) { $ p_1 $};
    \node at (4, -1.3) { $ p_2 $};
    } \normalsize
    \end{tikzpicture}}} \quad
    + \quad 2 \times
    \vcenter{\hbox{\begin{tikzpicture}[scale=0.6]
    { \scriptsize
    \draw[scalplusreverse] (0,-1) -- (2,1);
    \draw[scalplus] (4,-1) -- (2,1);
    \draw[Aplus] (2,1) -- (2,3);
    \draw[Aplus] (2,1) to [out=135,in=45] (0.75,0.75) to [out=-135, in=135] (0.5,-0.5);
    \node at (2, 3.3) { $ \mu, p_3 $ };
    \node at (1.1, 1.6) { $ \lambda, k_1 $ };
    \node at (2.7, 1.4) { $ \rho, k_2 $ };
    \node at (0, -1.3) { $ p_1 $};
    \node at (4, -1.3) { $ p_2 $};
    } \normalsize
    \end{tikzpicture}}} \right.
    \\
    &  \hspace{5cm} \left.
   + \quad \mathrm{e}^{\,\mathrm{i}\,k_1\cdot \theta\, p_3} \times
    \vcenter{\hbox{ { \scriptsize \begin{tikzpicture}[scale=0.6]
    \draw[scalplusreverse] (0, -1) -- (1, 0);
    \draw[scalplusreverse] (1, 0) -- (2, 1);
    \draw[scalplus] (4, -1) -- (3, 0);
    \draw[scalplus] (3, 0) -- (2, 1);
    \draw[Aplus] (2, 1) -- (2, 3);
    \draw[Aplus] (1, 0) -- (3, 0);
    \node at (2, 3.3) { $ \mu, p_3 $ };
    \node at (0, -1.3) { $ p_1 $};
    \node at (4, -1.3) { $ p_2 $};
    \node at (1.4, -0.5) {  $ \rho, k_1 $};
    \end{tikzpicture}}} } \normalsize \quad
    \right)
 \end{split}
\end{equation}

Using the expressions for the diagrams below, this reads explicitly
\begin{equation} \label{eq:3vertex1loop}
\begin{split}
&    \mathsf{\tilde G}^{\hbar,{\rm int}}_{\phi, \phi^{\dagger}, A_{\mu}}(p_1, p_2, p_3)_{\text{\tiny 1PI}}^\swone  \\[4pt]
& \hspace{1cm} =  g^3 \,
    (\mathrm{i}\, \hbar)^3\,
      \e^{-\frac\ii2\,\sum\limits_{i<j}\,p_i\cdot\theta\,p_j}  \, 
    \mathsf{\tilde{G}}(p_1) \,
    \mathsf{\tilde{G}}(p_2) \,
    \mathsf{\tilde{D}}(p_3) \,\mathsf{\tilde\Pi}_{\rho \mu}(p_3) \, (2\pi)^4\,\delta(p_1+p_2-p_3)
    \\
    & \hspace{2cm} \, \times \bigg(
    \int_k\, 
    (k-2p_1)^\lambda \,
    (k+2p_2 )^\alpha \,
    (-2k+p_1-p_2 )^\rho\, \\
    & \hspace{7.5cm} \times \mathsf{\tilde G}(k-p_1) \,\mathsf{\tilde G}(k+p_2) \, \mathsf{\tilde D}(k) \, \mathsf{\tilde\Pi}_{\lambda\alpha}(k)
    \\
    & \hspace{3cm} - 2\,
    \int_k\,
    \eta^{\lambda \rho} \,
    \left(
    (k + 2p_2)^{\alpha} \,
    \mathsf{\tilde{G}}(k+p_2) \,
    \mathsf{\tilde{D}}(k)\, \mathsf{\tilde\Pi}_{\alpha \lambda}( k) \right. \\
    & \hspace{7.5cm} \left.
    - \,
    (k+2p_1)^\alpha\,
    \mathsf{\tilde{G}}(k+p_1)\,
    \mathsf{\tilde{D}}(k)\, \mathsf{\tilde\Pi}_{\lambda \alpha}(k)
    \right) \bigg) \ .
\end{split}
\end{equation}

\paragraph{Diagrams.}

We close this section by summarising the explicit expressions for the diagrams which appeared in our calculation of the three-vertex function above:
\begin{subequations}
\begin{equation}
 \begin{split}
  \vcenter{\hbox{\begin{tikzpicture}[scale=0.6]
    { \scriptsize
    \draw[scalplusreverse] (0,-1) -- (2,1);
    \draw[scalplus] (4,-1) -- (2,1);
    \draw[Aplus] (2,1) -- (2,3);
    \draw[Aplus] (2,1) to [out=45,in=135] (3.25,0.75) to [out=-45, in=45] (3.5,-0.5);
    \node at (2, 3.3) { $ \mu, p_3 $ };
    \node at (1.3, 1.4) { $ \lambda, k_1 $ };
    \node at (2.9, 1.6) { $ \rho, k_2 $ };
    \node at (0, -1.3) { $ p_1 $};
    \node at (4, -1.3) { $ p_2 $};
    } \normalsize
    \end{tikzpicture}}} \quad &=
     g^3\  \e^{-\frac\ii2\,\sum\limits_{i<j}\,p_i\cdot\theta\,p_j}  \ (2\pi)^4\, \delta(p_1 + p_2 + p_3)\\[-30pt]
     & \quad\,\times  \tilde{\sf G}(p_1)\, \tilde{\sf G}(p_2)\, \tilde{\sf D}(p_3)\, \tilde{\mathsf{\Pi}}_{\mu\lambda} (p_3)\,\int_k\, (k + 2 p_2)_{\sigma}\,
    \tilde{\sf D}(k)\, \tilde{\mathsf{\Pi}}^{\sigma\lambda} (k)\, \tilde{\sf G}(k + p_2)
 \end{split}
\end{equation}
\begin{equation}
 \begin{split}
  \vcenter{\hbox{\begin{tikzpicture}[scale=0.6]
    { \scriptsize
    \draw[scalplusreverse] (0,-1) -- (2,1);
    \draw[scalplus] (4,-1) -- (2,1);
    \draw[Aplus] (2,1) -- (2,3);
    \draw[Aplus] (2,1) to [out=135,in=45] (0.75,0.75) to [out=-135, in=135] (0.5,-0.5);
    \node at (2, 3.3) { $ \mu, p_3 $ };
    \node at (1.1, 1.6) { $ \lambda, k_1 $ };
    \node at (2.7, 1.4) { $ \rho, k_2 $ };
    \node at (0, -1.3) { $ p_1 $};
    \node at (4, -1.3) { $ p_2 $};
    } \normalsize
    \end{tikzpicture}}} \quad &= -g^3\  \e^{-\frac\ii2\,\sum\limits_{i<j}\,p_i\cdot\theta\,p_j}  \  (2\pi)^4\,\delta(p_1 + p_2 + p_3) \\[-30pt]
    & \quad \, \times  \tilde{\sf G}(p_1)\, \tilde{\sf G}(p_2)\, \tilde{\sf D}(p_3)\, \tilde{\mathsf{\Pi}}_{\mu\lambda} (p_3)\,\int_k\, (k + 2 p_1)_{\sigma}\,
    \tilde{\sf D}(k)\, \tilde{\mathsf{\Pi}}^{\sigma\lambda} (k)\, \tilde{\sf G}(k + p_1)
 \end{split}
\end{equation}
\begin{equation}
 \begin{split}
\vcenter{\hbox{\begin{tikzpicture}[scale=0.6]
{ \scriptsize
    \draw[scalplusreverse] (0,-1) -- (1,0);
    \draw[scalplusreverse] (1,0) -- (2,1);
    \draw[scalplus] (4,-1) -- (2,1);
    \draw[Aplus] (2,1) -- (2,3);
    \draw[Aplus] (1,0) to [out=75,in=30] (-0.3,1.5) to [out=-135, in=-165] (1,0);
    \node at (2, 3.3) { $ \mu, p_3 $ };
    \node at (0, -1.3) { $ p_1 $};
    \node at (4, -1.3) { $ p_2 $};
    } \normalsize
    \end{tikzpicture}}}  \quad &= g^3\  \e^{-\frac\ii2\,\sum\limits_{i<j}\,p_i\cdot\theta\,p_j}  \  (2\pi)^4\,\delta(p_1 + p_2 + p_3) \\[-30pt]
    & \quad \, \times  \tilde{\sf G}(p_1)^2\, \tilde{\sf G}(p_2)\, \tilde{\sf D}(p_3)\, \tilde{\mathsf{\Pi}}_{\mu\sigma} (p_3)\, (p_2 -  p_1)^{\sigma}\,\int_k\,
    \tilde{\sf D}(k)\, \tilde{\mathsf{\Pi}}^{\lambda}_{\lambda} (k)
 \end{split}
\end{equation}
\begin{equation}
 \begin{split}
\vcenter{\hbox{\begin{tikzpicture}[scale=0.6]
 { \scriptsize
    \draw[scalplusreverse] (0,-1) -- (2,1);
    \draw[scalplus] (4,-1) -- (3,0);
    \draw[scalplus] (3,0) -- (2,1);
    \draw[Aplus] (2,1) -- (2,3);
    \draw[Aplus] (3,0) to [out=-15,in=-40] (4.3,1.5) to [out=130, in=105] (3,0);
    \node at (2, 3.3) { $ \mu, p_3 $ };
    \node at (0, -1.3) { $ p_1 $};
    \node at (4, -1.3) { $ p_2 $};
    } \normalsize
    \end{tikzpicture}}}  \quad &=  g^3\  \e^{-\frac\ii2\,\sum\limits_{i<j}\,p_i\cdot\theta\,p_j}  \  (2\pi)^4\,\delta(p_1 + p_2 + p_3) \\[-30pt]
    & \quad \, \times  \tilde{\sf G}(p_1)\, \tilde{\sf G}(p_2)^2\, \tilde{\sf D}(p_3)\, \tilde{\mathsf{\Pi}}_{\mu\sigma} (p_3)\, (p_2 -  p_1)^{\sigma}\,\int_k\,
    \tilde{\sf D}(k)\, \tilde{\mathsf{\Pi}}^{\lambda}_{\lambda} (k)
 \end{split}
\end{equation}
\begin{equation}
 \begin{split}
\vcenter{\hbox{\begin{tikzpicture}[scale=0.6]
 { \scriptsize
    \draw[scalplusreverse] (0,-1) -- (2,1);
    \draw[scalplus] (4,-1) -- (2,1);
    \draw[Aplus] (2,1) -- (2,2);
    \draw[Aplus] (2,2) -- (2,3);
    \draw[scalplus] (2,2) to [out=120,in=90] (0,2) to [out=-90, in=-120] (2,2);
    \node at (2, 3.3) { $ \mu, p_3 $ };
    \node at (0, -1.3) { $ p_1 $};
    \node at (4, -1.3) { $ p_2 $};
    } \normalsize
    \end{tikzpicture}}}  \quad &=  g^3\  \e^{-\frac\ii2\,\sum\limits_{i<j}\,p_i\cdot\theta\,p_j}  \  (2\pi)^4\,\delta(p_1 + p_2 + p_3) \\[-30pt]
    & \quad \, \times  \tilde{\sf G}(p_1)\, \tilde{\sf G}(p_2)\, \tilde{\sf D}(p_3)\, \tilde{\mathsf{\Pi}}_{\mu\rho} (p_3)\, \tilde{\sf D}(p_3)\, \tilde{\mathsf{\Pi}}^{\rho\sigma} (p_3)\, (p_2 -  p_1)_{\sigma}\,\int_k\,
    \tilde{\sf{G}} (k)
 \end{split}
\end{equation}
\begin{equation}
 \begin{split}
\vcenter{\hbox{\begin{tikzpicture}[scale=0.6]
{ \scriptsize
     \draw[scalplusreverse] (0,-1) -- (2,1);
    \draw[scalplus] (4,-1) -- (2,1);
    \draw[Aplus] (2,1) -- (2,3);
    \draw[Aplus] (2.5,0.5) to [out=45,in=135] (3.25,0.5) to [out=-45, in=45] (3.5,-0.5);
       \node at (2, 3.3) { $ \mu, p_3 $ };
    \node at (0, -1.3) { $ p_1 $};
    \node at (4, -1.3) { $ p_2 $};
    } \normalsize
    \end{tikzpicture}}}  \quad &= - g^3\  \e^{-\frac\ii2\,\sum\limits_{i<j}\,p_i\cdot\theta\,p_j}  \  (2\pi)^4\,\delta(p_1 + p_2 + p_3) \\[-35pt]
    & \hspace{1cm} \, \times  \tilde{\sf G}(p_1)\, \tilde{\sf G}(p_2)^{2}\, \tilde{\sf D}(p_3)\, \tilde{\mathsf{\Pi}}_{\mu\lambda} (p_3)\,  (p_2 -  p_1)^{\lambda}\\
    & \hspace{2cm} \times \int_k\, (k + 2p_2)^{\rho}\, (k + 2 p_2)^{\sigma}\, \tilde{\sf{G}} (k + p_2)\, \tilde{\sf D}(k)\, \tilde{\mathsf{\Pi}}_{\rho\sigma} (k)
 \end{split}
\end{equation}
\begin{equation}
 \begin{split}
\vcenter{\hbox{\begin{tikzpicture}[scale=0.6]
{ \scriptsize
     \draw[scalplusreverse] (0,-1) -- (2,1);
    \draw[scalplus] (4,-1) -- (2,1);
    \draw[Aplus] (2,1) -- (2,3);
    \draw[Aplus] (1.5,0.5) to [out=135,in=45] (0.75,0.5) to [out=-135, in=135] (0.5,-0.5);
    \node at (2, 3.3) { $ \mu, p_3 $ };
    \node at (0, -1.3) { $ p_1 $};
    \node at (4, -1.3) { $ p_2 $};
    } \normalsize
    \end{tikzpicture}}}  \quad &=  - g^3\  \e^{-\frac\ii2\,\sum\limits_{i<j}\,p_i\cdot\theta\,p_j}  \  (2\pi)^4\,\delta(p_1 + p_2 + p_3) \\[-35pt]
    & \hspace{1cm} \, \times  \tilde{\sf G}(p_1)^2\, \tilde{\sf G}(p_2)\, \tilde{\sf D}(p_3)\, \tilde{\mathsf{\Pi}}_{\mu\lambda} (p_3)\,  (p_2 -  p_1)^{\lambda}\\
     &\hspace{2cm}\times\int_k\, (k + 2p_1)^{\rho}\, (k + 2 p_1)^{\sigma}\, \tilde{\sf{G}} (k + p_1)\, \tilde{\sf D}(k)\, \tilde{\mathsf{\Pi}}_{\rho\sigma} (k)
 \end{split}
\end{equation}
\begin{equation}
 \begin{split}
\vcenter{\hbox{\begin{tikzpicture}[scale=0.6]
{ \scriptsize
     \draw[scalplusreverse] (0,-1) -- (2,1);
    \draw[scalplus] (4,-1) -- (2,1);
    \draw[Aplus] (2,1) -- (2,2);
    \draw[scalplus] (2,2) arc (-90:270:0.5);
    \draw[Aplus] (2,3) -- (2,4);
    \node at (2, 4.3) { $ \mu, p_3 $ };
    \node at (0, -1.3) { $ p_1 $};
    \node at (4, -1.3) { $ p_2 $};
    } \normalsize
    \end{tikzpicture}}} \quad &=  - g^3\  \e^{-\frac\ii2\,\sum\limits_{i<j}\,p_i\cdot\theta\,p_j}  \  (2\pi)^4\,\delta(p_1 + p_2 + p_3) \\[-45pt]
    & \hspace{1cm} \, \times  \tilde{\sf G}(p_1)\, \tilde{\sf G}(p_2)\, \tilde{\sf D}(p_3)\, \tilde{\mathsf{\Pi}}_{\mu\rho} (p_3)\, \tilde{\sf D}(p_3)\, \tilde{\mathsf{\Pi}}_{\lambda\sigma} (p_3)\,  (p_2 -  p_1)^{\lambda}\\
     &\hspace{2cm}\times\int_k\, (2k - p_3)^{\rho}\, (2k - p_3)^{\sigma}\, \tilde{\sf{G}} (k)\, \tilde{\sf G}(p_3 - k)
 \end{split}
\end{equation}
\begin{equation}
 \begin{split}
  \mathrm{e}^{\,\mathrm{i}\,p_3 \cdot \theta\, k} \, \times
\vcenter{\hbox{\begin{tikzpicture}[scale=0.6]
 { \scriptsize
    \draw[scalplusreverse] (0, -1) -- (1, 0);
    \draw[scalplusreverse] (1, 0) -- (2, 1);
    \draw[scalplus] (4, -1) -- (3, 0);
    \draw[scalplus] (3, 0) -- (2, 1);
    \draw[Aplus] (2, 1) -- (2, 3);
    \draw[Aplus] (1, 0) -- (3, 0);
    \node at (2, 3.3) { $ \mu, p_3 $ };
    \node at (0, -1.3) { $ p_1 $};
    \node at (4, -1.3) { $ p_2 $};
    \node at (1.4, -0.5) {  $ \rho, k $};
    } \normalsize
    \end{tikzpicture}}} \quad &= - g^3\  \e^{-\frac\ii2\,\sum\limits_{i<j}\,p_i\cdot\theta\,p_j}  \  (2\pi)^4\,\delta(p_1 + p_2 + p_3) \\[-35pt]
    & \hspace{1cm} \times  \tilde{\sf G}(p_1)\, \tilde{\sf G}(p_2)\, \tilde{\sf D}(p_3)\, \tilde{\mathsf{\Pi}}_{\mu\lambda} (p_3)   \\
     &\hspace{2cm}\times\int_k\, (p_2 -  p_1 + 2 k)^{\lambda}\, (k + 2p_2)^{\sigma}\, (k - 2p_1)^{\rho}\\
     &\hspace{3cm}\times \tilde{\sf{G}} (p_1-k)\, \tilde{\sf G}(k + p_2)\, \tilde{\sf D}(k)\, \tilde{\mathsf{\Pi}}_{\rho\sigma} (k)
 \end{split}
\end{equation}
\end{subequations}

\section{Ward-Takahashi Identities for Braided Gauge Symmetries} 
\label{sec:braided_gauge_symmetries}

In \cref{sub:photon_polarisation} we showed that the one-loop photon vacuum polarisation tensor $\Pi_\star(p)^\swone$ of braided scalar quantum electrodynamics is transverse: $p_\mu\, {\Pi}_{\star}^{ \mu \nu}(p)^\swone = 0$. This reflects the Ward-Takahashi identity for the braided $\sU(1)$ gauge symmetry at the level of the vacuum polarisations. In this final section we use the braided $L_\infty$-structure and the homological perturbation lemma to extend this Ward-Takahashi identity to all loop orders for correlators involving the electric current.

Our homotopy algebraic formulation of the Ward-Takahashi identities is similar in spirit to other approaches which have appeared recently in the literature.
Linearised Ward identities for tree-level scattering amplitudes in Yang-Mills theory are established in~\cite{Bonezzi:2023xhn} within the $L_\infty$-algebra framework. Scattering amplitudes are well understood in terms of a quasi-isomorphism onto the minimal model of the corresponding $L_\infty$-algebra $\CCL$, which in particular contains the asymptotically free fields (see e.g.~\cite{Arvanitakis:2019ald,Macrelli:2019afx,Jurco:2019yfd,Gomez2021,Szabo:2023cmv}). When one of the fields in a scattering amplitude is taken to be trivial in the cohomology $H^1(L)$, then the amplitude vanishes as a direct consequence of the homotopy Jacobi identities in the minimal model~\cite{Bonezzi:2023xhn}. The physical interpretation is that longitudinal external states do not contribute to the S-matrix. 

Anomalous Ward-Takahashi identities for correlation functions involving Noether currents for general global symmetries of field theories are derived in~\cite{Konosu:2024dpo} within the framework of quantum $A_\infty$-algebras. The derivation follows from an equivalent form of the homotopy algebraic Schwinger-Dyson equations from~\cite{Okawa:2022sjf,Konosu2023,Konosu:2023rkm}, which are based on the homological perturbation lemma. The anomaly matches with known anomalies in examples.

We stress that our approach is fundamentally different: it is based on the underlying local BRST cohomology within our braided BV formalism. Combined with the braided homological perturbation lemma, this leads to a `master identity' (see \eqref{eq:masterid}) which is functionally similar to the algebraic Schwinger-Dyson equation of~\cite[eq.~(3.45)]{Konosu:2024dpo}. Our Ward-Takahashi identity (see \eqref{eq:Ward-Takahashi identity}) is then analogous to that of~\cite[eq.~(4.45)]{Konosu:2024dpo}. However, in contrast to~\cite{Konosu:2024dpo}, our derivation leads to an interpretation of the anomaly in terms of the BV Laplacian of the extended BRST functional for the matter fields.

\subsection{Becchi-Rouet-Stora-Tyutin Transformations} 
\label{sub:brst_symmetries}

In the BV formalism, the braided gauge symmetries of the classical Maurer-Cartan theory from \cref{sub:braided_qed} are promoted to nilpotent global symmetries of the extended BV theory from \cref{sub:perturbative_calculations_finally1}.
Braided gauge transformations come from the BRST part 
of the extended BV functional \eqref{eq:BVactionext}, evaluated in \eqref{eq:Sbrst2} for the quadratic vertex and in \eqref{eq:S3BRST} for the cubic vertex.
This induces the nilpotent braided derivation of degree 1 given by
\begin{equation}\label{eq:Qbrst}
    Q_{\BRST} := \big\{ \mathcal{S}_{\BRST}^\swtwo + \CS_\BRST^\swthree \,,\, - \big\}_\star : \mathrm{Sym}_\star L[2] \longrightarrow (\mathrm{Sym}_\star L[2])[1] \ ,
\end{equation}
which is in fact a \emph{strict} derivation by momentum conservation.

Acting on a basis of antifields for $L[2]^0$ in momentum space, the BRST differential \eqref{eq:Qbrst} describes the braided gauge transformations \eqref{eq:bsqedgaugetransfo}:
\begin{equation} \label{eq:QBRSTmomentum}
    Q_{\BRST} 
    \begin{pmatrix}
        \mathtt{e}^{p} \\[0.3em]
        \bar{\mathtt{e}}^{p} \\[0.3em]
        \mathtt{v}_\mu\, \mathtt{e}^p
    \end{pmatrix}
    =
    \begin{pmatrix}\displaystyle
        -\mathrm{i}\,\int_{k_1,k_2}\, (2 \pi)^4\, \delta(k_1 + k_2 - p) \,
        \mathrm{e}^{\, \frac{\mathrm{i}}{2} \,k_1 \cdot \theta \,k_2} \
        \mathtt{e}^{k_1} \odot_\star \widetilde{\tte}^{k_2}
        \\[1.3em] \displaystyle
        \mathrm{i} \,\int_{k_1,k_2}\, (2 \pi)^4 \,\delta(k_1 + k_2 - p) \,
        \mathrm{e}^{\, \frac{\mathrm{i}}{2}\, k_1 \cdot \theta\, k_2} \ \bar{\mathtt{e}}^{k_1} 
        \odot_\star \widetilde{\tte}^{k_2}
        \\[1.3em]
        \tfrac{\mathrm{i}}{g}\, p_\mu \ \mathtt{\widetilde{e}}^p
    \end{pmatrix}
   \ ,
\end{equation}
as expected from the $L_\infty$-structure of \cref{sub:braided_qed}.
A short calculation verifies $(Q_\BRST)^2 = 0$.

The classical extended action functional $\mathcal{S}_{\mathrm{cl}} := \mathcal{S}_0 + \mathcal{S}_{\mathrm{int}}\in(\mathrm{Sym}_\star L[2])^0$ is invariant under BRST transformations if
\begin{equation}\label{eq:QbrstS}
    Q_{{\BRST}}\, \mathcal{S}_{\rm cl} =0\ ,
\end{equation}
that is, it is braided gauge invariant. We verify \eqref{eq:QbrstS} explicitly, using the vertices of the action functional from  \eqref{eq:3-vertex diagram} and \eqref{eq:4vertex}, along with the BRST transformations \eqref{eq:QBRSTmomentum}.

Recall that the free part of the action functional \eqref{eq:BVactionext} is given by 
\begin{subequations}\label{variation_action}
\begin{equation}
\begin{split}
    \mathcal{S}_{0} = 
    \frac{1}{2} \,
    \int_{k_1,k_2}\,
    (2 \pi)^4\, \delta(k_1 + k_2) \, &\bigg(
    \big(k_1^2 -m^2\big) \
    \bar{\mathtt{e}}^{k_1} \odot_\star \mathtt{e}^{k_2}  
    +
    \big(k_1^2 -m^2\big) \
    \mathtt{e}^{k_1} \odot_\star \bar{\mathtt{e}}^{k_2} 
    \\
    & \qquad +
    \big(- k_1^{2}\, \eta^{\mu \nu} + k_{1}^{\mu}\, k_{1}^{\nu}\big) \
    \mathtt{v}_\mu\, \mathtt{e}^{k_1} \odot_\star \mathtt{v}_{\nu}\, \mathtt{e}^{k_2}
    \bigg) \ .
\end{split}
\end{equation}
Applying the BRST transformations, we find
\begin{equation}
\begin{split}
    Q_{\BRST}\, \mathcal{S}_{0} &= \ii\,
    \int_{k_1,k_2,k_3}\, (2\pi)^4 \,
    \delta(k_1+k_2 + k_3) \,
    \big(k_2^2-m^2\big) \,
    \mathrm{e}^{ \,\frac{\mathrm{i}}{2} \, k_2 \cdot \theta \, k_3} 
    \\
    & \hspace{7cm} \times 
    \mathtt{\widetilde{e}}^{k_1} \odot_\star 
    \left(
    \mathtt{e}^{k_2} \odot_\star \bar{\mathtt{e}}^{k_3}
    - 
    \bar{\mathtt{e}}^{k_2}
    \odot_\star \mathtt{e}^{k_3} \right)
    \\[4pt]
    &= \ii\,
    \int_{k_1,k_2,k_3}\, (2\pi)^4 \,
    \delta(k_1+k_2 + k_3) \,
    \big(k_2^2-k_3^2\big) \,
    \mathrm{e}^{ \,\frac{\mathrm{i}}{2} \, k_2 \cdot \theta \, k_3} \
    \mathtt{\widetilde{e}}^{k_1} \odot_\star 
    \mathtt{e}^{k_2} \odot_\star \bar{\mathtt{e}}^{k_3}
    \ ,
\end{split}
\end{equation}
where the contributions from the last entry of \eqref{eq:QBRSTmomentum} vanish identically because of the projector $k^2\,\tilde{\mathsf{\Pi}}^{\mu\nu}(k)=k^2\,\eta^{\mu\nu}-k^\mu\,k^\nu$ onto transverse momentum.

The cubic variation splits into an order $g^0$ piece, in symmetric degree~3, and an order $g^1$ piece, in symmetric degree~4.
One finds the non-vanishing BRST transformation
\begin{equation}
\begin{split}\label{eq:QS3}
    Q_{\BRST}\, \mathcal{S}^\swthree_{\mathrm{int}}
    &= -\ii\,\int_{k_1,k_2,k_3} \, 
    (2 \pi)^4 \, \delta(k_1 + k_2 + k_3)  \,
    \big(k_2^2 - k_3^2\big) \,
    \mathrm{e}^{ \,\frac{\mathrm{i}}{2} \, k_2 \cdot \theta \, k_3} \
    \mathtt{\widetilde{e}}^{k_1}
    \odot_\star \mathtt{e}^{k_2}
    \odot_\star \bar{\mathtt{e}}^{k_3}
    \\
    & \quad \,
    - \mathrm{i}\, g\,\int_{q_1,q_2} \ \int_{k_1,k_2,k_3}\,
    (2 \pi)^8 \, \delta(k_1 + k_2 +k_3)  \,
    (k_2 - k_3)^\mu \,
    \mathrm{e}^{\,\frac{\mathrm{i}}{2}\,  k_2 \cdot \theta\, k_3+\frac{\mathrm{i}}{2}\, q_1 \cdot \theta\, q_2} \ \mathtt{v}_\mu\, \mathtt{e}^{k_1} 
    \\
    & \hspace{5cm}
    \odot_\star
    \left(
    \delta(q_1+q_2 - k_2) \
    \mathtt{e}^{q_1} \odot_\star \mathtt{\widetilde{e}}^{q_2} \odot_\star \bar{\mathtt{e}}^{k_3} \right. \\ 
    & \hspace{7cm} \left.
    -\, \delta(q_1 + q_2 - k_3) \
    \mathtt{e}^{k_2} \odot_\star \bar{\mathtt{e}}^{q_1} \odot_\star {\mathtt{\widetilde{e}}}^{q_2} 
    \right)\ .
\end{split}
\end{equation}

The variation of the quartic part of the action functional is found to be
\begin{equation}
\begin{split}
    Q_{\BRST}\,\mathcal{S}^\swfour_{\mathrm{int}} &= \ii\,g\,
    \int_{k_1,k_2,k_3,k_4} \, (2\pi)^4\,\delta(k_1+k_2+k_3+k_4) \\
    &\hspace{5cm} \times \left(
    k_1^\mu\, \mathtt{\widetilde{e}}^{k_1}
    \odot_\star 
    \mathtt{e}^{k_2} 
    \odot_\star 
    \bar{\mathtt{e}}^{k_3} 
    \odot_\star 
    \mathtt{v}_{\mu}\,\mathtt{e}^{k_4} \right. \\
    &\hspace{7cm} \left.
    +\,
    \mathtt{v}_{\mu}\,\mathtt{e}^{k_1} 
    \odot_\star 
    \mathtt{e}^{k_2} 
    \odot_\star 
    \bar{\mathtt{e}}^{k_3} 
    \odot_\star 
    k_4^\mu\,\mathtt{\widetilde{e}}^{k_4} 
    \right)\ .
\end{split}
\end{equation}
The term of symmetric degree~5 contained in $Q_{\textrm{\tiny BRST}}\, \mathcal{S}^\swfour_{\mathrm{int}}$, of order $g^2$, vanishes after a simple relabelling of momenta.

Summing the contributions in \eqref{variation_action}, we verify $Q_\BRST\, \mathcal{S}_{\rm cl} = 0$.
\end{subequations}

\subsection{Extended Electric Current} 
\label{sub:conserved_current}

Braided gauge invariance of the Maurer-Cartan functional \eqref{eq:MCbSQED} is equivalent to the weak conservation of the braided electric current \eqref{eq:braided_current}. We show that one can associate to the braided gauge invariant extended action functional $\mathcal{S}_{\rm cl}$ a symmetrised current $\mathcal{J}^\star \in (\mathrm{Sym}_\star L[1]^*)^0$, whose correlators are also conserved.
The extended electric current $\CJ^\star$ is the lift of the original braided Noether current $J^\star \in \Omega^1(\mathbbm{R}^{1,3})$ to the symmetric algebra $\mathrm{Sym}_\star L[1]^*$. 
Like the interaction terms of the extended BV functional, its form is much simpler than that of the original description, due to the braided symmetrisation.

The symmetric current 
\begin{align}
\mathcal{J}^\star = \mathcal{J}_2^\star + \mathcal{J}^\star_3 \ \in \  (\mathrm{Sym}_\star L[2])^{0}
\end{align}
is given in momentum space by the sum of the two contributions
\begin{align}\label{eq:current1}
    \mathcal{J}_2^{\star \mu}(p_1,p_2) 
    &=  
    \mathrm{e}^{\,\frac{\mathrm{i}}{2}\, p_1 \cdot \theta\, p_2} \ 
    \left(
    p_1^\mu\,
    \mathtt{e}^{p_1}
    \odot_\star \bar{\mathtt{e}}^{p_2}
    - 
    {\mathtt{e}}^{p_1}
    \odot_\star p_2^\mu\,
    \bar{\mathtt{e}}^{p_2}
    \right) \ , \\[4pt]
    \label{eq:current2}
    \mathcal{J}^{\star \mu}_3(p_1,p_2,p_3) 
    &= - 2\, g\, \eta^{\mu \nu}\,
    \mathrm{e}^{\,\frac{\mathrm{i}}{2}\, \sum\limits_{i<j}\, p_i \cdot \theta \, p_j} \ 
    \mathtt{e}^{p_1} 
    \odot_\star 
    \bar{\mathtt{e}}^{p_2} 
    \odot_\star 
    \mathtt{v}_{\nu}\,\mathtt{e}^{p_3}\ .
\end{align}
We shall demonstrate that the extended current $\CJ^\star$ is dual to the classical braided Noether current $J^\star \in \Omega^1(\mathbbm{R}^{1,3})$, written in its non-covariant form \eqref{eq:non-covariantcurrent}. 

For this, we must first introduce a pairing between the respective symmetric algebras.
We define the pairing between braided symmetric elements of degree $n$
\begin{equation}
\begin{split}
    \langle-,-\rangle_\star:\big(\mathrm{Sym}_\star (L[1])^*\big)^n \otimes \big(\mathrm{Sym}_\star L[1]\big)^n\longrightarrow \mathbbm{R}
    \end{split}
    \end{equation}
    as
    \begin{equation}
    \begin{split}
   \langle \xi^{1} \odot_\star \cdots \odot_\star \xi^{n}\,,\, \tau_{1} \odot_\star \cdots \odot_\star \tau_{n}\rangle_\star :=
    \sum_{\sigma \in S_n}\, \pm \, \langle \xi^{1},\tau_{\sigma_\RR(n)}\rangle_\star
    \cdots \langle\xi^{n},\tau_{{\sigma_\RR(1)}}\rangle_\star \ ,
\end{split}
\end{equation}
for elements $\xi^{i} \in L[1]^*$ and $\tau_{i} \in L[1]$, with $i=1,\dots,n$. Here $S_n$ is the symmetric group of degree $n$, and the permutation $\sigma_\RR$ acts by braided graded transposition.
For example, on quadratic symmetric powers the pairing is
\begin{equation}
\begin{split}
    \langle\xi^1 \odot_\star \xi^2 \,,\, \tau_1 \odot_\star \tau_2\rangle_\star
    &= \langle\xi^1,\tau_2\rangle_\star \; \langle\xi^2,\tau_1\rangle_\star
    \pm
    \langle\xi^1,\mathsf{R}_\alpha (\tau_1)\rangle_\star\;\langle\mathsf{R}^\alpha (\xi^2),\tau_2\rangle_\star \ .
\end{split}\,
\end{equation}

Consider the superfield $\mathcal{A} \in L^1$ of \eqref{VSpace}.
By shifting the degree using the suspension isomorphism $s: L \longrightarrow L[1]$, this defines a degree 0 element which has momentum space decomposition given by
\begin{equation}
    s \mathcal{A} = 
    \begin{pmatrix}
        \displaystyle \int_k\, \widehat{\phi}(k) \ \mathtt{e}_k
        \\[1.3em] \displaystyle \int_k\, \widehat{\bar\phi}(k) \ \bar{\mathtt{e}}_k
        \\[1.3em] \ \displaystyle \int_k\,  \widehat{A}{}_\mu(k) \ \mathtt{v}^\mu\, \mathtt{e}_k
    \end{pmatrix} \  \in \ L[1]^0 \ .
\end{equation}
The degree is carried by basis elements $\mathtt{e}_k$, $\mathtt{\bar e}_k$ and  $\mathtt{v}^\mu\, \mathtt{e}_k$ of degree 0, which are dual to their respective antifield bases as explained in \cref{sub:bv_action}.

Then the pairing of a quadratic basis antifield $\mathtt{e}^{p_1} \odot_\star \mathtt{\bar e}^{p_2}$ with symmetrised superfields is
\begin{equation}
\begin{split}
    \langle\mathtt{e}^{p_1} \odot_\star \bar{\mathtt{e}}^{p_2}\,,\,
    \tfrac{1}{2}\, s \mathcal{A} \odot_\star s\mathcal{A}\rangle_\star
   & = \frac{1}{2}\, \int_{k_1,k_2} \,\Big(
    \langle\mathtt{e}^{p_1},{\mathtt{e}}_{k_2}\rangle_\star\,
    \widehat{\phi}(k_2)
    \
    \langle\bar{\mathtt{e}}^{p_2}, \bar{\mathtt{e}}_{k_1}\rangle_\star\,
    \widehat{\bar \phi}(k_1)\\
    & \hspace{2.5cm}
    +
   \langle \mathtt{e}^{p_1}, \mathsf{R}_\alpha ({\mathtt{e}}_{k_1})\rangle_\star\,
    \widehat{\phi}(k_1)
    \
    \langle\mathsf{R}^\alpha (\bar{\mathtt{e}}^{p_2}),\bar{\mathtt{e}}_{k_2}\rangle_\star\,
    \widehat{\bar \phi}(k_2)\Big) \ .
\end{split}
\end{equation}
Using the dual pairings $\langle\mathtt{e}^{p},\mathtt{e}_k\rangle_\star = (2 \pi)^4\, \delta(p-k) = \langle \bar{\mathtt{e}}^{p},\bar{\mathtt{e}}_k\rangle_\star$, and multiplying with the phase factor entering the quadratic part of the extended current \eqref{eq:current1}, we find 
\begin{equation}
    \mathrm{e}^{\,\frac{\mathrm{i}}{2}\, p_1 \cdot \theta\, p_2} \, \langle\mathtt{e}^{p_1} \odot_\star \bar{\mathtt{e}}^{p_2}\,,\,
    \tfrac{1}{2}\, s \mathcal{A} \odot_\star s\mathcal{A}\rangle_\star
    = 
    \tfrac{1}{2}\,
    \big(
    \mathrm{e}^{\,\frac{\mathrm{i}}{2}\, p_1 \cdot \theta\, p_2}\,
    \widehat{\phi}(p_1)\,
    \widehat{\bar\phi}(p_2)
    + 
    \mathrm{e}^{-\frac{\mathrm{i}}{2}\, p_1 \cdot \theta\, p_2}\,
    \widehat{\phi}(p_1)\,
    \widehat{\bar\phi}(p_2)
    \big) \ ,
\end{equation}
which we recognise as the momentum space coordinate function of $\frac12\,(\phi \star \bar \phi + \bar \phi \star \phi)$. 
Multiplying by the appropriate momenta, we recover the quadratic part of \eqref{eq:non-covariantcurrent}.

Similarly, with the dual pairing on photon states $\langle\mathtt{v}_\nu\, \tte^k, \mathtt{v}^\mu\, \mathtt{e}_p\rangle_\star = (2 \pi)^4 \,  \delta_\mu^\nu \, \delta(k - p)$ we find that the pairing with symmetric degree 3 symmetrised superfields of the cubic extended current \eqref{eq:current2} is
\begin{equation}
\begin{split}
&   \mathrm{e}^{\,\frac{\mathrm{i}}{2}\, \sum\limits_{i<j}\, p_i \cdot \theta\, p_j}\,
    \langle\mathtt{e}^{p_1} 
    \odot_\star 
    \bar{\mathtt{e}}^{p_2} 
    \odot_\star 
    \mathtt{v}_{\mu}\,\mathtt{e}^{p_3}\,,\,
    \tfrac{1}{6}\,s \mathcal{A} \odot_\star s \mathcal{A} \odot_\star s \mathcal{A}\rangle_\star \\[4pt]
    & \hspace{6cm} 
    =\frac{1}{6}\,
    \sum_{\sigma \in S_3}\, \mathrm{e}^{\,\frac{\mathrm{i}}{2}\, \sum\limits_{i < j}\, p_{\sigma(i)} \cdot \theta\, p_{\sigma(j)}} \,
    \widehat{\phi}(p_1)\,
    \widehat{\bar\phi}(p_2)\,
    \widehat{A}_\mu(p_3) \ .
\end{split}
\end{equation}
We recognise this as the momentum space coordinate function of the cubic part of \eqref{eq:non-covariantcurrent}.

Therefore the extended electric current is related to the classical Noether current if we pair each term $\mathcal{J}_{2}^{\star}$ and $ \mathcal{J}_{3}^{\star}$ with the appropriate symmetric power of the superfield $s \mathcal{A}$.
If we introduce a symmetric exponential from the superfield $s \mathcal{A} \in L[1]^0$ as
\begin{equation}
    \mathrm{e}^{s \mathcal{A}}_0 := \sum_{n\geq1}\,\frac1{n!}\,(s \mathcal{A})^{\odot_\star n}  \ ,
\end{equation}
understood as living in a suitable completion of $(\mathrm{Sym}_\star L[1])^0$, then the classical Noether current is recovered by composing the pairing with the projection onto the subspace $\mathbbm{R}$.
We denote this projection by $\mathrm{pr}_\FR$, which restricts the pairing between components of like degrees. Then
\begin{equation}
    \mathrm{pr}_{\FR} \, \langle \mathcal{J}^\star,\mathrm{e}_0^{s \mathcal{A}}\rangle_\star = J^\star \ ,
\end{equation}
where $J^\star$ is the braided Noether current \eqref{eq:braided_current}.

This shows that the extended current is the lift to the symmetric algebra of the classical electric current.
The braided gauge invariance of the weakly conserved current \eqref{eq:braided_current} is also lifted to the invariance of the conserved current $\CJ^\star$ in the symmetric algebra, which can be expressed simply as
\begin{equation}\label{eq:gaugeinvcurrent}
Q_\BRST\,\mathcal{J}^{\star \mu} = 0\ .
\end{equation}

Similarly to the classical Noether current, the braided extended current appears in the BRST transformation of the scalar fields alone, which is generated by the operator $Q_\BRST^\swthree:=\{ \mathcal{S}^\swthree_\BRST, - \}_\star$.
Indeed, by isolating such contributions from \eqref{variation_action} we find
\begin{equation} \label{eq:matterBRSTaction}
\begin{split}
Q^\swthree_\BRST\,\mathcal{S}_{\rm cl} &= 
    \int_{k_1,k_2,k_3}\, (2\pi)^4 \, \delta(k_{1} + k_2 + k_3)\, (-\mathrm{i}\, k_{1\, \mu}) \ \mathtt{\widetilde{e}}^{k_1} \odot_\star \mathcal{J}_2^\star{}^\mu(k_2, k_3) \\
    & \quad \, + \int_{k_1,k_2,k_3,k_4}\, (2\pi)^4\, \delta(k_{1} + k_2 + k_3 + k_4)\, (-\mathrm{i}\, k_{1\, \mu}) \ \mathtt{\widetilde{e}}^{k_1} \odot_\star \mathcal{J}_3^\star{}^\mu(k_2, k_3,k_4) \ .
\end{split}
\end{equation}
The remaining BRST transformation of the photon field is generated by the operator $\{ \mathcal{S}^\swtwo_\BRST, - \}_\star$. The decomposition of \eqref{eq:QbrstS} in this way is the extended version of the decomposition of the classical braided Noether identity \eqref{eq:braidednoetherbsqed}, wherein we recognise the operator $-\ii\,k_{1\,\mu}$ as the codifferential $\dd^\dag$ in momentum space after resolving the delta-distributions.

\subsection{Anomalous Ward Identity} 
\label{sub:ward_takahashi_identities}

We show how the Ward identity is an immediate consequence of the homological perturbation lemma. 

\paragraph{Master Identity.}

Consider the strong deformation retract \eqref{eq:sdr_full} defined by the small perturbation $\boldsymbol{\delta}_{\hbar,\mathrm{int}}$ from \eqref{eq:deltahbarint}, the classical free projection $\sP_0$ onto \smash{$\mathbbm{R} \simeq \big(\mathrm{Sym}_\star H^\bullet( L[2])\big)^0$} from \eqref{eq:trivial project}, and the canonical inclusion $\mathsf{I}_0$ of $\mathbbm{R}$ into $\mathrm{Sym}_\star L[2]$. In the remainder of this section we make extensive use of  the key identity
\begin{align}\label{lem: Pell1delta}
\mathsf{P}_{\hbar,\mathrm{int}}' \circ (\boldsymbol\ell^\star_1 + \boldsymbol{\delta}_{\hbar,\mathrm{int}} ) = 0 \ ,
        \end{align}
        where $\mathsf{P}_{\hbar,\mathrm{int}}' := \mathsf{P}_0 + \mathsf{P}_{\hbar,\mathrm{int}}$.
        
To prove \eqref{lem: Pell1delta}, we use the homological perturbation lemma which asserts that the deformed projection $\mathsf{{P}}_{\hbar,\mathrm{int}}'$ remains a cochain map:
    \begin{equation}
        \mathsf{P}_{\hbar,\mathrm{int}}' \circ (\boldsymbol\ell^\star_1 + \boldsymbol{\delta}_{\hbar,\mathrm{int}} )
        = \boldsymbol{{\delta}}_{\hbar,\mathrm{int}}' \circ  \mathsf{P}_{\hbar,\mathrm{int}}' \ .
    \end{equation}
The homological perturbation lemma also provides explicit formulas for the maps
    \begin{equation}
        \boldsymbol{{\delta}}_{\hbar,\mathrm{int}}' = \mathsf{P}_0 \, \big(\mathrm{id}_{\mathrm{Sym}_\star L[2]} - \boldsymbol{\delta}_{\hbar,\mathrm{int}}\, \mathsf{H}\big)^{-1} \, \boldsymbol{\delta}_{\hbar,\mathrm{int}}\, \mathsf{I}_0
        \qquad \mbox{and} \qquad
        \mathsf{{P}}_{\hbar,\mathrm{int}}' = \mathsf{P}_0 \,\big(\mathrm{id}_{\mathrm{Sym}_\star L[2]} - \boldsymbol{\delta}_{\hbar,\mathrm{int}}\, \mathsf{H}\big)^{-1}  \ .
    \end{equation}
    As $\mathsf{{P}}_{\hbar,\mathrm{int}}'$ is based around the trivial projection $\mathsf{P}_0$ to $\mathbbm{R}$, and since the perturbation $\boldsymbol{\delta}_{\hbar,\mathrm{int}}$ acts trivially on the ground field $\mathbbm{R}\subset\mathrm{Sym}_\star L[2]$, i.e. $\boldsymbol{\delta}_{\hbar,\mathrm{int}}(\mathbbm{1})=0$, it follows that the composition $\boldsymbol{{\delta}}_{\hbar,\mathrm{int}}' \circ \mathsf{{P}}_{\hbar,\mathrm{int}}' = 0$, which implies \eqref{lem: Pell1delta}.
    
Among other things, this identity encompasses the algebraic Schwinger-Dyson equations of~\cite{Okawa:2022sjf,Konosu2023,Konosu:2023rkm,Konosu:2024dpo,Bogdanovic:2024jnf}.
For this, we first notice that the homological perturbation lemma implies \smash{$\mathsf{{P}}_{\hbar,\mathrm{int}}'\circ \mathsf{H} = 0$}, since $\sP_0\circ\sH=0$ and $\mathsf{H}^2 =  0$, whereas \smash{$\mathsf{{P}}_{\hbar,\mathrm{int}}'\circ \mathsf{I}_0 =\sP_0\circ\mathsf{I}_0 = \mathrm{id}_{\mathrm{Sym}_\star H^\bullet(L[2])}$} using $\mathsf{H}\circ \mathsf{I}_0 = 0$.
Consider the precomposition of \eqref{lem: Pell1delta} with the free contracting homotopy $\mathsf{H}$. This implies $\mathsf{{P}}_{{\hbar,\mathrm{int}}}' \circ (\boldsymbol\ell^\star_1 + \boldsymbol{\delta}_{\hbar,\mathrm{int}} ) \circ \mathsf{H} = 0$, and therefore
\begin{equation}
\mathsf{{P}}_{{\hbar,\mathrm{int}}}' \circ \boldsymbol\ell^\star_1 \circ \mathsf{H} =- \mathsf{{P}}_{{\hbar,\mathrm{int}}}' \circ  \boldsymbol{\delta}_{\hbar,\mathrm{int}}  \circ \mathsf{H} \ .
\end{equation}
Using the homotopy equivalence equation \eqref{eq:HEdata} of the free theory this implies
\begin{align}
    \mathsf{{P}}_{\hbar,\mathrm{int}}' \circ\big(
    - \mathsf{H}\circ \boldsymbol\ell_1^\star -\mathrm{id}_{\mathrm{Sym}_\star L[2]} + \mathsf{I}_0\circ \mathsf{P_0} \big)
    & = -\mathsf{{P}}_{\hbar,\mathrm{int}}'\circ \boldsymbol{\delta}_{\hbar,\mathrm{int}}\circ \mathsf{H} \ ,
\end{align}
and therefore
\begin{equation}\label{eq:SDE}
    \mathsf{{P}}_{\hbar,\mathrm{int}}' = \mathsf{P}_0 + \mathsf{{P}}_{{\hbar,\mathrm{int}}}' \circ  \boldsymbol{\delta}_{\hbar,\mathrm{int}}  \circ \mathsf{H} \ .
\end{equation}
We recognise \eqref{eq:SDE} as the algebraic Schwinger-Dyson equation, as derived in \cite{Bogdanovic:2024jnf}.

We study correlation functions by using this identity applied to insertions of fields in $L[2]^{-1}$.
Then by \eqref{lem: Pell1delta}, we get the identity
\begin{equation}\label{eq:masterid}
    \mathsf{{P}}_{\hbar, \mathrm{int}}' \circ ( \{\mathcal{S}_{\mathrm{cl}}, -\}_\star + \mathrm{i}\, \hbar\, \mathsf{\Delta}_{\textrm{\tiny BV}}) = 0 \ .
\end{equation}
With $\mathsf{P}_0$ projecting all elements with non-zero symmetric degree to $0$, the correlation functions are unchanged if one considers $\mathsf{P}_{\hbar,\mathrm{int}}$ as previously, or the full projection $\mathsf{{P}}_{\hbar,\mathrm{int}}'$.
In short, elements in the image of the quantum differential
\begin{equation}
    Q_{\hbar,\mathrm{int}} = \boldsymbol\ell_1^\star + \boldsymbol{\delta}_{\hbar,\mathrm{int}} = \{ \mathcal{S}_{\mathrm{cl}}, -\}_\star + \mathrm{i}\, \hbar\, \mathsf{\Delta}_{\textrm{\tiny BV}}
\end{equation}
have vanishing correlation functions coming from application of $\mathsf{P}_{\hbar,\mathrm{int}}$.

\paragraph{Consistency of Braided Gauge Symmetries.} 
\label{par:ward_identity}

Recall that a gauge symmetry in a conventional quantum field theory is non-anomalous if the correlation functions are all invariant under gauge transformations implemented by the conserved charge operator and the vacuum expectation values of the associated Noether current operator are conserved to all orders in perturbation theory. In braided scalar QED, the action of braided gauge symmetries does not follow the traditional properties~\cite{DimitrijevicCiric:2021jea,Giotopoulos:2021ieg}. Hence the requirement that it be anomaly-free is non-trivial and needs to be checked thoroughly.

In braided scalar QED, the current $\mathcal{J}^\star \in (\mathrm{Sym}_\star L[2])^0$ is non-anomalous if its correlators are coclosed off-shell, which translates to checking 
\begin{equation}\label{eq:divfreecurrent_SYM}
  \partial_\mu\,\Big(\, \sum_{k=1}^\infty\, \mathsf{P}_0\, (\boldsymbol{\delta}_{\hbar,{\rm int}}\, \mathsf{H}\big)^k\,(\mathcal{J}^{\star \mu})\Big) = 0\ ,
\end{equation}
for the usual deformation retract \eqref{eq:sdr_full}. We verify this to all orders in homological perturbation theory.

For this, we apply the identity \eqref{eq:masterid} to a functional of degree $-1$ living in the image of the BRST operator.
To obtain the Ward identity, we consider the scalar field transformation generated by $ Q_\BRST^\swthree $ on ghost fields,
\begin{equation}
    Q_\BRST ^\swthree\, \mathtt{\widetilde{e}}_k = \{ \mathcal{S}_\BRST^\swthree, \mathtt{\widetilde{e}}_k \}_\star \ \in \ L[2]^{-1} \ ,
\end{equation}
where $\mathtt{\widetilde{e}}_k \in (L[2])^{-2}$ is the momentum basis for the ghost fields.
Recall the duality between ghost field/antifield pairs in momentum space given by $\{ \mathtt{\widetilde{e}}^{k_1}, \mathtt{\widetilde{e}}_{k_2} \}_\star = (2 \pi)^4\,\delta(k_1 - k_2)$, with dual basis $\mathtt{\widetilde{e}}^k \in L[2]^1$.
Applying the identity \eqref{eq:masterid} to $Q_\BRST^\swthree\, \mathtt{\widetilde{e}}_k$ yields
\begin{equation}
\begin{split}
    \mathsf{{P}}_{\hbar, \mathrm{int}}' \circ ( \{\mathcal{S}_{\mathrm{cl}}, -\}_\star + \mathrm{i}\, \hbar\, \mathsf{\Delta}_{\textrm{\tiny BV}})
    \circ \{ \mathcal{S}_\BRST^\swthree, \mathtt{\widetilde{e}}_k \}_\star  &= 0 \ .
\end{split}
\end{equation}

We use the braided Jacobi identity to get
\begin{equation}
\begin{split}\label{eq:ScSbrstC}
\hspace{-0.5cm}
    \big\{ \mathcal{S}_{\mathrm{cl}}\,,\, \{ \mathcal{S}_\BRST^\swthree, \mathtt{\widetilde{e}}_k \}_\star \big\}_\star
    &= 
    -
    \big\{ \mathsf{R}_\alpha( \mathcal{S}_{\BRST}^\swthree)\,,\, \{ \mathsf{R}^\alpha(\mathcal{S}_{\mathrm{cl}}), \mathtt{\widetilde{e}}_k\}_\star \big\}_\star \\
    & \hspace{1cm}
    -
    \big\{ \mathsf{R}_{\alpha}\, \mathsf{R}_{\beta}(\mathtt{\widetilde{e}}_k) \,,\, \{ \mathsf{R}^{\alpha}(\mathcal{S}_{\mathrm{cl}}), \mathsf{R}^{\beta}(\mathcal{S}_\BRST^\swthree) \}_\star \big\}_\star
    =- \{ Q_\BRST^\swthree\, \mathcal{S}_\mathrm{cl}, \mathtt{\widetilde{e}}_k \}_\star\ .
\end{split}
\end{equation}
For the first equality we used the fact that $\mathcal{S}_\BRST^\swthree$ and $\mathcal{S}_{\mathrm{cl}}$ have degree $0$, while $\widetilde{\tte}_k$ has degree $-2$, so the grading factor is always 1 in the braided Jacobi identity of \cite{Nguyen:2021rsa}.
In the second equality we used the fact that the actions of $R$-matrices on $\mathcal{S}_\mathrm{cl}$ and $\mathcal{S}_\BRST^\swthree$ trivialise by translation invariance, and that the classical action functional does not contain ghost antifields, so $\left\{ \mathcal{S}_{\mathrm{cl}}, \mathtt{\widetilde{e}}_k \right\}_\star = 0$ and the first term vanishes.
Finally, the pairing with the classical action functional can be written as $\{\mathcal{S}_\mathrm{cl}, \mathcal{S}_{\BRST}^\swthree\}_\star = \{\mathcal{S}_{\BRST}^\swthree, \mathcal{S}_\mathrm{cl}\}_\star = Q_\BRST^\swthree\, \mathcal{S}_\mathrm{cl}$.

The BV Laplacian is a strict graded derivation of the antibracket, hence
\begin{equation}
    \mathsf{\Delta}_{\textrm{\tiny BV}} \,\{ \mathcal{S}_\BRST^\swthree, \mathtt{\widetilde{e}}_k \}_\star
    = \{ \mathsf{\Delta}_{\textrm{\tiny BV}}\, \mathcal{S}_\BRST^\swthree, \mathtt{\widetilde{e}}_k \}_\star
    + \{ \mathcal{S}_{\BRST}^\swthree, \mathsf{\Delta}_{\textrm{\tiny BV}}\, \mathtt{\widetilde{e}}_k \}_\star\ \ .
\end{equation} 
The second term vanishes since $\mathsf{\Delta}_{\textrm{\tiny BV}}\, \mathtt{\widetilde{e}}_k = 0$.

Putting everything together, we have derived the \textit{anomalous} Ward identity
\begin{equation}\label{eq:anomalousWard}
    \mathsf{{P}}_{\hbar,\mathrm{int}}'\, \{ Q_\BRST^\swthree\, \mathcal{S}_{\mathrm{cl}}, \mathtt{\widetilde{e}}_k \}_\star
    = \mathrm{i}\, \hbar\, \mathsf{{P}}_{\hbar,\mathrm{int}}' \,\{ \mathsf{\Delta}_{\textrm{\tiny BV}}\, \mathcal{S}_{\BRST}^\swthree, \mathtt{\widetilde{e}}_k \}_\star\ .
\end{equation}
The first term is related to the codifferential of the extended electric current derived in \eqref{eq:matterBRSTaction} through the pairing with the ghost field $\mathtt{\widetilde{e}}_k$:
\begin{equation}\label{eq:currentwithoutghost}
\begin{split}
    &\{ Q_\BRST^\swthree\, \mathcal{S}_{\mathrm{cl}}, \mathtt{\widetilde{e}}_k \}_\star = 
    \int_{k_1, k_2}\, (2\pi)^4\, \delta(k + k_1 + k_2)\, (-\mathrm{i}\, k_{\mu})\, \mathcal{J}^{\star \mu}_2(k_1,k_2)\\[5pt]
    &\hspace{4cm}
    +
    \int_{k_1,k_2, k_3}\, (2 \pi)^4\,\delta(k + k_1+k_2 + k_3)\, (-\mathrm{i}\, k_{\mu})\, \mathcal{J}^{\star \mu}_3(k_1,k_2,k_3) \ .
\end{split}
\end{equation}

The resulting identity links the off-shell conservation of a current to the BV Laplacian of the BRST functional.
Braided scalar QED is non-anomalous since $\mathsf{\Delta}_{\textrm{\tiny BV}}\,\mathcal{S}_{\BRST}^\swthree = 0$, which can be seen using the decomposition of the ghost three-point vertex derived in \eqref{eq:S3BRST}:
The contributions from the two vertices cancel each other out.
This result enforces the current conservation in momentum space representation
\begin{equation}
\begin{split}\label{eq:currentcorrelation conservation}    
    &\int_{k_1, k_2}\, (2\pi)^4\, \delta(k + k_1 + k_2)\, (-\mathrm{i}\, k_{\mu})\, \mathsf{{P}}_{\hbar,\mathrm{int}}'\,\big(
    \mathcal{J}^{\star \mu}_2(k_1,k_2)\big) \\
&\hspace{2cm}
    +
    \int_{k_1,k_2, k_3}\, (2 \pi)^4\,\delta(k + k_1+k_2 + k_3)\, (-\mathrm{i}\, k_{\mu})\, \mathsf{{P}}_{\hbar,\mathrm{int}}'\,\big(\mathcal{J}^{\star \mu}_3(k_1,k_2,k_3)\big) = 0
\end{split}
\end{equation}
 at all orders of perturbation theory.
Thus the braided gauge symmetry is non-anomalous. The vanishing of the left-hand side of \eqref{eq:currentcorrelation conservation} is a trivial consequence of momentum conservation in correlation functions, which sets the total momentum of each current component to zero.

\subsection{Anomalous Ward-Takahashi Identities}

We extend the result of \cref{sub:ward_takahashi_identities} to correlation functions of the extended electric current together with field insertions.
Let $\mathcal{O} \in (L[2]^0)^{\odot_\star n}$ be a symmetric polynomial of antifields only. 
Then $\mathcal{O}$ trivially satisfies $\mathsf{\Delta}_{\textrm{\tiny BV}}\, \mathcal{O} = 0$ and $\left\{ \mathtt{\widetilde{e}}_k\,,\, \mathcal{O} \right\}_\star=0$ for a ghost basis field $\mathtt{\widetilde{e}}_k \in L[2]^{-2}$.
We also have $\left\{ \mathcal{S}_\mathrm{cl}, \mathcal{O} \right\}_\star = 0$, since the classical action functional acts non-trivially only on fields in $L[2]^{-1}$ and $\mathcal{O}$ is an antifield.

We apply the identity \eqref{eq:masterid} to functionals of the form
\begin{equation}\label{eq:QcO}
    Q_\BRST ^\swthree \, \mathtt{\widetilde{e}}_k\, \odot_\star \mathcal{O} 
    = \{ \mathcal{S}_\BRST^{\swthree}, \mathtt{\widetilde{e}}_k \}_\star \odot_\star \mathcal{O} \ \in \ (\mathrm{Sym}_\star L[2])^{-1}\ .
\end{equation}
We first need to evaluate the image of this term under the quantum differential, which is given by
\begin{equation}
    \big(\left\{ \mathcal{S}_\mathrm{cl}, - \right\}_\star + \mathrm{i}\, \hbar\, \mathsf{\Delta}_{\textrm{\tiny BV}}\big) \circ (Q_\BRST^\swthree\, \mathtt{\widetilde{e}}_k\, \odot_\star \mathcal{O})\ .
\end{equation}
We split the calculation into two terms.

The antibracket is a braided graded derivation over $\odot_\star$, where we use the grading factors of \cite{Nguyen:2021rsa} with $\mathcal{S}_\mathrm{cl}$, $\mathcal{S}_\BRST^\swthree$ and $\mathcal{O}$ of degree $0$, while $Q_\BRST^\swthree\, \mathtt{\widetilde{e}}_k$ has degree $-1$.
We consider the term
\begin{equation}
\begin{split}
\hspace{-1cm}
    \{ \mathcal{S}_{\mathrm{cl}}, Q_\BRST^\swthree\, \mathtt{\widetilde{e}}_k \odot_\star \mathcal{O} \}_\star
    &= \{ \mathcal{S}_{\mathrm{cl}}, Q_\BRST^\swthree\, \mathtt{\widetilde{e}}_k \} \odot_\star \mathcal{O}
    - 
    \mathsf{R}_\alpha (Q_\BRST^\swthree\, \mathtt{\widetilde{e}}_k) \odot_\star \big\{ \mathsf{R}^\alpha (\mathcal{S}_{\mathrm{cl}}), \mathcal{O} \big\}_\star
    \\[4pt]
    &= -\{ Q_\BRST^\swthree\, \mathcal{S}_{\mathrm{cl}}, \mathtt{\widetilde{e}}_k \}_\star \odot_\star \mathcal{O}\ .
\end{split}
\end{equation}
In the second equality we used the fact that $\left\{ \mathcal{S}_{\mathrm{cl}}, - \right\}_\star$ acts trivially on braided symmetric tensor products of antifields, together with \eqref{eq:ScSbrstC}.

For the next term, we recall that the BV Laplacian is a strict graded derivation over $\odot_\star$ up to an antibracket contribution.
We then have
\begin{equation}
\begin{split}
\mathsf{\Delta}_{\textrm{\tiny BV}}\big(\big\{ \mathcal{S}_{\BRST}^\swthree, \mathtt{\widetilde{e}}_k \}_\star \odot_\star \mathcal{O}\big)
    &= \mathsf{\Delta}_{\textrm{\tiny BV}}\, \{ \mathcal{S}_{\BRST}^\swthree, \mathtt{\widetilde{e}}_k \}_\star  \odot_\star \mathcal{O} 
    - \{ \mathcal{S}_{\BRST}^\swthree,\mathtt{\widetilde{e}}_k \}_\star \odot_\star \mathsf{\Delta}_{\textrm{\tiny BV}}\, \mathcal{O} \\
    & \hspace{2cm}
    + \big\{ \{ \mathcal{S}_\BRST^\swthree, \mathtt{\widetilde{e}}_k \}_\star , \mathcal{O} \big\}_\star \\[4pt]
&    = \{ \mathsf{\Delta}_{\textrm{\tiny BV}}\, \mathcal{S}_\BRST^\swthree, \mathtt{\widetilde{e}}_k \}_\star \odot_\star \mathcal{O}
    + \{ Q_\BRST^\swthree\, \mathtt{\widetilde{e}}_k, \mathcal{O} \}_\star\ .
\end{split}
\end{equation}
In the second equality we used $\mathsf{\Delta}_{\textrm{\tiny BV}}\, \mathcal{O} = 0 = \mathsf{\Delta}_{\textrm{\tiny BV}}\, \mathtt{\widetilde{e}}_k$.
The final term is evaluated using the compatibility of the BRST differential with the antibracket, which follows from the Jacobi identity:
\begin{equation}
\begin{split}
    Q_\BRST^\swthree \, \{ \mathtt{\widetilde{e}}_k, \mathcal{O} \}_\star 
    &= \{ Q_\BRST^\swthree\, \mathtt{\widetilde{e}}_k, \mathcal{O} \}_\star
    + \{ \mathtt{\widetilde{e}}_k, Q_\BRST^\swthree\, \mathcal{O} \} \ .
\end{split}
\end{equation}
Since $\left\{ \mathtt{\widetilde{e}}_k, \mathcal{O} \right\}_\star = 0$, this implies $\{ Q_\BRST^\swthree\, \mathtt{\widetilde{e}}_k, \mathcal{O} \}_\star = - \{ \mathtt{\widetilde{e}}_k, Q_\BRST^\swthree\, \mathcal{O} \}_\star$.

Altogether, applying the identity \eqref{eq:masterid} to functionals of the form \eqref{eq:QcO} implies
\begin{equation}\label{eq:Ward-Takahashi identity}
    \mathsf{{P}}_{\hbar,\mathrm{int}}' \big(\{ Q_\BRST^\swthree\, \mathcal{S}_{\mathrm{cl}}, \mathtt{\widetilde{e}}_k \}_\star \odot_\star \mathcal{O}\big) 
    + \mathrm{i}\, \hbar\, \mathsf{{P}}_{\hbar,\mathrm{int}}'
    \big(\{ \mathtt{\widetilde{e}}_k, Q_\BRST^\swthree\, \mathcal{O} \}_\star\big)
    =  \mathrm{i}\, \hbar\, \mathsf{{P}}_{\hbar,\mathrm{int}}'\big(
    \{ \mathsf{\Delta}_{\textrm{\tiny BV}}\, \mathcal{S}_{\BRST}^\swthree, \mathtt{\widetilde{e}}_k \}_\star \odot_\star \mathcal{O}\big)\ .
\end{equation}
This is the anomalous Ward-Takahashi identity.
In the present case, the anomaly vanishes since $\mathsf{\Delta}_{\textrm{\tiny BV}}\, \mathcal{S}_\BRST^\swthree = 0$, and we obtain in momentum space
\begin{equation}\label{eq:WTInonanomalous}
    \mathsf{{P}}_{\hbar,\mathrm{int}}' \big(\{ Q_\BRST^\swthree\, \mathcal{S}_{\mathrm{cl}}, \mathtt{\widetilde{e}}_k \}_\star \odot_\star \mathcal{O}\big) 
    =-
    \mathrm{i}\, \hbar\, \mathsf{{P}}_{\hbar,\mathrm{int}}'
    \big(\{ \mathtt{\widetilde{e}}_k, Q_\BRST^\swthree\, \mathcal{O} \}_\star\big)
     \ ,
\end{equation}
where the left-hand side is related to the extended electric current through \eqref{eq:currentwithoutghost}. We consider various explicit examples below.

\paragraph*{Improper Three-Vertex Function.}

In textbook quantum field theory, an amputated correlator obtained by applying the Maxwell operator to an external on-shell photon leg of a correlation function using the LSZ reduction formula is called an \textit{improper} vertex function~\cite{Srednicki}. 
    Our main example is the derivation of the braided Ward-Takahashi identity for the improper scalar-scalar-photon vertex, sometimes called the electromagnetic vertex.
    
Following \cite{Srednicki}, a vertex amputated of a single on-shell photon leg may be calculated by replacing the Maxwell operator $(\eta^{\mu \nu}\,\square - \partial^\mu\, \partial^\nu)\, A_\nu$ in the LSZ formula with the associated Noether current $g\,J^{\mu}$ inside the correlation function, since $\mathcal{F}^{ \mu}_A = \square\, A^\mu - \partial^\mu\, \partial_\nu\, A^\nu - g\, J^{ \mu} = 0$ is the equation of motion for an on-shell photon field $A\in L[2]^{-1}$ in the unextended theory. 
In our framework of braided homological perturbation theory, we lift the equation of motion $\mathcal{F}^{\star \mu}_A = \ell_1^\star( A)^\mu  - g\, J^{\star \mu} = 0$ to the symmetric algebra using the procedure discussed in \cref{sub:conserved_current}. This shows how correlation functions containing the symmetric current $g\,\mathcal{J}^{\star \mu}$ from \eqref{eq:current1} and \eqref{eq:current2} can be interpreted as improper vertices.

We now show that the identity \eqref{eq:WTInonanomalous} is exactly the relation required to derive the braided deformation of the commutative Ward identity for the electromagnetic vertex~\cite{Srednicki}.
    Consider the insertion of the scalar pair $\mathcal{O} = \mathtt{e}^{p_1} \odot_\star \mathtt{\bar e}^{p_2}$ in the braided Ward-Takahashi identity giving
    \begin{equation}\label{eq:WTI_improper3vertex}
    \mathsf{{P}}'_{\hbar,\mathrm{int}}\big( \{ Q_\BRST^\swthree\, \mathcal{S}_{\mathrm{cl}}, \mathtt{\widetilde{e}}_k \}_\star \odot_\star \mathtt{e}^{p_1} \odot_\star \mathtt{\bar e}^{p_2}\big) 
    = - \mathrm{i}\, \hbar\, \mathsf{{P}}'_{\hbar,\mathrm{int}} \, \big\{ \mathtt{\widetilde{e}}_k, Q_\BRST^\swthree\, (\mathtt{e}^{p_1} \odot_\star \mathtt{\bar e}^{p_2}) \big\}_\star \ .
    \end{equation}
    Let us first study the left-hand side of \eqref{eq:WTI_improper3vertex}.
    Using \eqref{eq:currentwithoutghost}, we simply append a pair of scalars to find
    \begin{equation}
    \begin{split}\label{eq:divergenceimpropervertex}
        &\{ Q_\BRST^\swthree\, \mathcal{S}_{\mathrm{cl}}, \mathtt{\widetilde{e}}_k \}_\star 
        \odot_\star \mathtt{e}^{p_1} \odot_\star \mathtt{\bar e}^{p_2} \\[4pt]
      & \hspace{1cm}  =  \int_{k_1, k_2}\, (2\pi)^4\, \delta(k + k_1 + k_2)\, (-\mathrm{i}\, k_{\mu})\, \mathcal{J}^{\star \mu}_2(k_1,k_2) \odot_\star \mathtt{e}^{p_1} \odot_\star \mathtt{\bar e}^{p_2}\\
        &\hspace{2cm}
        +
        \int_{k_1,k_2, k_3}\, (2 \pi)^4\,\delta(k + k_1 + k_2 + k_3)\, (-\mathrm{i}\, k_{\mu})\, \mathcal{J}^{\star \mu}_3(k_1,k_2,k_3)\odot_\star \mathtt{e}^{p_1} \odot_\star \mathtt{\bar e}^{p_2} \ .
    \end{split}
    \end{equation}
    
    We introduce the one-form $C^{\star} \in \Omega^1\big((\mathbbm{R}^{1,3})^{\times 3}\big)$ with momentum space components
\begin{equation}
    \begin{split}\label{eq:LHSwardtaka}
 \hspace{-1mm}  C^{\star \mu}(k,p_1,p_2) & := 
        \int_{k_1, k_2}\, (2\pi)^4\, \delta(k + k_1 + k_2)\, \mathsf{P}_{\hbar,\mathrm{int}}' 
        \big(
        g\,\mathcal{J}^{\star \mu}_2(k_1,k_2) \odot_\star \mathtt{e}^{p_1} \odot_\star \mathtt{\bar e}^{p_2}\big)
        \\
        &\quad \
        +
        \int_{k_1, k_2, k_3}\, (2 \pi)^4\,\delta(k + k_1 + k_2 + k_3)\, \mathsf{P}_{\hbar,\mathrm{int}}' \big(
        g\,\mathcal{J}^{\star \mu}_3(k_1,k_2,k_3)\odot_\star \mathtt{e}^{p_1} \odot_\star \mathtt{\bar e}^{p_2} \big)
        \ ,
    \end{split}
    \end{equation}
whose codifferential $-\ii\,k_\mu\,C^{\star\mu}(k,p_1,p_2)$ is the projection under $\mathsf{P}'_{\hbar,\mathrm{int}}$ of the right-hand side of \eqref{eq:divergenceimpropervertex}, times a factor of the electromagnetic coupling $g$.
    The integrand is a sum of a four-point and a five-point correlation function, which due to momentum conservation imposed by  the map $\mathsf{P}'_{\hbar,\mathrm{int}}$ comes with an overall delta-distribution $(2 \pi)^4\, \delta(p_1 + p_2 - k)$. Hence \eqref{eq:LHSwardtaka} can be viewed as a three-point diagram.

We now turn to the right-hand side of the Ward-Takahashi identity \eqref{eq:WTI_improper3vertex}.
    We first use the dual pairing between ghost field/antifields given by $\left\{ \mathtt{\widetilde{e}}_k, \mathtt{\widetilde{e}}^p \right\}_\star = (2 \pi)^4\, \delta(k-p)$, together with the matter BRST transformation from \eqref{eq:Qbrst}, to express
    \begin{equation}
    \begin{split}
        \{ \mathtt{\widetilde{e}}_{k}, Q_\BRST^\swthree\, \mathtt{e}^{p_1} \}_\star
        &= -\ii\,\int_{k_1,k_2}\, (2 \pi)^4\, \delta(k_1 + k_2 - p_1)\,  \mathtt{e}^{k_1}\, \{ \mathtt{\widetilde{e}}_k, \mathtt{\widetilde{e}}^{k_2}\}_\star\, \mathrm{e}^{-\mathrm{i}\, k_1 \cdot \theta\, k}\,
        \mathrm{e}^{\,\frac{\mathrm{i}}{2}\, k_1 \cdot \theta\, k_2}
        \\[4pt]
        &= - \mathrm{i}\, \mathrm{e}^{\,\frac{\mathrm{i}}{2}\, k \cdot \theta\, p_1}\, \mathtt{e}^{p_1-k}  \ ,
    \end{split}
    \end{equation}
    and similarly $\{ \mathtt{\widetilde{e}}_k, Q_\BRST^\swthree\, \mathtt{\bar e}^{p_2} \}_\star =  \mathrm{i}\, \mathrm{e}^{\,\frac{\mathrm{i}}{2}\, k\cdot \theta\, p_2 }\, \mathtt{\bar e}^{p_2-k}$.
    Since $\left\{ \mathtt{\widetilde{e}}_k, \mathtt{e}^{p_1} \right\}_\star = 0$, i.e. the ghost field acts trivially on antifields in $L[2]^0$, we get
    \begin{equation}
    \begin{split}
        \big\{ \mathtt{\widetilde{e}}_k, Q^\swthree_\BRST\, (\mathtt{e}^{p_1} \odot_\star \mathtt{\bar e}^{p_2}) \big\}_\star
        = \mathrm{i}\, \mathrm{e}^{\,\mathrm{i}\, k \cdot \theta\, p_1} \,
        \big( -\mathrm{e}^{-\frac{\mathrm{i}}{2}\, k \cdot \theta\, p_1}\,
        \mathtt{e}^{p_1-k} \odot_\star \mathtt{\bar e}^{p_2}
        +\mathrm{e}^{\,\frac{\mathrm{i}}{2}\, k\cdot \theta\, p_2 }\,
        \mathtt{e}^{p_1} \odot_\star  \mathtt{\bar e}^{p_2-k}
        \big)\ ,
    \end{split}
    \end{equation}
    where an extra $R$-matrix appears when commuting $\mathtt{\widetilde{e}}_k$ with $\mathtt{e}^{p_1}$.
    
    It is now just a matter of taking the interacting projection $\mathsf{P}'_{\hbar,\mathrm{int}}$ to find the right hand side of the Ward-Takahashi identity from
    \begin{equation}\label{eq:RHSwardtaka}
    \begin{split}
        \mathsf{{P}}'_{\hbar,\mathrm{int}} \big(\{ \mathtt{\widetilde{e}}_k, Q^\swthree_\BRST\, (\mathtt{e}^{p_1} \odot_\star \mathtt{\bar e}^{p_2}) \}_\star\big)
        &= \mathrm{i}\,\mathrm{e}^{\,\mathrm{i}\, k \cdot \theta\, p_1} \,
         \big(\mathrm{e}^{\,\frac{\mathrm{i}}{2}\, k \cdot \theta\, p_2}\,
        \mathsf{\tilde{G}}^{\hbar,\mathrm{int}}_{\phi, \bar \phi}(p_1, p_2 - k) \\
        & \hspace{4cm} - \mathrm{e}^{-\frac{\mathrm{i}}{2}\, k \cdot \theta\, p_1}\,
        \mathsf{\tilde{G}}^{\hbar,\mathrm{int}}_{\phi, \bar \phi}(p_1 - k, p_2)
 \big) \ ,
 \end{split}
    \end{equation}
    where the full scalar propagator is defined as $\mathsf{\tilde{G}}^{\hbar,\mathrm{int}}_{\phi, \bar \phi}(p_1, p_2) := \mathsf{P}'_{\hbar,\mathrm{int}}(\mathtt{e}^{p_1} \odot_\star \mathtt{\widetilde{e}}^{p_2})$.
By combining \eqref{eq:LHSwardtaka} and \eqref{eq:RHSwardtaka} into the braided Ward-Takahashi identity \eqref{eq:WTI_improper3vertex} we find
    \begin{equation}
    \begin{split}\label{eq:takahashi_notinverse}
        & k_\mu\, C^{\star \mu}(k,p_1,p_2) 
        = \ii\,\hbar\, g\,\mathrm{e}^{- \frac{\mathrm{i}}{2}\, p_1 \cdot \theta\, p_2}\,
        \big( 
        \mathsf{\tilde{G}}^{\hbar,\mathrm{int}}_{\phi, \bar \phi}(p_1, p_2 - k)
        -
        \mathsf{\tilde{G}}^{\hbar,\mathrm{int}}_{\phi, \bar \phi}(p_1 - k, p_2)
        \big)\ ,
    \end{split}
    \end{equation}
where we used momentum conservation. This is the braided deformation of the commutative Ward identity for the electromagnetic vertex, which we have obtained as a direct consequence of the homological perturbation lemma.

The identity \eqref{eq:takahashi_notinverse} involves all diagrams. We can extract a simpler identity for the fully amputated exact three-point vertex function $\Gamma_\star^\mu(p_1,p_2)$, which involves only 1PI contributions and recovers the textbook-type analytical expression. This is
thought of as the one-form $C^{\star}$ amputated of its two remaining external scalar legs which come from combining the explicit scalar antifields on the right-hand side of \eqref{eq:LHSwardtaka} with the scalar antifields in the current $\CJ^\star$ to generate propagators. For this, we dress the  full propagator \smash{$\mathsf{\tilde{G}}^{\hbar,\mathrm{int}}_{\phi, \bar \phi}(p_1, p_2)$} with the scalar self-energy $\Sigma_\star(p)$ at all loops:
    \begin{equation}
        \mathsf{\tilde{G}}^{\hbar,\mathrm{int}}_{\phi, \bar \phi}(p_1, p_2)
        := - \mathrm{i}\, \hbar\, (2 \pi)^4\, \delta(p_1+p_2)\ 
        \frac{1}{p_1^2 - m^2 - \Sigma_\star(p_1)} \ ,
    \end{equation}
and write
\begin{align}
C^{\star\mu}(k,p_1,p_2) = (-\ii\,\hbar)^2\,(2\pi)^4\,\delta(p_1+p_2-k) \ \frac{\Gamma_\star^\mu(p_1,p_2)}{\big(p_1^2 - m^2 - \Sigma_\star(p_1)\big)\,\big(p_2^2 - m^2 - \Sigma_\star(p_2)\big)} \ .
\end{align}

Upon substitution into \eqref{eq:takahashi_notinverse} we then find
\begin{align} \label{eq:WTamputated}
 (p_1+p_2)_\mu\,\Gamma_\star^\mu(p_1,p_2) = g \, \e^{-\frac\ii2\,p_1\cdot\theta\,p_2}\, \big[\big(p_1^2 - \Sigma_\star(p_1)\big) - \big(p_2^2 - \Sigma_\star(p_2)\big)\big] \ .
\end{align}
The identity \eqref{eq:WTamputated} involves only 1PI diagrams.
It relates the full amputated three-vertex function $\Gamma_\star^\mu(p_1,p_2) $ to the full scalar propagator \smash{$\big(p^2-m^2-\Sigma_\star(p)\big)^{-1}$},
which in the commutative limit $\theta=0$ matches Takahashi's identity in ordinary scalar QED.

We can easily illustrate the identity \eqref{eq:WTamputated} to leading orders in homological perturbation theory using the explicit results of \cref{sub:scalar_self_energy} and \cref{app:3_point_vertex_correction}, understood formally throughout with reflection symmetric cutoff regularisation of all loop integrals.
At tree-level, from \eqref{eq:3vertextree} we find
\begin{align}
\Gamma_\star^\mu(p_1,p_2)^\swzero = g \, \e^{-\frac\ii2\,p_1\cdot\theta\,p_2} \, (p_1-p_2)^\mu \ ,
\end{align}
whose codifferential is
\begin{align}
(p_1+p_2)_\mu\,\Gamma_\star^\mu(p_1,p_2)^\swzero = g\,\e^{-\frac\ii2\,p_1\cdot\theta\,p_2}\,\big(p_1^2-p_2^2\big) \ .
\end{align}
Since $\Sigma_\star(p)^\swzero=0$ from \eqref{eq:bare scal prop}, this is indeed \eqref{eq:WTamputated} at tree-level.

At one-loop order, after simplifying the integrands in \eqref{eq:3vertex1loop} using the transverse projector $\tilde{\mathsf{\Pi}}_{\lambda\alpha}(q)$ we find
\begin{equation}
\begin{split}
    \Gamma^\mu_\star(p_1,p_2)^\swone 
    & = - 4 \, \mathrm{i}\, \hbar \,g^3 \,
    \e^{-\frac\ii2\,p_1\cdot\theta\,p_2}  \, 
    \bigg(
    \int_q\, 
    p_1^\lambda \,
    p_2^\alpha \,
    (-2q+p_1-p_2 )^\mu\, \\
    & \hspace{6.5cm} \times \mathsf{\tilde G}(q-p_1) \,\mathsf{\tilde G}(q +p_2) \, \mathsf{\tilde D}(q) \, \mathsf{\tilde\Pi}_{\lambda\alpha}(q)
    \\
    & \hspace{4cm} + 
    \int_q\,
    \eta^{\lambda \mu} \,
    \left(
    p_1^{\alpha} \,
    \mathsf{\tilde{G}}(q + p_1) \,
    \mathsf{\tilde{D}}(q)\, \mathsf{\tilde\Pi}_{\lambda \alpha}(q) \right. \\
    & \hspace{6.5cm} \left.
    - \,
    p_2^\alpha\,
    \mathsf{\tilde{G}}(q + p_2)\,
    \mathsf{\tilde{D}}(q)\, \mathsf{\tilde\Pi}_{\lambda \alpha}(q)
    \right) \bigg) \ .
\end{split}
\end{equation}
We contract this with the external momentum $(p_1 + p_2)_\mu$ and use the identity
\begin{align}
(p_1+p_2)\cdot (-2 q + p_1 - p_2) = \mathsf{\tilde{G}}(p_1 - q)^{-1} - \mathsf{\tilde{G}}(p_2 + q)^{-1} \ .
\end{align} 
Using the reflection symmetry $q \longmapsto -q$ of the integral, together with the fact that all propagators are even functions, we then find that the cross-terms proportional to $p_1^\lambda \, p_2^{\alpha}$ and $p_1^\alpha\, p_2^\lambda$ cancel between the two integrals.

The codifferential of the three-vertex at one-loop reduces to
\begin{equation}
\begin{split}
    (p_1+p_2)_\mu\, \Gamma^{\mu}_\star(p_1,p_2)^\swone
    &=
    - 4\, \mathrm{i}\, \hbar \,g^3 \,
    \e^{-\frac\ii2\,p_1\cdot\theta\,p_2} \\
    & \hspace{2cm} \times
    \int_q\,
    \left(
    p_1^\lambda\, p_1^\alpha\,
    \mathsf{\tilde{G}}(q + p_1) \,
    - \,
    p_2^\lambda\, p_2^\alpha\,
    \mathsf{\tilde{G}}(q + p_2)
    \right) \, 
    \mathsf{\tilde{D}}(q)\, \mathsf{\tilde\Pi}_{\lambda\alpha }(q) 
    \ .
\end{split}
\end{equation}
By rewriting the one-loop scalar self-energy from \eqref{eq:scal_Self_energy} as
\begin{equation}
    \Sigma_\star(p)^\swone = 4\,\mathrm{i} \,\hbar\, g^2\, \int_q\, 
    \, p^{\lambda}\, p^{\alpha}\, \tilde{\sf{G}} (q + p)\, \tilde{\sf D}(q)\, \tilde{\mathsf{\Pi}}_{\lambda\alpha} (q) + 3\,\ii\,\hbar\,g^2\,\int_q \, \mathsf{\tilde{D}}(q) 
    \ ,
\end{equation}
it follows that
\begin{equation}
    (p_1+p_2)_\mu\, \Gamma^{\mu}_\star(p_1,p_2)^\swone
    = -g\, \e^{-\frac\ii2\,p_1\cdot\theta\,p_2}  \, 
    \big(\Sigma_\star(p_1)^\swone - \Sigma_\star(p_2)^\swone\big)\ .
\end{equation}
This is indeed \eqref{eq:WTamputated} at one-loop order.

\paragraph{Photon Propagator.} 

Consider now the Ward-Takahashi identity \eqref{eq:WTInonanomalous} for an insertion of a single photon antifield $\mathcal{O} = \mathtt{v}_\nu\, \mathtt{e}^{p_2} \in L[2]^0$.
Since the cubic part of the BRST operator acts trivially on photon antifields, this example yields the simple identity
\begin{equation}\label{eq:WT_photonprop}
    \mathsf{P}'_{\hbar,\mathrm{int}} \big(\{ Q_\BRST^\swthree\, \mathcal{S}_{\mathrm{cl}}, \mathtt{\tilde{e}}_{p_1} \}_\star \odot_\star \mathtt{v}_\nu\, \mathtt{e}^{p_2}\big)  = 0\ .
\end{equation}
Introducing the rank two tensor $D^\star$ on $\mathbbm{R}^{1,3}\times\FR^{1,3}$ with momentum space components
\begin{equation}
\begin{split}
    D^{\star \mu}_\nu(p_1,p_2)
    & := 
        \int_{k_1, k_2}\, (2\pi)^4\, \delta(p_1 + k_1 + k_2)\, \mathsf{P}_{\hbar,\mathrm{int}}' 
        \big(
        g\,\mathcal{J}^{\star \mu}_2(k_1,k_2) \odot_\star \mathtt{v}_\nu\, \mathtt{e}^{p_2}\big)
        \\
        &\quad \
        +
        \int_{k_1, k_2, k_3}\, (2 \pi)^4\,\delta(p_1 + k_1 + k_2 + k_3)\, \mathsf{P}_{\hbar,\mathrm{int}}' \big(
        g\,\mathcal{J}^{\star \mu}_3(k_1,k_2,k_3)\odot_\star \mathtt{v}_\nu\,\mathtt{e}^{p_2} \big)
        \ ,
\end{split}
\end{equation}
then \eqref{eq:WT_photonprop} is written as
\begin{equation}\label{eq:WTI_photonprop}
    p_{1\,\mu} \, D^{\star \mu}_{\nu}(p_1,p_2) = 0\ .
\end{equation}

From \eqref{eq:current1} and \eqref{eq:current2} it follows that $D^\star$ is of order $g^2$. By interpreting it as a two-point correlation function with one external photon leg amputated, we write
\begin{equation}
    {D}^{\star \mu}_\nu(p_1,p_2) := \ii\,\hbar\,  (2 \pi)^4\, \delta(p_1 - p_2)\, \Pi_\star^{\mu\lambda}(p_1) \, \bigg(\frac1{\eta - \tilde{\mathsf{D}}(p_1)\,\tilde{\mathsf{\Pi}}(p_1)\,\Pi_\star(p_1)}\bigg)_{\lambda\rho} \, \tilde{\mathsf{D}}(p_2)\,\tilde{\mathsf{\Pi}}^\rho_\nu(p_2)\ ,
\end{equation}
where $\Pi_\star(p)$ is the photon vacuum polarisation tensor, which receives contributions from only 1PI diagrams and which we calculated at one-loop order in \cref{sub:photon_polarisation}.
Then the identity \eqref{eq:WTI_photonprop} implies that the photon vacuum polarisation is transverse at all loop orders:
\begin{equation}
    p_\mu\, \Pi_\star^{\mu \nu}(p) = 0\ .
\end{equation}
In other words, the Ward identity holds for braided scalar QED at all loop orders.

%

\paragraph{More General Identities.} 

The calculations above can be readily extended to insertions of arbitrary numbers of conjugate pairs of scalar antifields as well as of photon antifields:
\begin{align}
    \mathcal{O} =\CO_{(m,n)} := \mathtt{v}_{\mu_2}\, \mathtt{e}^{p_2} \odot_\star \cdots \odot_\star \mathtt{v}_{\mu_m}\, \mathtt{e}^{p_m}
    \odot_\star
    \mathtt{e}^{q_1} \odot_\star \cdots \odot_\star \mathtt{e}^{q_n}
    \odot_\star
    \mathtt{\bar e}^{\bar q_1} \odot_\star \ldots \odot_\star \mathtt{\bar e}^{\bar q_n}\ .
\end{align}
We use $Q^\swthree_\BRST$ acting as a strict derivation trivially on photon antifields, and resolving the $R$-matrices on $n$ insertions of scalar antifields as well as on a product of conjugate scalar antifields.

Introduce the rank $m$ tensor $D_{(m,n)}^\star$ on $(\mathbbm{R}^{1,3})^{\times (m+2n)}$ with components
\begin{equation}
\begin{split}
 &   D^{\star \mu_1}_{(m,n)\mu_2\cdots\mu_m}(p_1,\dots,p_m;q_1,\dots,q_n;\bar q_1,\dots,\bar q_n) \\[4pt]
& \hspace{2cm}    := 
        \int_{k_1, k_2}\, (2\pi)^4\, \delta(p_1 + k_1 + k_2)\, \mathsf{P}_{\hbar,\mathrm{int}}' 
        \big(
        g\,\mathcal{J}^{\star \mu}_2(k_1,k_2) \odot_\star \CO_{(m,n}\big)
        \\
        &\hspace{3cm}
        +
        \int_{k_1, k_2, k_3}\, (2 \pi)^4\,\delta(p_1 + k_1 + k_2 + k_3)\, \mathsf{P}_{\hbar,\mathrm{int}}' \big(
        g\,\mathcal{J}^{\star \mu}_3(k_1,k_2,k_3)\odot_\star \CO_{(m,n)} \big)
        \ ,
\end{split}
\end{equation}
with labels of incoming momenta $\{p_i\}_{i=1}^m$, $\{q_i\}_{i=1}^n$ and $\{\bar q_i\}_{i=1}^n$  for photons, scalars and conjugate scalars respectively. Its codifferential \smash{$-\ii\,p_{1\,\mu_1}\, D^{\star \mu_1}_{(m,n)\mu_2\cdots\mu_m}$} is the left-hand side of the Ward-Takahashi identity times a factor of $g$. Below we will use the shorthand  $p_{1\cdots k}=p_1+\cdots+p_k$ for sums of momenta.

We obtain a relation between $D^\star_{(m,n)}$ and a sum of exact correlation functions of lower order.
Denote the exact correlation function of $m$ photons and $n$ conjugate pairs of scalars as
\begin{equation}
\begin{split}
& \tilde{\mathsf{G}}^\star_{(m,n) \mu_1\cdots \mu_m}(p_1, \dots,p_m;q_1, \ldots,q_n;\bar q_1,\ldots,\bar q_n) \\[4pt]
    &\hspace{2cm} :=
    \mathsf{P}'_{\hbar,\mathrm{int}}\big(\mathtt{v}_{\mu_1}\, \mathtt{e}^{p_1} \odot_\star \cdots \odot_\star \mathtt{v}_{\mu_m}\, \mathtt{e}^{p_m}
    \odot_\star
    \mathtt{e}^{q_1} \odot_\star \cdots \odot_\star \mathtt{e}^{q_n}
    \odot_\star
    \mathtt{\bar e}^{\bar q_1} \odot_\star \cdots \odot_\star \mathtt{\bar e}^{\bar q_n}\big)\ .
\end{split}
\end{equation}
Then the general Ward-Takahashi identity \eqref{eq:WTInonanomalous} reads as
\begin{equation}
\small
\begin{split}\label{eq:General_WTI}
    &
     p_{1\,\mu_1}\,D^{\star \mu_1}_{(m,n)\mu_2 \cdots \mu_m }
    (p_1, \ldots, p_m; q_1, \ldots, q_n; \bar q_1, \ldots, \bar q_n) \\[4pt]
&    = - \ii\,\hbar\, g\,  \mathrm{e}^{\,\mathrm{i}\, p_1 \cdot \theta\, p_{2 \cdots m}} \, \bigg(
    \sum_{j=1}^n\, \mathrm{e}^{\,\mathrm{i}\, p_1 \cdot \theta\, ( q_{1 \cdots (j-1)} + \frac{1}{2}\, q_j )} \\
    & \hspace{5cm} \times
    \tilde{\mathsf{G}}^\star_{(m-1,n)\mu_2 \cdots \mu_m }
    (p_2, \ldots, p_m; q_1, \ldots, q_j - p_1, \ldots, q_n; \bar q_1, \ldots, \bar q_n) \\
    &\hspace{3.5cm}
    - \mathrm{e}^{\mathrm{i}\, p_1 \cdot \theta\, q_{1 \cdots n}} \,
    \sum_{j=1}^n\, \mathrm{e}^{\mathrm{i}\, p_1 \cdot \theta\, ( \bar q_{1 \cdots (j-1)} + \frac{1}{2}\, \bar q_j)} \\
    & \hspace{5cm} \times
    \tilde{\mathsf{G}}^\star_{(m-1,n)\mu_2 \cdots \mu_m }
    (p_2, \ldots, p_m; q_1, \ldots, q_n; \bar q_1,\ldots, \bar q_j-p_1, \ldots, \bar q_n)
    \bigg) \ .
\end{split}
\normalsize
\end{equation}

Graphically, the general form of the non-anomalous Ward-Takahashi identity \eqref{eq:General_WTI} is
\begin{equation}
\small
\begin{split}\label{eq:WTI_generaldiagram}
    &  p_{1\,\mu_1}\,
    \left( \
    \begin{tikzpicture}[scale=0.8,baseline]
    \scriptsize
        \coordinate (C) at (0,0);
        \draw[Aplus] ($(C) + (200:1) $) node[left]{$\mu_1 \ $} node{$\bullet$} -- (C);
        \draw[Aplus] ($(C) + (150:2) $) node[left]{$p_2,\mu_2$} -- (C);
        \node at ($(C) + (125:1.7)$){$.^{.^.}$};
        \draw[Aplus] ($(C) + (100:2) $) node[above]{$p_m,\mu_m$} -- (C);
        \draw[scalplus] (C) -- ($(C) + (60:2) $) node[above]{$\ q_1$};
        \node at ($(C) + (45:1.7)$){$._{._.}$};
        \draw[scalplus] (C) -- ($(C) + (30:2) $) node[right]{$q_n$};
        \draw[scalplusreverse] (C) -- ($(C) + (-30:2) $) node[right]{$\bar q_1$};
        \node at ($(C) + (-45:1.7)$){$.^{.^.}$};
        \draw[scalplusreverse] (C) -- ($(C) + (-60:2) $) node[below]{$\bar q_n$};
        \draw[fill=white] (C) circle[radius=.8] node{$D^\star$};
    \end{tikzpicture} 
    \ \right) 
    \\[4pt]
    & \hspace{1cm}
    = - \ii\, \hbar\,g\, \mathrm{e}^{\,\mathrm{i}\, p_1 \cdot \theta\, p_{2 \cdots m}} \, 
    \left(\
    \sum_{j=1}^n\, \mathrm{e}^{\,\mathrm{i}\, p_1 \cdot \theta\, ( q_{1 \cdots (j-1)} + \frac{1}{2}\, q_j )} \
    \begin{tikzpicture}[scale=0.7,baseline]
    \scriptsize
        \coordinate (C) at (0,0);
        \draw[Aplus] ($(C) + (210:2) $) node[left]{$p_2,\mu_2$} -- (C);
        \node at ($(C) + (175:1.7)$){$\vdots$};
        \draw[Aplus] ($(C) + (150:2) $) node[left]{$p_m,\mu_m$} -- (C);
        \draw[scalplus] (C) -- ($(C) + (120:2) $);
        \node at ($(C) + (105:1.8)$){$\ldots$};
        \draw[scalplus] (C) -- ($(C) + (90:2) $) node[above]{$q_j - p_1$};
        \node at ($(C) + (75:1.8)$){$\ldots$};
        \draw[scalplus] (C) -- ($(C) + (60:2) $);
        \draw[scalplusreverse] (C) -- ($(C) + (-60:2) $) node[below]{$\bar q_1$};
        \node at ($(C) + (-90:1.7)$){$\ldots$};
        \draw[scalplusreverse] (C) -- ($(C) + (-120:2) $) node[below]{$\bar q_n$};
        \draw[fill=white] (C) circle[radius=.8] node{$\tilde{\mathsf{G}}^\star$};
    \end{tikzpicture} 
    \right. \\
    & \hspace{5cm} \left.
    - \, \mathrm{e}^{\,\mathrm{i}\, p_1 \cdot \theta\, q_{1 \cdots n}} \, 
    \sum_{j=1}^n\, \mathrm{e}^{\,\mathrm{i}\, p_1 \cdot \theta\, ( \bar q_{1 \cdots (j-1)} + \frac{1}{2}\, \bar q_j)} \
    \begin{tikzpicture}[scale=0.7,baseline]
    \scriptsize
        \coordinate (C) at (0,0);
        \draw[Aplus] ($(C) + (210:2) $) node[left]{$p_2,\mu_2$} -- (C);
        \node at ($(C) + (175:1.7)$){$\vdots$};
        \draw[Aplus] ($(C) + (150:2) $) node[left]{$p_m,\mu_m$} -- (C);
        \draw[scalplus] (C) -- ($(C) + (120:2) $) node[above]{$q_1$};
        \node at ($(C) + (90:1.7)$){$\ldots$};
        \draw[scalplus] (C) -- ($(C) + (60:2) $) node[above]{$q_n$};
        \draw[scalplusreverse] (C) -- ($(C) + (-120:2) $);
        \node at ($(C) + (-105:1.7)$){$\ldots$};
        \draw[scalplusreverse] (C) -- ($(C) + (-90:2) $) node[below]{$\bar q_j - p_1$};
        \node at ($(C) + (-75:1.7)$){$\ldots$};
        \draw[scalplusreverse] (C) -- ($(C) + (-60:2) $);
        \draw[fill=white] (C) circle[radius=.8] node{$\tilde{\mathsf{G}}^\star$};
    \end{tikzpicture}
    \ \right) \ .
    \normalsize
\end{split}
\end{equation}

In this diagram we denoted with a bullet point on the external photon leg labelled by  $(\mu_1,p_1)$ the amputation by the free photon propagator.
The half-phases appearing on the right-hand side in these formulas are results of the operator $\{ \mathtt{\tilde{e}}_{p_1},  - \}_\star\,Q^\swthree_\BRST $ acting on the scalar legs and their conjugates.
The overall phase $\mathrm{e}^{\,\mathrm{i}\, p_1 \cdot \theta\, p_{2 \cdots m}}$ stems from $Q^\swthree_\BRST$ acting trivially on photon antifields, so the operator $\{ \mathtt{\tilde{e}}_{p_1},   - \}_\star\,Q^\swthree_\BRST$ simply jumps over the photon legs.
The full phases are the braiding due to this operator moving past each leg in succession.

\newpage

\label{Bibliography}
\lhead{\emph{Bibliography}}  
\bibliographystyle{ourstyle}  
\bibliography{bsqed.bib}

\end{document}